\documentclass[12pt]{article}
\pdfoutput=1
\usepackage[]{hyperref}
\usepackage[table]{xcolor}
\usepackage[]{putex}
\usepackage[utf8]{inputenc}

\usepackage{amsmath,amsthm,amssymb}
\usepackage{tikz}
\usepackage{tikz-cd}
\usetikzlibrary{decorations.markings}
\usetikzlibrary{calc}
\usepackage{subcaption}
\usepackage{enumerate}
\usepackage{enumitem}
\usepackage{ifthen}
\usepackage{xifthen}
\usepackage{cite}
\usepackage{empheq}
\usepackage[theorems,skins]{tcolorbox}

\newtcbox{\narrowtcboxmath}[1][]{nobeforeafter, math upper, tcbox raise base, 
          enhanced, rounded corners,left=0em, top=.5em, right=0em, bottom=.5em}

\numberwithin{equation}{section}

\newcommand\Uchan{\cellcolor{brown!30}}
\newcommand\Vchan{\cellcolor{blue!20}}
\newcommand\Dchan{\cellcolor{red!50}}

\usetikzlibrary{tikzmark, calc}
\tikzset{
    double color fill/.code 2 args={
        \pgfdeclareverticalshading[%
            tikz@axis@top,tikz@axis@middle,tikz@axis@bottom%
        ]{diagonalfill}{100bp}{%
            color(0bp)=(tikz@axis@bottom);
            color(50bp)=(tikz@axis@bottom);
            color(50bp)=(tikz@axis@middle);
            color(50bp)=(tikz@axis@top);
            color(85bp)=(tikz@axis@top)
        }
        \tikzset{shade, left color=#1, right color=#2, shading=diagonalfill}
    }
}
\newcommand\BGcell[4][0pt]{%
  \begin{tikzpicture}[overlay,remember picture]%
    \path[#4] ( $ (pic cs:#2) + (0\tabcolsep,2.6ex) $ ) rectangle ( $ (pic cs:#3) +  (1.22*\tabcolsep,-#1*\baselineskip-1ex) $ );
  \end{tikzpicture}%
}%
\newcounter{BGnum}
\newcommand\cellBG[3]{
    \multicolumn{1}{
        !{\BGcell{startBG\arabic{BGnum}}{endBG\arabic{BGnum}}{%
                #1}
            \tikzmark{startBG\arabic{BGnum}}}
            #2
        !{\tikzmark{endBG\arabic{BGnum}}}}
        {#3} 
      \addtocounter{BGnum}{1}
}
\usepackage{dcolumn, array, booktabs}

\usetikzlibrary{arrows.meta}

\newcommand\tikzmarkAlt[2]{%
\tikz[remember picture,overlay] 
\node[inner sep=0pt,outer sep=2pt] (#1){#2};%
}
\newcommand\link[3]{%
\begin{tikzpicture}[remember picture, overlay, >=stealth, shorten >= 1pt]
  \draw[-, thick, #3, postaction={decorate,decoration={markings,
        mark=between positions 0 and 1 step 19pt with {\draw[thin,#3, fill=#3] circle (2.5pt);} }}] (#1.north) to  (#2.north);
\end{tikzpicture}%
}
\newcommand\linkhor[3]{%
\begin{tikzpicture}[remember picture, overlay, >=stealth, shorten >= 1pt]
  \draw[-, thick, #3, postaction={decorate,decoration={markings,
        mark=between positions 0 and 1 step 23pt with {\draw[thin,#3, fill=#3] circle (2.5pt);} }}] (#1.north) to  (#2.north);
\end{tikzpicture}%
}
\newcommand\linkarrow[3]{%
\begin{tikzpicture}[remember picture, overlay, >=stealth, shorten >= 1pt]
  \draw[->, dotted, thick, #3] (#1) to  (#2);
\end{tikzpicture}%
}
 \newcommand\linkarcCW[3]{%
  \begin{tikzpicture}[overlay, remember picture, yshift=.25\baselineskip, shorten >=.5pt, shorten <=.5pt] %
     \draw[{Circle[#3,length=5pt]}-{Circle[#3,length=8pt]}, thick, #3, postaction={decorate,decoration={markings,
        mark= at position 0.5 with {\arrow{>}} }}] (#1) to [bend left]  (#2);
  \end{tikzpicture}%
  }
  \newcommand\linkarcCWterminal[3]{%
  \begin{tikzpicture}[overlay, remember picture, yshift=.25\baselineskip, shorten >=.5pt, shorten <=.5pt]%
     \draw[{Circle[#3,length=5pt]}-{Circle[#3,length=8pt]}, thick, #3, postaction={decorate,decoration={markings,
        mark= at position 0.5 with {\arrow{>}} }}] (#1) to [bend left]  (#2);
  \end{tikzpicture}%
  }
 \newcommand\linkarcACW[3]{%
  \begin{tikzpicture}[overlay, remember picture, yshift=.25\baselineskip, shorten >=.5pt, shorten <=.5pt]
     \draw[{Circle[#3,length=5pt]}-, thick, #3, postaction={decorate,decoration={markings,
        mark= at position 0.5 with {\arrow{>}} }}] (#1) to [bend right]  (#2);
  \end{tikzpicture}%
  }

\makeatletter
\DeclareRobustCommand*{\bfseries}{%
  \not@math@alphabet\bfseries\mathbf
  \fontseries\bfdefault\selectfont
  \boldmath
}
\makeatother
\hypersetup{
  colorlinks=true,
  linkcolor=blue,
  citecolor=green!60!black,
  urlcolor=cyan!80!black,
  linktoc=all
  }

\input{glyphtounicode}
\def\musepic#1{\vcenter{\hbox{\usebox{#1}}}}

\newsavebox{\figNCombChannel}
\savebox{\figNCombChannel}{%
\begin{tikzpicture}[scale=.95]
    \tikzstyle{vint}=[draw,scale=0.3,color=blue,fill=blue,circle]
    \coordinate (y1) at (-3/2,0);
    \coordinate (y2) at (-1,0);
    \coordinate (x3) at (-1,1);
    \coordinate (y3) at (0,0);
    \coordinate (z3) at (0,1/2);
    \coordinate (x4) at (0,1);
    \coordinate (y4) at (1,0);
    \coordinate (z4) at (1,1/2);
    \coordinate (x5) at (1,1);
    \coordinate (y5) at (2,0);
    \coordinate (x6) at (2,1);
    \coordinate (y6) at (5/2,0);
	\draw[very thick, line cap=round, dash pattern=on 0 off 8] (z3)--(z4);
	\draw[thick] (y1)--(y2)--(x3);
	\draw[thick, red] (y2)--(y3);
	\draw[thick] (y3)--(x4);
	\draw[thick, red, dashed] (y3)--(y4);
	\draw[thick] (y4)--(x5);
	\draw[thick, red] (y4)--(y5);
	\draw[thick] (y5)--(x6);	
	\draw[thick] (y5)--(y6);
	\draw (y1) node[anchor=east] {$\Delta_{1}$};
	\draw (x3) node[anchor=south] {$\Delta_2$};
	\draw (x4) node[anchor=south] {$\Delta_3$};
	\draw (x5) node[anchor=south] {$\Delta_{n-2}$};
	\draw (x6) node[anchor=south] {$\Delta_{n-1}$};
	\draw (y6) node[anchor=west] {$\Delta_{n}$};
	\draw[red] ($(y2)!0.5!(y3)$) node[anchor=north] {$\Delta_{{\delta_1}}$};
	\draw[red] ($(y4)!0.5!(y5)$) node[anchor=north] {$\Delta_{{\delta_{n-3}}}$};
	\draw (y2) node[vint] {};
	\draw (y3) node[vint] {};
	\draw (y4) node[vint] {};
	\draw (y5) node[vint] {};
\end{tikzpicture}}
\newsavebox{\figSevenChannel}
\savebox{\figSevenChannel}{%
\begin{tikzpicture}[scale=.8]
    \tikzstyle{vint}=[draw,scale=0.3,color=blue,fill=blue,circle]
    \coordinate (y1) at (-3/2,0);
    \coordinate (y2) at (-1,0);
    \coordinate (x3) at (-1,1);
    \coordinate (y3) at (0,0);
    \coordinate (z3) at (0,1/2);
    \coordinate (x4) at (0,1);
    \coordinate (y4) at (1,0);
    \coordinate (z4) at (1,1/2);
    \coordinate (x5) at (1,1);
    \coordinate (y5) at (2,0);
    \coordinate (x6) at (2,1);
    \coordinate (y6) at (5/2,0);
    \coordinate (w3) at (1/2,3/2);
    \coordinate (w4) at (3/2,3/2);
	\draw[thick] (y1)--(y2)--(x3);
	\draw[thick, red] (y2)--(y3);
	\draw[thick] (y3)--(x4);
	\draw[thick, red] (y3)--(y4);
	\draw[thick, red] (y4)--(x5);
	\draw[thick, red] (y4)--(y5);
	\draw[thick] (y5)--(x6);	
	\draw[thick] (y5)--(y6);
	\draw[thick] (w3)--(x5)--(w4);
	\draw (y1) node[anchor=east] {$\Delta_{1}$};
	\draw (x3) node[anchor=south] {$\Delta_2$};
	\draw (x4) node[anchor=south] {$\Delta_3$};
	\draw[red] ($(y4)!0.5!(x5)$) node[anchor=west] {$\Delta_{\delta_3}$};
	\draw (w3) node[anchor=south] {$\Delta_{4}$};
	\draw (w4) node[anchor=south] {$\Delta_{5}$};
	\draw (x6) node[anchor=south] {$\Delta_{6}$};
	\draw (y6) node[anchor=west] {$\Delta_{7}$};
	\draw[red] ($(y2)!0.5!(y3)$) node[anchor=north] {$\Delta_{{\delta_1}}$};
	\draw[red] ($(y3)!0.5!(y4)$) node[anchor=north] {$\Delta_{\delta_2}$};
	\draw[red] ($(y4)!0.5!(y5)$) node[anchor=north] {$\Delta_{{\delta_{4}}}$};
	\draw (y2) node[vint] {};
	\draw (y3) node[vint] {};
	\draw (y4) node[vint] {};
	\draw (y5) node[vint] {};
    \draw (x5) node[vint] {};
\end{tikzpicture}}
\newsavebox{\figNOPEChannel}
\savebox{\figNOPEChannel}{%
\begin{tikzpicture}[scale=.9]
    \tikzstyle{vint}=[draw,scale=0.3,color=blue,fill=blue,circle]
    \coordinate (y1) at (-3/2,-1/2);
    \coordinate (y2) at (-1,0);
    \coordinate (x3) at (-3/2,1/2);
    \coordinate (y3) at (0,0);
    \coordinate (z3) at (0,1/2);
    \coordinate (x4) at (0,1);
    \coordinate (w1) at (-1/2,3/2);
    \coordinate (w2) at (1/2,3/2);
    \coordinate (y4) at (2,0);
    \coordinate (z4) at (2,1/2);
    \coordinate (x5) at (2,1);
    \coordinate (w3) at (3/2,3/2);
    \coordinate (w4) at (5/2,3/2);
    \coordinate (y5) at (3,0);
    \coordinate (x6) at (7/2,1/2);
    \coordinate (y6) at (7/2,-1/2);
    \draw[very thick, line cap=round, dash pattern=on 0 off 15] (z3)--(z4);
	\draw[thick] (y1)--(y2)--(x3);
	\draw[thick] (w1)--(x4)--(w2);
	\draw[thick, red] (y2)--(y3)--(x4);
	\draw[thick, red, dashed] (y3)--(y4);
	\draw[thick, red] (y4)--(x5);
	\draw[thick] (w3)--(x5)--(w4);
	\draw[thick, red] (y4)--(y5);
	\draw[thick] (y5)--(x6);	
	\draw[thick] (y5)--(y6);
	\draw (y1) node[anchor=east] {$\Delta_{1}$};
	\draw (x3) node[anchor=east] {$\Delta_2$};
	\draw (w1) node[anchor=south] {$\Delta_{3}$};
	\draw (w2) node[anchor=south] {$\Delta_4$};
	\draw (w3) node[anchor=south] {$\Delta_{n-3}$};
	\draw (w4) node[anchor=south] {$\Delta_{n-2}$};
	\draw[red] ($(y3)!0.5!(x4)$) node[anchor=east] {$\Delta_{\delta_2}$};
	\draw[red] ($(y4)!0.5!(x5)$) node[anchor=west] {$\Delta_{\delta_{n-4}}$};
	\draw (x6) node[anchor=west] {$\Delta_{n-1}$};
	\draw (y6) node[anchor=west] {$\Delta_{n}$};
	\draw[red] ($(y2)!0.5!(y3)$) node[anchor=north] {$\Delta_{{\delta_1}}$};
	\draw[red] ($(y4)!0.5!(y5)$) node[anchor=north] {$\Delta_{{\delta_{n-3}}}$};
	\draw (y2) node[vint] {};
	\draw (y3) node[vint] {};
	\draw (y4) node[vint] {};
	\draw (y5) node[vint] {};
	\draw (x4) node[vint] {};
    \draw (x5) node[vint] {};
\end{tikzpicture}}
\newsavebox{\figSevenMixedChannelMomentum}
\savebox{\figSevenMixedChannelMomentum}{%
\begin{tikzpicture}[scale=1]
    \tikzstyle{vint}=[draw,scale=0.3,color=blue,fill=blue,circle]
    \coordinate (y1) at (-3/2,0);
    \coordinate (y2) at (-1,0);
    \coordinate (x3) at (-1,1);
    \coordinate (y3) at (0,0);
    \coordinate (z3) at (0,1/2);
    \coordinate (x4) at (0,1);
    \coordinate (y4) at (1,0);
    \coordinate (z4) at (1,1/2);
    \coordinate (x5) at (1,1);
    \coordinate (w1) at (1/2,3/2);
    \coordinate (w2) at (3/2,3/2);
    \coordinate (y5) at (2,0);
    \coordinate (x6) at (2,1);
    \coordinate (y6) at (5/2,0);
	\draw[thick, decoration={markings, mark=at position 0.6 with {\arrow{>}}}, postaction={decorate}] (y1)--(y2);
	\draw[thick, decoration={markings, mark=at position 0.6 with {\arrow{<}}}, postaction={decorate}] (y2)--(x3);
	\draw[thick, red] (y2)--(y3);
	\draw[thick, decoration={markings, mark=at position 0.6 with {\arrow{<}}}, postaction={decorate}] (y3)--(x4);
	\draw[thick, red] (y3)--(y4);
	\draw[thick, decoration={markings, mark=at position 0.6 with {\arrow{>}}}, postaction={decorate}] (w1)--(x5);
	\draw[thick, decoration={markings, mark=at position 0.6 with {\arrow{<}}}, postaction={decorate}] (x5)--(w2);
	\draw[thick, red] (y4)--(x5);
	\draw[thick, red] (y4)--(y5);
	\draw[thick, decoration={markings, mark=at position 0.6 with {\arrow{<}}}, postaction={decorate}] (y5)--(x6);	
	\draw[thick, decoration={markings, mark=at position 0.6 with {\arrow{<}}}, postaction={decorate}] (y5)--(y6);
	\draw (y1) node[anchor=east] {$p_{1}$};
	\draw (x3) node[anchor=south] {$p_2$};
	\draw (x4) node[anchor=south] {$p_3$};
	\draw (w1) node[anchor=south] {$p_{4}$};
	\draw (w2) node[anchor=south] {$p_{5}$};
	\draw (x6) node[anchor=south] {$p_{6}$};
	\draw (y6) node[anchor=west] {$p_{7}$};
	\draw[red] ($(y2)!0.5!(y3)$) node[anchor=north] {$s_1$};
	\draw[red] ($(y3)!0.5!(y4)$) node[anchor=north] {$s_2$};
	\draw[red] ($(x5)!0.5!(y4)$) node[anchor=west] {$s_3$};
	\draw[red] ($(y4)!0.5!(y5)$) node[anchor=north] {$s_{4}$};
	\draw (y2) node[vint] {};
	\draw (y3) node[vint] {};
	\draw (y4) node[vint] {};
	\draw (y5) node[vint] {};
\end{tikzpicture}}
\newsavebox{\figNCombChannelMomentum}
\savebox{\figNCombChannelMomentum}{%
\begin{tikzpicture}[scale=1]
    \tikzstyle{vint}=[draw,scale=0.3,color=blue,fill=blue,circle]
    \coordinate (y1) at (-3/2,0);
    \coordinate (y2) at (-1,0);
    \coordinate (x3) at (-1,1);
    \coordinate (y3) at (0,0);
    \coordinate (z3) at (0,1/2);
    \coordinate (x4) at (0,1);
    \coordinate (y4) at (1,0);
    \coordinate (z4) at (1,1/2);
    \coordinate (x5) at (1,1);
    \coordinate (y5) at (2,0);
    \coordinate (x6) at (2,1);
    \coordinate (y6) at (5/2,0);
	\draw[very thick, line cap=round, dash pattern=on 0 off 8] (z3)--(z4);
	\draw[thick, decoration={markings, mark=at position 0.6 with {\arrow{>}}}, postaction={decorate}] (y1)--(y2);
	\draw[thick, decoration={markings, mark=at position 0.6 with {\arrow{<}}}, postaction={decorate}] (y2)--(x3);
	\draw[thick, red] (y2)--(y3);
	\draw[thick, decoration={markings, mark=at position 0.6 with {\arrow{<}}}, postaction={decorate}] (y3)--(x4);
	\draw[thick, red, dashed] (y3)--(y4);
	\draw[thick, decoration={markings, mark=at position 0.6 with {\arrow{<}}}, postaction={decorate}] (y4)--(x5);
	\draw[thick, red] (y4)--(y5);
	\draw[thick, decoration={markings, mark=at position 0.6 with {\arrow{<}}}, postaction={decorate}] (y5)--(x6);	
	\draw[thick, decoration={markings, mark=at position 0.6 with {\arrow{<}}}, postaction={decorate}] (y5)--(y6);
	\draw (y1) node[anchor=east] {$p_{1}$};
	\draw (x3) node[anchor=south] {$p_2$};
	\draw (x4) node[anchor=south] {$p_3$};
	\draw (x5) node[anchor=south] {$p_{n-2}$};
	\draw (x6) node[anchor=south] {$p_{n-1}$};
	\draw (y6) node[anchor=west] {$p_{n}$};
	\draw[red] ($(y2)!0.5!(y3)$) node[anchor=north] {$s_1$};
	\draw[red] ($(y4)!0.5!(y5)$) node[anchor=north] {$s_{n-3}$};
	\draw (y2) node[vint] {};
	\draw (y3) node[vint] {};
	\draw (y4) node[vint] {};
	\draw (y5) node[vint] {};
\end{tikzpicture}}
\newsavebox{\figNOPEChannelMomentum}
\savebox{\figNOPEChannelMomentum}{%
\begin{tikzpicture}[scale=.95]
    \tikzstyle{vint}=[draw,scale=0.3,color=blue,fill=blue,circle]
    \coordinate (y1) at (-3/2-2,-1/2);
    \coordinate (y2) at (-1-2,0);
    \coordinate (x3) at (-3/2-2,1/2);
    \coordinate (y3) at (0-2,0);
    \coordinate (z3) at (0,1/2);
    \coordinate (x4) at (0-2,1);
    \coordinate (w1) at (-1/2-2,3/2);
    \coordinate (w2) at (1/2-2,3/2);
    \coordinate (y4) at (2,0);
    \coordinate (z4) at (2,1/2);
    \coordinate (x5) at (2,1);
    \coordinate (w3) at (3/2,3/2);
    \coordinate (w4) at (5/2,3/2);
    \coordinate (y5) at (3,0);
    \coordinate (x6) at (7/2,1/2);
    \coordinate (y6) at (7/2,-1/2);
    \coordinate (a3) at (0,0);
    \coordinate (a4) at (0,1);
    \coordinate (b1) at (-1/2,3/2);
    \coordinate (b2) at (1/2,3/2);
    \draw[very thick, line cap=round, dash pattern=on 0 off 15] (z3)--(z4);
	\draw[thick, decoration={markings, mark=at position 0.6 with {\arrow{>}}}, postaction={decorate}] (y1)--(y2);
	\draw[thick, decoration={markings, mark=at position 0.6 with {\arrow{<}}}, postaction={decorate}] (y2)--(x3);
	\draw[thick, decoration={markings, mark=at position 0.6 with {\arrow{>}}}, postaction={decorate}] (w1)--(x4);
	\draw[thick, decoration={markings, mark=at position 0.6 with {\arrow{<}}}, postaction={decorate}] (x4)--(w2);
	\draw[thick, decoration={markings, mark=at position 0.6 with {\arrow{>}}}, postaction={decorate}] (b1)--(a4);
	\draw[thick, decoration={markings, mark=at position 0.6 with {\arrow{<}}}, postaction={decorate}] (a4)--(b2);
	\draw[thick, red] (y2)--(y3)--(x4);
	\draw[thick, red, dashed] (a3)--(y4);
	\draw[thick, red] (y3)--(a3);
	\draw[thick, red] (y4)--(x5);
	\draw[thick, decoration={markings, mark=at position 0.6 with {\arrow{>}}}, postaction={decorate}] (w3)--(x5);
	\draw[thick, decoration={markings, mark=at position 0.6 with {\arrow{<}}}, postaction={decorate}] (x5)--(w4);
	\draw[thick, red] (y4)--(y5);
	\draw[thick, decoration={markings, mark=at position 0.6 with {\arrow{<}}}, postaction={decorate}] (y5)--(x6);
	\draw[thick, decoration={markings, mark=at position 0.6 with {\arrow{<}}}, postaction={decorate}] (y5)--(y6);
	\draw[thick, red] (a3)--(a4);
	\draw (y1) node[anchor=east] {$p_{1}$};
	\draw (x3) node[anchor=east] {$p_2$};
	\draw (w1) node[anchor=south] {$p_{3}$};
	\draw (w2) node[anchor=south] {$p_4$};
	\draw (b1) node[anchor=south] {$p_{5}$};
	\draw (b2) node[anchor=south] {$p_6$};
	\draw (w3) node[anchor=south] {$p_{n-3}$};
	\draw (w4) node[anchor=south] {$p_{n-2}$};
	\draw[red] ($(y3)!0.5!(x4)$) node[anchor=east] {$s_2$};
	\draw[red] ($(y4)!0.5!(x5)$) node[anchor=west] {$s_{n-4}$};
	\draw (x6) node[anchor=west] {$p_{n-1}$};
	\draw (y6) node[anchor=west] {$p_{n}$};
	\draw[red] ($(y2)!0.5!(y3)$) node[anchor=north] {$s_1$};
	\draw[red] ($(y4)!0.5!(y5)$) node[anchor=north] {$s_{n-3}$};
	\draw[red] ($(a3)!0.5!(a4)$) node[anchor=east] {$s_4$};
	\draw[red] ($(y3)!0.5!(a3)$) node[anchor=north] {$s_3$};
	\draw (y2) node[vint] {};
	\draw (y3) node[vint] {};
	\draw (y4) node[vint] {};
	\draw (y5) node[vint] {};
	\draw (x4) node[vint] {};
    \draw (x5) node[vint] {};
    \draw (a3) node[vint] {};
    \draw (a4) node[vint] {};
\end{tikzpicture}}
\newsavebox{\figNCombWitten}
\savebox{\figNCombWitten}{%
\begin{tikzpicture}[scale=.95]
    \tikzstyle{vint}=[draw,scale=0.55,color=teal,fill=green,circle];
    \draw[thick, color=black!30] (1/2,0) circle[radius=2];
    \coordinate (y1) at (-3/2,0);
    \coordinate (y2) at (-1,0);
    \coordinate (x3) at (-1,1.34);
    \coordinate (y3) at (0,0);
    \coordinate (z3) at (0,1/2);
    \coordinate (x4) at (0,1.95);
    \coordinate (y4) at (1,0);
    \coordinate (z4) at (1,1/2);
    \coordinate (x5) at (1,1.95);
    \coordinate (y5) at (2,0);
    \coordinate (x6) at (2,1.34);
    \coordinate (y6) at (5/2,0);
	\draw[very thick, line cap=round, dash pattern=on 0 off 8] (z3)--(z4);
	\draw[thick] (y1)--(y2)--(x3);
	\draw[thick] (y2)--(y3);
	\draw[thick] (y3)--(x4);
	\draw[thick, dashed] (y3)--(y4);
	\draw[thick] (y4)--(x5);
	\draw[thick] (y4)--(y5);
	\draw[thick] (y5)--(x6);	
	\draw[thick] (y5)--(y6);
	\draw (y1) node[anchor=east] {$\Delta_{1}$};
	\draw (x3) node[anchor=south east] {$\Delta_2$};
	\draw (x4) node[anchor=south] {$\Delta_3$};
	\draw (x5) node[anchor=south] {$\Delta_{n-2}$};
	\draw (x6) node[anchor=south west] {$\Delta_{n-1}$};
	\draw (y6) node[anchor=west] {$\Delta_{n}$};
	\draw ($(y2)!0.5!(y3)$) node[anchor=north] {$\Delta_{{\delta_1}}$};
	\draw ($(y4)!0.5!(y5)$) node[anchor=north] {$\Delta_{{\delta_{n-3}}}$};
	\draw (y2) node[vint] {};
	\draw (y3) node[vint] {};
	\draw (y4) node[vint] {};
	\draw (y5) node[vint] {};
\end{tikzpicture}}
\newsavebox{\figNOPEWitten}
\savebox{\figNOPEWitten}{%
\begin{tikzpicture}[scale=.95]
    \tikzstyle{vint}=[draw,scale=0.55,color=teal,fill=green,circle];
    \draw[thick, color=black!30] (1/2,0) circle[radius=2];
    \coordinate (y1) at (-1.414,-1/2);
    \coordinate (y2) at (-1,0);
    \coordinate (x3) at (-1.414,1/2);
    \coordinate (y3) at (0,0);
    \coordinate (z3) at (0,1/2);
    \coordinate (x4) at (0,1.3);
    \coordinate (y4) at (1,0);
    \coordinate (z4) at (1,1/2);
    \coordinate (x5) at (1,1.3);
    \coordinate (y5) at (2,0);
    \coordinate (x6) at (2.414,1/2);
    \coordinate (y6) at (2.414,-1/2);
    \coordinate (a1) at (.3,1.97);
    \coordinate (a2) at (-0.45,1.75);
    \coordinate (b1) at (.66,1.97);
    \coordinate (b2) at (1.41,1.75);
	\draw[very thick, line cap=round, dash pattern=on 0 off 8] (z3)--(z4);
	\draw[thick] (y1)--(y2)--(x3);
	\draw[thick] (y2)--(y3);
	\draw[thick] (y3)--(x4);
	\draw[thick, dashed] (y3)--(y4);
	\draw[thick] (y4)--(x5);
	\draw[thick] (y4)--(y5);
	\draw[thick] (y5)--(x6);	
	\draw[thick] (y5)--(y6);
	\draw[thick] (a1)--(x4)--(a2);
	\draw[thick] (b1)--(x5)--(b2);
	\draw (y1) node[anchor=east] {$\Delta_{1}$};
	\draw (x3) node[anchor=east] {$\Delta_2$};
	\draw (a1) node[anchor=south east] {$\Delta_4$};
	\draw (a2) node[anchor=south east] {$\Delta_3$};
	\draw (b1) node[anchor=south] {$\Delta_{n-3}$};
	\draw (b2) node[anchor=south west] {$\Delta_{n-2}$};
	\draw (x6) node[anchor=west] {$\Delta_{n-1}$};
	\draw (y6) node[anchor=west] {$\Delta_{n}$};
	\draw ($(y2)!0.5!(y3)$) node[anchor=north] {$\Delta_{{\delta_1}}$};
	\draw ($(y4)!0.5!(y5)$) node[anchor=north] {$\Delta_{{\delta_{n-3}}}$};
	\draw ($(y3)!0.5!(x4)$) node[anchor=east] {$\Delta_{{\delta_2}}$};
	\draw ($(y4)!0.5!(x5)$) node[anchor=west] {$\Delta_{{\delta_{n-4}}}$};
	\draw (y2) node[vint] {};
	\draw (y3) node[vint] {};
	\draw (y4) node[vint] {};
	\draw (y5) node[vint] {};
	\draw (x4) node[vint] {};
	\draw (x5) node[vint] {};
\end{tikzpicture}}
\newsavebox{\figSevenWitten}
\savebox{\figSevenWitten}{%
\begin{tikzpicture}[scale=.95]
    \tikzstyle{vint}=[draw,scale=0.55,color=teal,fill=green,circle];
    \draw[thick, color=black!30] (1/2,0) circle[radius=2];
    \coordinate (y1) at (-3/2,0);
    \coordinate (y2) at (-1,0);
    \coordinate (x3) at (-1,1.34);
    \coordinate (y3) at (0,0);
    \coordinate (z3) at (0,1/2);
    \coordinate (x4) at (0,1.95);
    \coordinate (y4) at (1,0);
    \coordinate (z4) at (1,1/2);
    \coordinate (x5) at (1,1.1);
    \coordinate (a1) at ($(1/2,0) + (84:2)$);
    \coordinate (a2) at ($(1/2,0) + (54:2)$);
    \coordinate (y5) at (2,0);
    \coordinate (x6) at (2,1.34);
    \coordinate (y6) at (5/2,0);
	\draw[thick] (y1)--(y2)--(x3);
	\draw[thick] (y2)--(y3);
	\draw[thick] (y3)--(x4);
	\draw[thick] (y3)--(y4);
	\draw[thick] (y4)--(x5);
	\draw[thick] (a1)--(x5)--(a2);
	\draw[thick] (y4)--(y5);
	\draw[thick] (y5)--(x6);	
	\draw[thick] (y5)--(y6);
	\draw (y1) node[anchor=east] {$\Delta_{1}$};
	\draw (x3) node[anchor=south east] {$\Delta_2$};
	\draw (x4) node[anchor=south] {$\Delta_3$};
	\draw (a1) node[anchor=south] {$\Delta_{4}$};
	\draw (a2) node[anchor=south] {$\Delta_{5}$};
	\draw (x6) node[anchor=south west] {$\Delta_{6}$};
	\draw (y6) node[anchor=west] {$\Delta_{7}$};
	\draw ($(y2)!0.5!(y3)$) node[anchor=north] {$\Delta_{{\delta_1}}$};
	\draw ($(y3)!0.5!(y4)$) node[anchor=north] {$\Delta_{{\delta_{2}}}$};
	\draw ($(y4)!0.5!(x5)$) node[anchor=east] {$\Delta_{{\delta_{3}}}$};
	\draw ($(y4)!0.5!(y5)$) node[anchor=north] {$\Delta_{{\delta_{4}}}$};
	\draw (y2) node[vint] {};
	\draw (y3) node[vint] {};
	\draw (y4) node[vint] {};
	\draw (y5) node[vint] {};
	\draw (x5) node[vint] {};
\end{tikzpicture}}

\newsavebox{\figEdge}

\newcommand{\DrawEdge}[2]{
\savebox{\figEdge}{%
\begin{tikzpicture}[scale=.95]
    \tikzstyle{vint}=[draw,scale=0.3,color=blue,fill=blue,circle]
    \coordinate (y1) at (0,0);
    \coordinate (y2) at (2,0);
	\draw[thick, red] (y1)--(y2);
	\draw[red] ($(y1)!0.5!(y2)$) node[anchor=north] {$#1$};
	\draw[red] ($(y1)!0.5!(y2)$) node[anchor=south] {$#2$};
    \draw (y1) node[vint] {};
    \draw (y2) node[vint] {};
\end{tikzpicture}
    }
\musepic{\figEdge}
}
\newsavebox{\figVertex}

\newcommand{\DrawVertex}[6]{
\sbox{\figVertex}{%
    \begin{tikzpicture}[scale=.8]
    \tikzstyle{vint}=[draw,scale=0.3,color=blue,fill=blue,circle]
    \coordinate (y4) at (1,1/4);
    \coordinate (x5) at (1,1);
    \coordinate (w3) at (1/2,7/4);
    \coordinate (w4) at (3/2,7/4);
	\ifthenelse{\isempty{#6}}{\draw[thick] (y4)--(x5);}{\draw[thick,red] (y4)--(x5);}
	\ifthenelse{\isempty{#2}}{\draw[thick] (w3)--(x5);}{\draw[thick,red] (w3)--(x5);}
	\ifthenelse{\isempty{#4}}{\draw[thick] (w4)--(x5);}{\draw[thick,red] (w4)--(x5);}
	\draw[red] ($(y4)!0.5!(x5)$) node[anchor=west] {$#6$};
	\draw[red] ($(w3)!0.5!(x5)$) node[anchor= east] {$#2$};
	\draw[red] ($(w4)!0.5!(x5)$) node[anchor= west] {$#4$};
	\ifthenelse{\isempty{#6}}{\draw (y4) node[anchor=north] {$#5$};}{\draw[red] (y4) node[anchor=north] {$#5$};}
	\ifthenelse{\isempty{#2}}{\draw (w3) node[anchor=south] {$#1$};}{\draw[red] (w3) node[anchor=south] {$#1$};}	
	\ifthenelse{\isempty{#4}}{\draw (w4) node[anchor=south] {$#3$};}{\draw[red] (w4) node[anchor=south] {$#3$};}
    \draw (x5) node[vint] {};
\end{tikzpicture}
    }
\musepic{\figVertex}
}
\newsavebox{\figVertexPostMellin}

\newcommand{\DrawVertexPostMellin}[6]{
\sbox{\figVertexPostMellin}{%
    \begin{tikzpicture}[scale=.8]
    \tikzstyle{vint}=[draw,scale=0.3,color=blue,fill=blue,circle]
    \coordinate (y4) at (1,1/4);
    \coordinate (x5) at (1,1);
    \coordinate (w3) at (1/2,7/4);
    \coordinate (w4) at (3/2,7/4);
	\ifthenelse{\isempty{#6}}{\draw[thick] (y4)--(x5);}{\draw[thick,red] (y4)--(x5);}
	\ifthenelse{\isempty{#2}}{\draw[thick] (w3)--(x5);}{\draw[thick,red] (w3)--(x5);}
	\ifthenelse{\isempty{#4}}{\draw[thick] (w4)--(x5);}{\draw[thick,red] (w4)--(x5);}
	\ifthenelse{\isempty{#6}}{\draw (y4) node[anchor=north] {$#5$};}{\draw[red] (y4) node[anchor=north] {$#5$};}
	\ifthenelse{\isempty{#2}}{\draw (w3) node[anchor=south] {$#1$};}{\draw[red] (w3) node[anchor=south] {$#1$};}	
	\ifthenelse{\isempty{#4}}{\draw (w4) node[anchor=south] {$#3$};}{\draw[red] (w4) node[anchor=south] {$#3$};}
    \draw (x5) node[vint] {};
\end{tikzpicture}
    }
\musepic{\figVertexPostMellin}
}

\begin{document}

\title{Towards Feynman rules for conformal blocks}
\authors{Sarah Hoback$^1$\footnote{\tt sarahhoback98@gmail.com} \& Sarthak Parikh$^2$\footnote{\tt sparikh@caltech.edu}}
\institution{PC}{$^1$Department of Physics and Astronomy, Pomona College, Claremont, CA 91711, USA}
\institution{Caltech}{$^2$Division of Physics, Mathematics and Astronomy, California Institute of Technology,\cr\hskip0.06in Pasadena, CA 91125, USA}

\abstract{ 
We conjecture a simple set of ``Feynman rules'' for constructing  $n$-point global conformal blocks in any channel in $d$ spacetime dimensions, for external and exchanged scalar operators for arbitrary $n$ and $d$. 
The vertex factors are given in terms of Lauricella hypergeometric functions of one, two or three variables, and the Feynman rules furnish an explicit power-series expansion in powers of cross-ratios.
These rules are conjectured based on previously known results in the literature, which include four-, five- and six-point examples as well as the $n$-point comb channel blocks.
We prove these rules for all previously known cases, as well as for a seven-point block in a new topology and the even-point blocks in the ``OPE channel.''
The proof relies on holographic methods, notably the Feynman rules for Mellin amplitudes of tree-level AdS diagrams in a scalar effective field theory, and is easily applicable to any particular choice of a conformal block.
}

\maketitle

\setcounter{tocdepth}{2}

{\hypersetup{linkcolor=black}
\tableofcontents
}


\section{Introduction}
\label{INTRO}

Conformal blocks are theory-independent building blocks of conformal field theories (CFTs) which capture contributions to conformal correlators from entire conformal families of representations appearing in the intermediate channels of correlation functions.
Via the AdS/CFT correspondence, they play an important role in the gravitational context as well; for example they provide a basis for writing down any bulk Witten diagram.

Conformal blocks also play a crucial, central role in the revived conformal bootstrap program~\cite{Ferrara:1973yt,Polyakov:1974gs,Rattazzi:2008pe} (see also the recent review~\cite{Poland:2018epd} and references therein), which has led to significant advances in understanding properties of $d$-dimensional CFTs as well as holography. 
This has resulted in considerable interest in and a spate of new results for conformal blocks. 

However, until recently, much of the focus has been restricted to {\it four}-point conformal blocks~\cite{Ferrara:1971vh,Ferrara:1973vz,Ferrara:1974ny,Dolan:2000ut,Dolan:2003hv,Dolan:2011dv}.
A variety of techniques are now available for obtaining four-point global conformal blocks for arbitrary external and exchanged representations in the intermediate channels in various forms such as closed-form, integral or series representations. A partial list of methods includes various recursive techniques, shadow formalism, use of differential operators, dimensional reduction, integrability methods, and holographic geodesic diagram techniques~\cite{Dolan:2000ut,Dolan:2011dv,Zamolodchikov:1985ie,Kos:2013tga,Penedones:2015aga,Iliesiu:2015akf,Costa:2016xah,Costa:2016hju,Kravchuk:2017dzd,Zhou:2018sfz,Erramilli:2019njx,SimmonsDuffin:2012uy,Costa:2011mg,Costa:2011dw,Echeverri:2015rwa,Echeverri:2016dun,Karateev:2017jgd,Cuomo:2017wme,Isono:2017grm,Fortin:2016lmf,Fortin:2019fvx,Fortin:2019dnq,Fortin:2019gck,Sleight:2017fpc,Costa:2018mcg,Hogervorst:2016hal,Kaviraj:2019tbg, Besken:2016ooo,Bhatta:2016hpz,Bhatta:2018gjb,Isachenkov:2016gim,Schomerus:2016epl,Buric:2019rms,Buric:2019dfk,Hijano:2015zsa,Nishida:2016vds,Castro:2017hpx,Dyer:2017zef,Chen:2017yia,Gubser:2017tsi,Kraus:2017ezw,Tamaoka:2017jce,Nishida:2018opl,Das:2018ajg}.

The focus on four-point blocks is due in part to the fact that conformal bootstrap is typically implemented at the level of four-point correlators. This is expected to be sufficient for constraining the full CFT data as long as one includes crossing-symmetry constraints from all possible four-point correlators, including those with arbitrary representations at external legs. 
This can be non-trivial and computationally very costly to implement. 
An alternative to this approach may be an $n$-point bootstrap program restricted simply to external scalars~\cite{Rosenhaus:2018zqn}. Implementing this approach would necessarily require the knowledge of higher-point scalar conformal blocks in arbitrary channels.

Recently, bulk unitarity methods~\cite{Meltzer:2019nbs} have also clarified the role of higher-point tree-level AdS diagrams in four-point results beyond the planar limit, i.\ e.\ in understanding the properties of higher-loop corrections. 
Higher-point tree-level AdS diagrams in turn are easily expressible via a conformal block decomposition or via a spectral representation in terms of direct channel conformal blocks and leading OPE coefficients. 
Thus the knowledge of higher-point conformal blocks in arbitrary channels would be particularly useful in probing holography at higher-loops.

However, obtaining explicit representations for conformal blocks is a notoriously hard problem, even though in principle the blocks are fixed entirely by conformal symmetry.
The challenges are particularly pronounced in the case of higher-point blocks in $d$ spacetime dimensions, where until recently hardly any results were available. 
The $d$-dimensional five-point block was obtained using the shadow formalism in ref.~\cite{Rosenhaus:2018zqn} (see also refs.~\cite{Parikh:2019ygo,Goncalves:2019znr}). 
A holographic representation for the five-point block was worked out in ref.~\cite{Parikh:2019ygo}, and subsequently extended to the six-point block in the so-called ``OPE channel''~\cite{Jepsen:2019svc}, as well as to higher-point blocks in the the comb channel~\cite{Parikh:2019dvm}. Ref.~\cite{Parikh:2019dvm} also worked out an explicit power-series expansion for the $n$-point comb channel blocks.
CFT embedding space methods~\cite{Fortin:2016lmf,Fortin:2019dnq} have also been fruitful in yielding higher-point blocks \cite{Fortin:2020ncr}; notably providing a series expansion for the $n$-point comb channel block~\cite{Fortin:2019zkm}, and the six-point block in the OPE channel, referred to as the ``snowflake channel''~\cite{Fortin:2020yjz}.\footnote{See also ref.~\cite{Anous:2020vtw} for an application to two-dimensional six-point global blocks for stress tensor exchanges, and ref.~\cite{Pal:2020dqf} for obtaining representations of (higher-point) diagrams in two and four spacetime dimensions in terms of solutions to Lauricella systems for conformal groups $SL(2,\mathbb{C})$ and $SL(2,\mathbb{H})$. Recent progress in higher-point diagrams has also come via momentum space techniques~\cite{Albayrak:2018tam,Albayrak:2019asr,Albayrak:2019yve,Albayrak:2020isk}.}

While the recent burst of activity and progress in studying higher-point functions and conformal blocks is encouraging, the situation is far from settled. 
A particularly troubling aspect of going to higher-point blocks is that the number of possible inequivalent channels grows very rapidly with $n$, thus it seems highly inefficient and impractical to work out the associated conformal blocks on a case by case basis. 
What would be desirable is a set of Feynman-like rules which could be determined once and for all, that enable writing down any conformal block in any topology without having to do any computations.

\vspace{1em}
Motivated by these considerations, in this paper we will present a simple, conjectural prescription for writing down an {\it arbitrary} $d$-dimensional $n$-point scalar conformal block with scalar exchanges in {\it any given channel}.
Even though the blocks themselves are non-perturbative objects, we call them ``Feynman rules'' because they are reminiscent of Feynman rules for Mellin amplitudes~\cite{Fitzpatrick:2011ia, Paulos:2011ie,Nandan:2011wc}.
This conjecture was motivated by carefully studying the power-series expansions of all known examples of scalar conformal blocks in the literature, particularly as presented in refs.~\cite{Rosenhaus:2018zqn,Parikh:2019ygo,Jepsen:2019svc,Parikh:2019dvm,Fortin:2019zkm,Fortin:2020yjz}.

As a highly non-trivial check of these rules, we compare the predicted blocks belonging to an infinite family of blocks previously unknown in the literature against a first-principles derivation and find exact agreement. 
These are the $n$-point conformal blocks in the so-called ``OPE channel'' for arbitrary even $n$. 
We also test the rules in the case of a seven-point block in a topology different from the comb channel, which we simply refer to as the ``mixed channel,'' and find perfect agreement.

The key idea which enables us to compute these new families of blocks from first principles was  previously utilized in refs.~\cite{Parikh:2019ygo,Jepsen:2019svc,Parikh:2019dvm} to obtain the holographic duals of higher-point blocks.
To obtain a particular scalar conformal block, we start with a tree-level Witten diagram in a cubic $\phi^3$ effective field theory whose direct channel conformal block decomposition admits the desired block as its single-trace contribution.
We call such a Witten diagram the ``canonical Witten diagram'' for the block, and there is a unique choice for each conformal block.
Then the single-trace contribution to the canonical Witten diagram is given by the desired block {\it times} a set of known mean field theory OPE coefficients. 
Thus the key step is to obtain the single-trace projection of the Witten diagram, as this will immediately yield the conformal block.

Here, we appeal to Mellin space technology~\cite{Mack:2009gy,Mack:2009mi,Penedones:2010ue}, which serves a two-fold purpose. 
Firstly, in a large $N$ bulk theory, Mellin amplitudes are meromorphic functions with  poles corresponding precisely to the exchange of single-trace operators; this provides a convenient route to single-trace projections.
Secondly, Mellin amplitudes for all tree-level scalar Witten diagrams in scalar effective field theories are known (thanks to the Mellin space Feynman rules~\cite{Fitzpatrick:2011ia, Paulos:2011ie,Nandan:2011wc}); this enables us to obtain an explicit single-trace projection of {\it any} $n$-point canonical Witten diagram.
This method of projecting out the multi-trace exchanges to obtain the conformal block is quite general, efficient and constructive, so it can be used to work out any particular conformal block.
The main, and often only, computationally challenging step of this procedure will be the actual evaluation of all residual Mellin integrals, which is required to obtain an explicit power-series expansion for the block. 
However, in all examples we attempted we were able to systematically work out all such integrals merely by repeated, and often inductive, applications of the first Barnes lemma~\cite{Barnes1908}.

The outline for the rest of the paper is as follows:
In section~\ref{FEYNMAN} we propose the Feynman rules for conformal blocks, and in section~\ref{FEYNMANEXAMPLES} we illustrate how to apply them to obtain a seven-point block in the ``mixed channel,'' the $n$-point comb channel block, and the $n$-point OPE channel block.
In section~\ref{PROOF} we revisit all examples from section~\ref{FEYNMANEXAMPLES} and using the Mellin-space single-trace projection technique, we prove the Feynman rules in each case. 
We end with some discussion and future directions in section~\ref{DISCUSS}. Various technical details and computations are provided in the appendices.

\vspace{1em}

When this work was largely complete, we learned of parallel, independent work to appear by Fortin, Ma and Skiba~\cite{FMSToAppear}, which has partial overlap with some results of this paper.

\section{Feynman rules for conformal blocks}
\label{FEYNMAN}

\begin{figure}
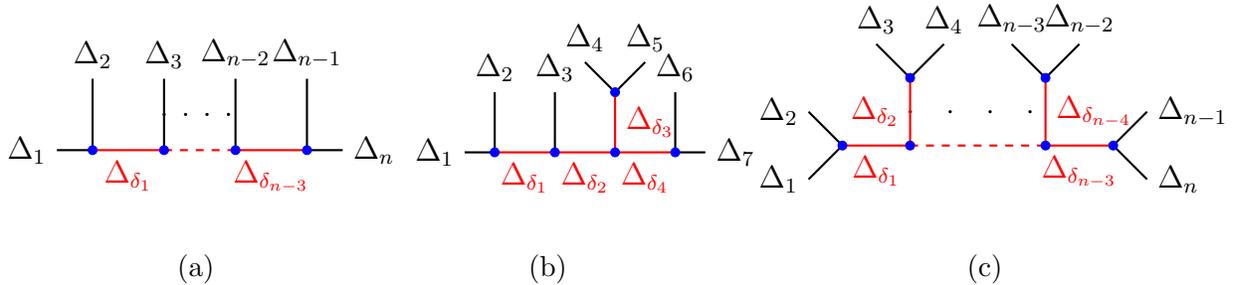

    \centering
    \begin{subfigure}[b]{0.32\textwidth}
         \[   \musepic{\figNCombChannel}  \]
        \caption{}
         \label{fig:CBexComb}
    \end{subfigure}
    \begin{subfigure}[b]{0.23\textwidth}
          \[  \musepic{\figSevenChannel}   \]
        \caption{}
         \label{fig:CBexMix}
    \end{subfigure}
    \hspace{5mm}
    \begin{subfigure}[b]{0.38\textwidth}
          \[  \musepic{\figNOPEChannel}    \]
        \caption{}
         \label{fig:CBexOPE}
    \end{subfigure}
    \caption{{\it Graphical representation of conformal blocks:} Any $n$-point block admits a unique representation as an unrooted binary tree with $n$ leaves, and consequently $n-3$ internal edges (colored {\Red red} to guide the eye) and $n-2$ internal nodes/vertices (marked in {\Blue blue}). All edges are labeled with conformal dimensions; the labels on external edges (edges attached to the leaves of the unrooted tree) are shown at the leaves for better presentation. { (a):} The graph shows an $n$-point  ``comb channel''  conformal block (for $n\geq 4$) for external scalar operators ${\cal O}_1(x_1), \ldots, {\cal O}_n(x_n)$ with conformal dimensions $\Delta_1,\ldots,\Delta_n$ and insertion coordinates $x_1,\ldots,x_n$ respectively, and exchanged scalar operators ${\cal O}_{\delta_1},\ldots,{\cal O}_{\delta_{n-3}}$ along the internal edges with conformal dimensions $\Delta_{\delta_1},\ldots,\Delta_{\delta_{n-3}}$, respectively. {(c):} The graph shows an $n$-point block in the ``OPE channel,'' for even $n \geq 6$. One can obtain the OPE channel topology by starting with an ${n \over 2}$-point comb channel block and attaching two external edges at every leaf to get the $n$-point OPE channel block. {(b):} The graph shows a $7$-point example in a ``mixed channel'' which is neither the comb nor the OPE channel.}
    \label{fig:CBex}
\end{figure}

Given any $n$-point conformal block, let the dimensions and insertion coordinates of the external operators be respectively, $\Delta_i$  and $x_i$ for $i=1,\ldots, n$. 
Let the dimensions of the exchanged operators be enumerated $\Delta_{\delta_i}$ for $i=1,\ldots,n-3$.
See figure~\ref{fig:CBex} for some examples of graphical representation of blocks in different channels as unrooted binary trees with $n$ leaves (and correspondingly $n-2$ internal vertices and $n-3$ internal edges), which will play a central role in the Feynman rules. 
Different inequivalent channels/topologies  correspond to different OPE structures which can contribute to a conformal correlation function.

The Feynman rules presented here give an expression for the desired conformal block in the desired channel as an $n(n-3)/2$-fold power series in powers of $n(n-3)/2$ independent cross-ratios built out of operator insertion positions $x_i$.\footnote{We assume sufficiently high $d$. 
Otherwise some of the cross-ratios will be dependent, but the prescription still works. 
In this case, however, a more efficient power series with fewer overall sums also exists.} 
The set of cross-ratios will be treated as input data fed into the rules to obtain the conformal block.
We will assume the independent cross-ratios are $u_i$ for $i=1,\ldots, n-3$ and $v_j$ for $j=1,\ldots, \binom{n-2}{2}$, such that in the $n-3$ separate OPE limits where $u_i \approx 0$ for a particular $i$, the leading $u_i$-dependent contribution to the conformal block is given by
\eqn{CBscaling}{
 W_n(x_i) \Big|_{ u_i \approx 0} \propto u_i^{\Delta_{\delta_i}/2}\,.
}
There is a choice in picking $n(n-3)/2$ independent cross-ratios subject to the constraint~\eno{CBscaling}, and also correspondingly a choice of the leg factor. 
The Feynman rules described here work for any such choice. 
From here on, we fix a choice.

Away from the limit~\eno{CBscaling}, the conformal block admits an expansion of the form
\eqn{CBexpansion}{
\tcboxmath{ W_n(x_i) = W_n^{0}(x_i) \left(\prod_{i=1}^{n-3} u_i^{\Delta_{\delta_i}/2} \right) g(u,1-v)\,, }
}
where $W_n^{0}(x_i)$, which will be referred to as the ``leg factor,'' depends only on position coordinates $x_i$ and external dimensions $\Delta_i$.  The function $g(u,1-v)$ is expressed as a power series in $u_i$ and $(1-v_j)$ for all $i,j$, with the leading behaviour $g(u,1-v) = 1 + O(u_i, 1-v_j)$. 
This function sums all descendant contributions to the conformal block. 
The Feynman rules provide a prescription for writing down this function in terms of ``edge factors''$E_i$ and ``vertex factors''  $V_i$  associated respectively with each internal edge and internal vertex of the unique unrooted binary tree representation of the desired conformal block (see e.g.\ figure~\ref{fig:CBex}): 
\eqn{Feynman}{
\tcboxmath{ g(u,1-v) =  \sum_{k_i, j_{rs}=0}^\infty \left[\left( \prod_{i=1}^{n-3} {u_i^{k_i} \over k_i!} \right) \left(\prod_{(rs)}^{\binom{n-2}{2}} {\left(1-v_{rs}\right)^{j_{rs}} \over j_{rs}!} \right) \left( \prod_{i=1}^{n-3} E_i \right) \left(\prod_{i=1}^{n-2} V_i \right) \right], }
}
The position-independent edge and vertex factors depend solely on the external and exchanged conformal dimensions, as well as the non-negative integral parameters being summed over, $k_i$ and $j_{rs}$, where $i=1,\ldots,n-3$ and the $(rs)$ index takes $\binom{n-2}{2}$ values. 
(For convenience we have also re-enumerated the $v_j$ cross-ratios as $v_{rs}$; the precise mapping will be explained shortly.)
They are determined as follows:
\begin{itemize}
    \item Label each {\it internal}  edge with an index $i$ running from $1$ to $n-3$, such that the conformal dimension of the exchanged operator running along the edge is twice the exponent of the cross-ratio $u_i$ appearing in~\eno{CBscaling}. Associate to each such edge an integral parameter $k_i$ and a factor of
    \eqn{EdgeDef}{
    \tcboxmath{ E_i := { (\Delta_{\delta_i} - h +1)_{k_i} \over (\Delta_{\delta_i})_{2k_i + \ell_{\delta_i}} }\,, }
    }
    where $\Delta_{\delta_i}$ is the conformal dimension of the exchanged operator running along the edge, and $\ell_{\delta_i}$ is an integral parameter associated with the conformal dimension $\Delta_{\delta_i}$ to be determined later. 
     Here $(a)_b \equiv \Gamma(a+b)/\Gamma(a)$ is the Pochhammer symbol, and we have defined
     \eqn{hDef}{
     h := d/2\,.
     }
    We  refer to the parameters $k_i$  as ``single-trace parameters,'' and the parameters $\ell_{\delta_i}$ as ``post-Mellin parameters.'' 
    The single-trace parameter $k_i$ also appears in the series expansion~\eno{Feynman} as the exponent of the cross-ratio $u_i$.
     The post-Mellin parameters are specified entirely in terms of specific positive linear combinations of the parameters $j_{rs}$ appearing in~\eno{Feynman}, which we call ``Mellin parameters.''
     The precise relation between the two will be discussed in section~\ref{INTEGERS}. In that section, we will also present an alternate prescription for assigning the appropriate single-trace parameter to each internal edge.
    
    \item Label each {\it internal} (i.e.\ cubic) vertex with an index $i$ running from $1$ to $n-2$.\footnote{This indexing is not to be confused with the indexing of edges described above.}
     Let the number of incident internal edges on it be denoted $M$.
     It is clear that $M$ can only be $1, 2$, or $3$.
     Let the conformal dimensions attached to the edges be $\Delta_{a}, \Delta_{b}$, and $\Delta_{c}$.
     Consider first a vertex with $M=3$ (i.e.\ with all incident edges internal).
     To this vertex, assign a factor of
    \begingroup\makeatletter\def\f@size{11.3}\check@mathfonts
    \begin{empheq}[box=\narrowtcboxmath]{align} 
    \label{VertexDef}
V_i &:= (\Delta_{ab,c})_{k_{ab,c}+{1\over 2}\ell_{ab,c}} (\Delta_{ac,b})_{k_{ac,b}+{1\over 2}\ell_{ac,b}} (\Delta_{bc,a})_{k_{bc,a}+{1\over 2}\ell_{bc,a}} \cr 
     &  \times F_A^{(3)}\!\left[\Delta_{abc,}- h; \{-k_a, -k_b, -k_c \}; \left\{\Delta_a -h+1, \Delta_b -h+1,\Delta_c -h+1 \right\}; 1,1,1 \right] 
\end{empheq}
\endgroup

    where $F_A^{(3)}$ is the Lauricella function of three variables, defined in~\eno{LauricellaDef}.
    Here $k_a, k_b$, and $k_c$ are the respective single-trace parameters associated with each internal edge above, and $\ell_a, \ell_b$, and $\ell_c$ are the post-Mellin parameters associated with $\Delta_a, \Delta_b$ and $\Delta_c$, respectively.
    Here and below, we are using the shorthand,
    \eqn{Deltaijk}{
    \Delta_{i_1\ldots i_m,i_{m+1} \ldots i_n} := {1\over 2} \left( \Delta_{i_1} + \cdots + \Delta_{i_m} - \Delta_{i_{m+1}} - \cdots - \Delta_{i_n} \right)
    }
    for conformal dimensions $\Delta_i$, whereas for  single-trace parameters and post-Mellin parameters we are using
    \eqn{kellijk}{
    k_{i_1 \ldots i_m, i_{m+1} \ldots i_{n}} &:= k_{i_1} + \cdots + k_{i_m} - k_{i_{m+1}} - \cdots - k_{i_n}  \cr 
    \ell_{i_1 \ldots i_m, i_{m+1} \ldots i_{n}} &:= \ell_{i_1} + \cdots + \ell_{i_m} - \ell_{i_{m+1}} - \cdots - \ell_{i_n} \,.
    }
     
    For a vertex with $M=2$ (respectively, $M=1$), one (respectively, two) of the incident edges is an external edge. 
    So far, external edges  have not been assigned a single-trace parameter.
    It is convenient to view an external edge as an edge with its single-trace parameter set to zero.
    Then the vertex factor continues to be given by~\eno{VertexDef}, but with the associated single-trace parameter(s) set to zero. 
    
    It is worth noting that the Lauricella function $F_A^{(3)}$ in~\eno{VertexDef} with say, $k_a=0$ reduces to the Lauricella function of two variables, 
    \eqn{}{ 
    F_A^{(2)}\left[\Delta_{abc,}- h; \{ -k_b, -k_c \}; \left\{ \Delta_b -h+1,\Delta_c -h+1 \right\}; 1,1 \right].
    }
    Likewise if two of the attached edges are external, with say, $k_a = k_b=0$, then the Lauricella function reduces further to the Lauricella function of one variable,
    \eqn{}{ 
    F_A^{(1)}\left[\Delta_{abc,}- h; \{ -k_c \}; \left\{ \Delta_c -h+1 \right\}; 1 \right].
    }
    In appendix~\ref{LAURICELLA} we list some identities relating these Lauricella functions to other known functions.
\end{itemize}

Modulo the relation between Mellin and post-Mellin parameters which will be explained in section~\ref{INTEGERS}, this concludes the complete set of Feynman rules for writing down  an explicit power series expansion of any scalar $n$-point conformal block with scalar exchanges in {\it any} channel. 

Readers familiar with series expansions of conformal blocks may feel puzzled by the apparent cross-ratio-basis independence of the series coefficients appearing in the expansion~\eno{Feynman}.
However, the explicit form of the edge and vertex factors does in fact depend on the choice of basis of cross-ratios; this dependence is encoded in the correct pairing between the single-trace parameters and cross-ratios as discussed above, as well as the  precise relation between Mellin and post-Mellin parameters, which we discuss next. 
At the end, as noted in~\eno{Feynman}, one sums over all single-trace and Mellin parameters.

\subsection{Mellin and post-Mellin parameters}
\label{INTEGERS}

Recall that Mellin variables~\cite{Mack:2009gy, Mack:2009mi} are complex-valued variables $\gamma_{ij}$ ($1 \leq i, j \leq n$) which are symmetric, $\gamma_{ij} = \gamma_{ji}$ with $\gamma_{ii} := -\Delta_i$,  satisfying the following $n$ constraints:
\eqn{MellinConstraints}{
 \sum_{j=1}^n \gamma_{ij} = 0 \qquad (i=1,\ldots,n) \,.
}
This leads to $n(n-3)/2$ independent components. 
These variables play a central role in the context of Mellin amplitudes of $n$-point bulk diagrams~\cite{Penedones:2010ue,Fitzpatrick:2011ia,Paulos:2011ie,Nandan:2011wc}, which will be reviewed in section~\ref{PROOF} in the proof of the proposed Feynman rules for conformal blocks.
The constraints above can be solved in terms of auxiliary momentum variables $p_i$ (for $i=1,\ldots, n$) such that $p_i \cdot p_j := \gamma_{ij}$ (thus individual $p_i$ are ``on-shell,'' i.e.\ $p_i \cdot p_i = -\Delta_i$), when ``momentum conservation,'' $\sum_{i=1}^n p_i = 0$ is imposed.
In this auxiliary space, the role of the $n(n-3)/2$ independent Mellin variables is played by a choice of $n(n-3)/2$ independent Mandelstam invariants $s_{i_1\ldots i_k}$ defined via
\eqn{MandelstamDef}{
s_{i_1\ldots i_k} := -(p_{i_1} + \cdots + p_{i_k})^2 = \sum_{j=i_1}^{i_k} \Delta_{j} - 2 \sum_{i_1 \leq r < s \leq i_k} \gamma_{rs}\,.
}

 In section~\ref{PROOF}, for working out the Feynman rules, we will be  interested in the following object which we call the ``Mellin product,''
\eqn{MellinProd}{
\prod_{1 \leq i < j \leq n} {1 \over (x_{ij}^2)^{\gamma_{ij}}}\,,
}
where $x_i$ (for $i=1,\ldots, n$) are boundary coordinates at which operators of conformal dimension $\Delta_i$ are inserted.
The reason why such an object appears will be clear in section~\ref{PROOF} where we obtain the conformal block Feynman rules starting from the Mellin representation of certain bulk Witten diagrams.
This product can be recast in terms of conformal cross-ratios built out of $x_i$ coordinates, as we now describe.

For any given choice of independent cross-ratios and a given channel, there is a canonical choice of $n(n-3)/2$ independent Mellin variables,
\eqn{MellinIndependent}{
\left\{ \gamma_{ij} : (ij) \in  {\cal U}_{\rm channel} \right\} \, \bigcup\, \left\{ \gamma_{ij} : (ij) \in  {\cal V}_{\rm channel} \right\}  \,,
}
which makes the Mellin product expressed in terms of cross-ratios physically intuitive. 
The sets ${\cal U}_{\rm channel}$ and ${\cal V}_{\rm channel}$, of cardinalities $n-3$ and $\binom{n-2}{2}$ respectively, will be defined shortly.
More precisely,
given a particular channel and any choice of conformal cross-ratios $\{ u_i, v_{rs} \}$  consistent with~\eno{CBscaling}-\eno{CBexpansion}, there exists a choice of independent Mellin variables $\gamma_{ij}$~\eno{MellinIndependent} such that the Mellin product can be re-expressed in terms of a product over powers of the given cross-ratios,\footnote{Without loss of generality, we are assuming the given cross-ratios have been enumerated such that the subscripts match, i.e.\ $u_i$ goes with $s_i$ and $v_{rs}$ goes with $\gamma_{rs}$ as shown in~\eno{ArbMellinProd}.}
\eqn{ArbMellinProd}{
\prod_{1 \leq i < j \leq n} {1 \over (x_{ij}^2)^{\gamma_{ij}}} =: W_{n}^0(x_i) \left(\prod_{i=1}^{n-3} u_i^{s_{i}/2} \right) \left( \prod_{(rs) \in {\cal V}_{\rm channel}} v_{rs}^{-\gamma_{rs}} \right),
}
where $W_n^0$ is the leg-factor for the given choice of cross-ratios in the particular channel. 
 The set $\{s_i\}$ is the set of $(n-3)$ independent Mandelstam invariants associated with the $(n-3)$ internal legs of the binary graph representation of the block.
In enumerating the Mandelstam invariants, we labeled the internal edges with an index $i=1,\ldots, n-3$ such that the Mandelstam invariant for the edge $i$, given by $s_i$, appears in the exponent of the cross-ratio $u_i$.  
Accordingly, we can assign the single-trace parameter associated with this internal edge, appearing in the summand of~\eno{Feynman}, the edge factor~\eno{EdgeDef} and the vertex factors~\eno{VertexDef}  to be $k_i$.

In the final product in~\eno{ArbMellinProd}, the set of Mellin variables appearing in the exponents determines precisely the set $\left\{ \gamma_{ij} : (ij) \in  {\cal V}_{\rm channel} \right\}$.
This will be taken to be the definition of ${\cal V}_{\rm channel}$.
The set ${\cal U}_{\rm channel}$ is then defined to be the set of pairs of indices such that $\{ \gamma_{ij} : (ij) \in {\cal U}_{\rm channel} \}$ gives the residual $n-3$ independent Mellin variables. 
It is worth noting that dependence in~\eno{ArbMellinProd} on the Mellin variables from this set  is encoded in the Mandelstam invariants $s_i$. 
We will  denote ${\cal D}_{\rm channel}$ to the set such that $\{ \gamma_{ij} : (ij) \in {\cal D}_{\rm channel} \}$ produces all dependent Mellin variables.
Of course, the union of all these sets gives 
\eqn{UnionAll}{
 {\cal U}_{\rm channel} \, \bigcup\, {\cal V}_{\rm channel} \, \bigcup\, {\cal D}_{\rm channel} = \left\{ (ij) : 1 \leq i < j \leq n\right\} .
} 

We define the set of {\it Mellin parameters} to be the set 
\eqn{MellinIntegers}{
{\cal J}_{\rm channel} := \left\{ j_{rs}  : (rs) \in {\cal V}_{\rm channel} \right\} ,
}
of cardinality $\binom{n-2}{2}$. 
Mellin parameters make a direct appearance in the summand of the Feynman prescription for conformal blocks~\eno{Feynman}, where they appear in the exponents of certain cross-ratios, as well as in the edge and vertex factors~\eno{EdgeDef} and~\eno{VertexDef} via the post-Mellin parameters $\ell_a$.
To obtain the full conformal block, one sums all Mellin parameters over all integral values from $0$ to $\infty$.
We now give the prescription to compute the post-Mellin parameters $\ell_a$ associated with the conformal dimensions $\Delta_a$ in terms of the Mellin parameters. 

For an {\it external} operator with conformal dimension $\Delta_i = - \gamma_{ii}$ inserted at position $x_i$, we define the associated post-Mellin parameter to be
\eqn{PostMellinExternal}{
\tcboxmath{ \ell_i : = \sum_{\substack{(rs) \in {\cal V}_{\rm channel} \\  r=i {\rm \ or\ } s=i}} j_{rs}  =  \sum_{\substack{j_{rs} \in {\cal J}_{\rm channel} \\  r=i {\rm \ or\ } s=i}} j_{rs}  \,. }
}
If the set $\{(rs) \in {\cal V}_{\rm channel} : r=i {\rm \ or\ } s=i\}$ is empty, then $\ell_i=0$.
Note that this definition implies that the sum over all post-Mellin parameters associated to external conformal dimensions evaluates to twice the sum over all Mellin parameters, 
\eqn{}{ 
\sum_{i=1}^n \ell_i = 2\sum_{{\cal J}_{\rm channel}} j_{rs}\,.
}

For {\it exchanged} operators of conformal dimensions $\Delta_{\delta_i}$, the prescription to compute the post-Mellin parameters proceeds iteratively as follows:
\begin{enumerate}
    \item First, at all internal vertices of the binary graph with precisely two external edges and one internal edge incident,  add the post-Mellin parameters associated with the external dimensions, and then {\it drop} all terms which are multiples of two (i.e.\ terms which are even for {\it all} integral values of the Mellin parameters).
    Assign this non-negative sum to be the post-Mellin parameter of the internal (exchanged) operator. 
    \eqn{PostMellinInternal1}{
    \tcboxmath{ \DrawVertexPostMellin{\Delta_1}{}{\Delta_2}{}{\Delta_{\delta_3}}{1}\!\!\!\! :  \quad   \ell_{\delta_3} \stackrel{2{\cal J}}{=} \ell_1 + \ell_2 \,, }
    }
    where the symbol $\stackrel{2{\cal J}}{=}$ means equality holds once one drops all terms which are {\it even for  all integral values} of Mellin parameters.
    For example, if $\ell_1 = j_{12} + j_{13} + j_{16}$ and $\ell_2 = j_{12} + j_{23} + j_{24}$, then $\ell_{\delta_3}\stackrel{2{\cal J}}{=} \ell_1 + \ell_2$ implies $\ell_{\delta_3} = j_{13} + j_{16} + j_{23} + j_{24}$.
    
    \item  If all internal post-Mellin parameters have {\it not} already been determined, pick any internal vertex where the post-Mellin parameters of precisely two of the edges are already known. 
    The post-Mellin parameter of the third edge is given by the sum of the other two post-Mellin parameters, after dropping terms which are even multiples of Mellin parameters, exactly as shown in~\eno{PostMellinInternal1}. For example, if two of the post-Mellin parameters are known at a vertex with three incident exchanged operators, then the third is determined as follows:    
    \eqn{PostMellinInternal2}{
    \tcboxmath{ \DrawVertexPostMellin{\Delta_{\delta_1}}{1}{\Delta_{\delta_2}}{1}{\Delta_{\delta_3}}{1}\!\!\!\! :  \quad   \ell_{\delta_3} \stackrel{2{\cal J}}{=} \ell_{\delta_1} + \ell_{\delta_2} \,. }
    }
    
    If there are multiple choices of vertices for fixing the unknown post-Mellin parameter of an internal edge, pick any. 
    The final assignment will be independent of this choice.
    
    \item Repeat step 2 until all internal conformal dimensions have been assigned a post-Mellin parameter.
    
\end{enumerate}
Note that this prescription guarantees that all post-Mellin parameters are written as positive linear combinations of Mellin parameters. Furthermore, at any internal vertex, the sum of any two of the post-Mellin parameters equals the third post-Mellin parameter up to terms which are even multiples of Mellin parameters. That is, if $\ell_a, \ell_b$ and $\ell_c$ are the post-Mellin parameters for conformal dimensions incident at a common vertex, then
\eqn{PostMellinPermute}{
\ell_a \stackrel{2{\cal J}}{=} \ell_b + \ell_c \qquad 
\ell_b \stackrel{2{\cal J}}{=} \ell_c + \ell_a \qquad 
\ell_c \stackrel{2{\cal J}}{=} \ell_a + \ell_b\,.
}

In the next section, we illustrate how to apply these rules to determine the $n$-point conformal block in the comb channel, the $n$-point conformal block in the OPE channel, and the seven-point block in the mixed channel (all depicted in figure~\ref{fig:CBex}).
In section~\ref{PROOF}, we will reproduce these blocks from first principles which serves as a highly non-trivial check of the Feynman rules.

\section{Examples}
\label{FEYNMANEXAMPLES}

In this section, we illustrate how to apply the Feynman rules to three classes of examples: the $n$-point conformal block in the {\it comb} channel and the {\it OPE} channel  for arbitrary $n$,\footnote{We remind the reader that the OPE channel in this paper is only defined for even $n$.} and the seven-point {\it mixed} channel block (see figure~\ref{fig:CBex} for their definitions). 
All known $d$-dimensional scalar conformal blocks  with scalar exchanges in the literature fall into one of the classes above. This includes the well-known four-point block, and the recently obtained five-point block~\cite{Rosenhaus:2018zqn}, $n$-point comb channel blocks~\cite{Parikh:2019dvm,Fortin:2019zkm} and the six-point OPE channel block~\cite{Fortin:2020yjz}. However, the seven-point example to be discussed next and the $n$-point OPE channel examples for $n \geq 8$ are new results.

We invite the reader to test their understanding of section~\ref{FEYNMAN} by applying the Feynman rules in the trivial case of the four-point block and rediscover the well-known series expansion, or the slightly less non-trivial though straightforward case of the five-point block. 
These are special cases of the $n$-point comb channel block which is discussed in section~\ref{NCOMB}.

\subsection{Seven-point mixed channel block}
\label{7MIXED}

In this section we work out the seven-point conformal block in the ``mixed channel'' shown in figure~\ref{fig:CBexMix}. We use the following independent cross-ratios as input data:
\eqn{7mixCR}{
\begin{gathered}
u_1 := { x_{12}^2x_{37}^2\over x_{17}^2 x_{23}^2} \qquad u_2 := {x_{23}^2 x_{57}^2 \over x_{25}^2 x_{37}^2} \qquad u_3:= {x_{45}^2 x_{27}^2 \over x_{25}^2 x_{47}^2} \qquad u_4 := {x_{67}^2 x_{25}^2 \over x_{27}^2 x_{56}^2} \cr 
v_{13}:= {x_{13}^2 x_{27}^2 \over x_{17}^2 x_{23}^2} \qquad v_{i6} := {x_{i6}^2 x_{57}^2 \over x_{i7}^2 x_{56}^2} \qquad (1\leq i \leq 4) \cr 
v_{ij} := {x_{ij}^2 x_{27}^2 x_{57}^2 \over x_{i7}^2 x_{j7}^2 x_{25}^2} \qquad \left((ij) \in \left\{(14),(15),(24),(34),(35)\right\} \right).
\end{gathered}
}
A convenient choice of dependent Mellin variables which turns the Mellin product~\eno{MellinProd} for the seven-point block in mixed channel into the form~\eno{ArbMellinProd} is 
\eqn{Dmix}{
\{\gamma_{ij} : (ij) \in {\cal D}_{\rm 7,mix} \} \quad {\rm where} \quad {\cal D}_{\rm 7,mix} = \left\{ (17),(25),(27),(37),(47),(56),(57) \right\} .
}
In terms of the independent Mellin variables, these can be expressed as
\eqn{DMixVars}{
\gamma_{17}&= \Delta_1 -\gamma_{12}-\gamma_{13}-\gamma_{14}-\gamma_{15}-\gamma_{16}
\cr 
\gamma_{25}
&= \Delta_{12345,67}-\gamma_{12}-\gamma_{13}-\gamma_{14}-\gamma_{15}-\gamma_{23}-\gamma_{24}-\gamma_{34}-\gamma_{35}-\gamma_{45}+\gamma_{67}
\cr 
\gamma_{27}
&= \Delta_{267,1345}+\gamma_{13}+\gamma_{14}+\gamma_{15}-\gamma_{26}+\gamma_{34}+\gamma_{35}+\gamma_{45}-\gamma_{67}
\cr 
\gamma_{37}&= \Delta_{3} - \gamma_{13} - \gamma_{23} - \gamma_{34} - \gamma_{35} -\gamma_{36}
\cr 
\gamma_{47}&= \Delta_4 -\gamma_{14}-\gamma_{24}-\gamma_{34}-\gamma_{45}-\gamma_{46}
\cr 
\gamma_{56}
&=\Delta_6 -\gamma_{16}-\gamma_{26}-\gamma_{36}-\gamma_{46}-\gamma_{67} \cr
\gamma_{57} &= \Delta_{57,12346} +\gamma_{12} +\gamma_{13}+\gamma_{14}+\gamma_{16}+\gamma_{23}+\gamma_{24}+\gamma_{26}+\gamma_{34}+\gamma_{36}+\gamma_{46} \,.
}
In terms of these, the Mellin product then takes the form
\eqn{7mixMellinProd}{
\prod_{1 \leq i < j \leq 7} {1 \over (x_{ij}^2)^{\gamma_{ij}}} =: W_{7,{\rm mix}}^0(x_i) \left(\prod_{i=1}^{4} u_i^{s_{i}/2} \right) \left( \prod_{(rs) \in {\cal V}_{\rm 7,mix}} v_{rs}^{-\gamma_{rs}} \right),
}
where the leg-factor turns can be expressed as
\eqn{7mixLegFactor}{
W_{7,{\rm mix}}^{0} &= \left({x_{27}^2 \over x_{12}^2 x_{17}^2 } \right)^{\Delta_1\over 2} \left({x_{17}^2 \over x_{12}^2 x_{27}^2 } \right)^{\Delta_2\over 2} \left({x_{27}^2 \over x_{23}^2 x_{37}^2 } \right)^{\Delta_3\over 2} \left({x_{57}^2 \over x_{45}^2 x_{47}^2 } \right)^{\Delta_4\over 2} \cr 
&\times \left({x_{47}^2 \over x_{45}^2 x_{57}^2 } \right)^{\Delta_5\over 2} \left({x_{57}^2 \over x_{56}^2 x_{67}^2 } \right)^{\Delta_6\over 2} \left({x_{56}^2 \over x_{57}^2 x_{67}^2 } \right)^{\Delta_7\over 2}\,,
}
and the set associated to the Mellin parameters over which the $\binom{5}{2}$-dimensional product runs in~\eno{7mixMellinProd}  is,
\eqn{Vmix}{
{\cal V}_{7, \rm mix} = \big\{(13), (14), (15), (16), (24), (26), (34), (35), (36), (46) \big\}\,.
}
The exponents $s_i$ in~\eno{7mixMellinProd} are given by
\eqn{7mixMandel}{
\begin{gathered}
s_1 = \Delta_1 + \Delta_2 - 2\gamma_{12} \qquad s_2 = \Delta_1 + \Delta_2 + \Delta_3 - 2\gamma_{12} - 2\gamma_{13} -2\gamma_{23} \cr 
s_3 = \Delta_4 + \Delta_5 - 2\gamma_{45} \qquad s_4 = \Delta_6 + \Delta_7 - 2\gamma_{67}\,,
\end{gathered}
}
which are indeed the Mandelstam invariants attached to the internal edges of the associated binary graph, as we now describe.
\begin{figure}
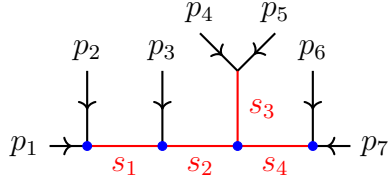

\centering
         \[  \musepic{\figSevenMixedChannelMomentum} \]
    \caption{{\it Auxiliary momenta and Mandelstam variable assignments:}  Graphical representation of the $7$-point conformal block in the mixed channel from figure~\ref{fig:CBexMix} with auxiliary momenta and Mandelstam variables shown.}
    \label{fig:CB7mom}
\end{figure}
In the auxiliary momentum space, one assigns an incoming momentum to each external edge of the unrooted binary tree representation of the conformal block, such that the sum over all momenta is zero (see figure~\ref{fig:CB7mom}). 
Let $p_i$ be the momentum attached to the external edge labelled with conformal dimension $\Delta_i$, with the on-shell condition $p_i^2 = - \Delta_i$ and momentum conservation. 
Then the Mandelstam invariants associated to each internal leg are 
\eqn{7mixMandelMom}{
\begin{gathered}
s_1 = -(p_1+p_2)^2 \qquad s_2= -(p_1+p_2+p_3)^2 \cr 
s_3 = -(p_4+p_5)^2 \qquad s_4 = -(p_6+p_7)^2\,.
\end{gathered}
}
Using~\eno{MandelstamDef}, it is easy to see this gives back~\eno{7mixMandel}.

With the cross-ratios and the set ${\cal V}_{\rm 7,mixed}$ in place, the only computational task remaining is determining the post-Mellin parameters.  Recall that the Mellin parameters form the set~\eno{MellinIntegers} 
and the post-Mellin parameters for the external conformal dimensions/edges are given by~\eno{PostMellinExternal}.
Explicitly, for the present choice of cross-ratios, this yields
\eqn{7PostMellinExternal}{
\ell_1 &= j_{13}+j_{14}+j_{15}+j_{16} \qquad \ell_2= j_{24}+j_{26} \cr 
\ell_3 &= j_{13}+j_{34}+j_{35}+j_{36} \qquad \ell_4 = j_{14}+j_{24}+j_{34}+j_{46}
\cr
\ell_5 &= j_{15}+j_{35} \qquad \qquad \quad \qquad \ell_6 = j_{16}+j_{26}+j_{36}+j_{46} \qquad\quad  \ell_7 = 0\,.
}
Now one can solve for the post-Mellin parameters for the internal edges/exchanged dimensions using the algorithm described around \eno{PostMellinInternal1}-\eno{PostMellinInternal2}.
For illustrative purposes, we work it out explicitly for each internal leg below.
\begin{enumerate}
    \item First we consider all vertices with precisely two incident external edges and one incident internal edge:
             \eqn{}{
            \DrawVertexPostMellin{\Delta_1}{}{\Delta_2}{}{\Delta_{\delta_1}}{1} &: \quad  \ell_{\delta_1} \stackrel{2{\cal J}}{=} \ell_1 + \ell_2 \Rightarrow \ell_{\delta_1} = j_{13} + j_{14} + j_{15} + j_{16} + j_{24} + j_{26} \cr 
            \DrawVertexPostMellin{\Delta_4}{}{\Delta_5}{}{\Delta_{\delta_3}}{1} &: \quad  \ell_{\delta_3} \stackrel{2{\cal J}}{=} \ell_4 + \ell_5 \Rightarrow \ell_{\delta_3} = j_{14} + j_{15} + j_{24} + j_{34} + j_{35} + j_{46} \cr 
            \DrawVertexPostMellin{\Delta_6}{}{\Delta_7}{}{\Delta_{\delta_4}}{1} &: \quad   \ell_{\delta_4} \stackrel{2{\cal J}}{=} \ell_6 + \ell_7 \Rightarrow \ell_{\delta_4} = j_{16} + j_{26} + j_{36} + j_{46}\,.
            }
    \item Finally, to determine $\ell_{\delta_2}$, one can choose to look at one of two possible vertices. We will work it out using both to demonstrate choice-independence. From one choice of a vertex, we get
            \eqn{}{ 
             \DrawVertexPostMellin{\Delta_3}{}{\Delta_{\delta_1}}{1}{\Delta_{\delta_2}}{1} : \quad  \ell_{\delta_2} \stackrel{2{\cal J}}{=} \ell_3 + \ell_{\delta_1} \Rightarrow \ell_{\delta_2} =  j_{14} + j_{15} + j_{16} + j_{24} + j_{26} + j_{34} + j_{35} + j_{36}\,.
             }
    On the other hand, the choosing the following vertex yields,
            \eqn{}{ 
             \DrawVertexPostMellin{\Delta_{\delta_3}}{1}{\Delta_{\delta_4}}{1}{\Delta_{\delta_2}}{1} : \quad  \ell_{\delta_2} \stackrel{2{\cal J}}{=} \ell_{\delta_3} + \ell_{\delta_4} \Rightarrow \ell_{\delta_2} =  j_{14} + j_{15} + j_{16} + j_{24} + j_{26} + j_{34} + j_{35} + j_{36}\,.
             }
        As promised, the assignments agree.
\end{enumerate}

Now, using~\eno{EdgeDef} and~\eno{VertexDef}, we can write down the internal  edge and  vertex factors for the conformal block. As described in section~\ref{INTEGERS}, to each internal edge associated with the Mandelstam invariant $s_i$ (see figure~\ref{fig:CB7mom} and equation~\eno{7mixMandel}), assign the single-trace parameter $k_i$. Then, the ($7-3=4$) edge factors are
\eqn{7mixEfac}{
 \DrawEdge{\Delta_{\delta_1}}{k_1} &: \quad  E_1 =  
{(\Delta_{\delta_1} - h +1)_{k_{1}} \over  (\Delta_{\delta_1})_{2k_1 + j_{13}+j_{14}+j_{15}+j_{16}+j_{24}+j_{26}} }
\cr 
 \DrawEdge{\Delta_{\delta_2}}{k_2}  &: \quad E_2 =  
{(\Delta_{\delta_1} - h +1)_{k_{2}} \over  (\Delta_{\delta_2})_{2k_2 + j_{14}+j_{15}+j_{16}+j_{24}+j_{26}+j_{34}+j_{35}+j_{36}} }
\cr 
 \DrawEdge{\Delta_{\delta_3}}{k_3} &: \quad E_3 = {(\Delta_{\delta_1} - h +1)_{k_{3}} \over  (\Delta_{\delta_3})_{2k_3 +  j_{14}+j_{15}+j_{24}+j_{34}+j_{35}+j_{46} } }
\cr 
\DrawEdge{\Delta_{\delta_4}}{k_4} &: \quad E_4  = {(\Delta_{\delta_4} - h +1)_{k_{4}} \over  (\Delta_{\delta_4})_{2k_4 + j_{16}+j_{26}+j_{36}+j_{46}} }\,,
}
and the ($7-2=5$)  internal vertices of the unrooted binary tree, listed here:
\eqn{7mixVFig}{
\begin{gathered}
V_1 : \DrawVertex{\Delta_1}{}{\Delta_2}{}{\Delta_{\delta_1}}{k_1} \qquad 
V_2 : \DrawVertex{\Delta_4}{}{\Delta_5}{}{\Delta_{\delta_3}}{k_3} \qquad 
V_3 : \DrawVertex{\Delta_6}{}{\Delta_7}{}{\Delta_{\delta_4}}{k_4} \cr
V_4 : \DrawVertex{\Delta_{\delta_1}}{k_1}{\Delta_{\delta_2}}{k_2}{\Delta_3}{} \qquad 
V_5 : \DrawVertex{\Delta_{\delta_2}}{k_2}{\Delta_{\delta_3}}{k_3}{\Delta_{\delta_4}}{k_4}\,,
\end{gathered}
}
give the following vertex factors:
\eqn{7mixVfac}{
V_1 &= 
(\Delta_{12,\delta_1})_{-k_1} 
(\Delta_{2\delta_1,1})_{k_1 + j_{24}+j_{26}} 
(\Delta_{1\delta_1,2})_{k_1 + j_{13}+j_{14}+j_{15}+j_{16}}
\cr
&\quad  \times F_A^{(1)}\left[\Delta_{12\delta_1,}- h; \{-k_1\}; \left\{\Delta_{\delta_1} -h+1\right\}; 1 \right]
\cr
V_2&=
(\Delta_{45,\delta_3})_{-k_3} 
(\Delta_{5\delta_3,4})_{k_3 + j_{15}+j_{35}} 
(\Delta_{4\delta_3,5})_{k_3 + j_{14}+j_{24}+j_{34}+j_{46}}
\cr
&\quad  \times F_A^{(1)}\left[\Delta_{45\delta_3,}- h; \{-k_3 \}; \left\{\Delta_{\delta_3} -h+1\right\};1\right]
\cr
V_3&=
(\Delta_{67,\delta_4})_{-k_4} 
(\Delta_{7\delta_4,6})_{k_4}
(\Delta_{6\delta_4,7})_{k_4 + j_{16}+j_{26}+j_{36}+j_{46}}
\cr
&\quad  \times F_A^{(1)}\left[\Delta_{67\delta_4,}- h; \{-k_4 \}; \left\{\Delta_{\delta_4} -h+1\right\}; 1 \right]
\cr
V_4 &=
(\Delta_{3 \delta_1,\delta_2})_{k_1-k_2 + j_{13}} 
(\Delta_{3\delta_2, \delta_1})_{k_2-k_1 + j_{34}+j_{35}+j_{36}} 
(\Delta_{\delta_1\delta_2,3})_{k_1+k_2+j_{14}+j_{15}+j_{16}+j_{24}+j_{26}} 
\cr
&\quad  \times F_A^{(2)}\left[\Delta_{3\delta_1\delta_2,}- h; \{-k_1, -k_2\}; \left\{\Delta_{\delta_1} -h+1, \Delta_{\delta_2} -h+1 \right\}; 1,1\right]
\cr
V_5&=
(\Delta_{ \delta_3\delta_4,\delta_2})_{k_3+k_4-k_2+j_{46}} 
(\Delta_{\delta_2 \delta_4,\delta_3})_{k_2+k_4-k_3+j_{16}+j_{26}+j_{36}} 
\cr
&\quad \times
(\Delta_{\delta_2 \delta_3,\delta_4})_{k_2+k_3-k_4+j_{14}+j_{15}+j_{24}+j_{34}+j_{35}} 
\cr
& \quad  \times F_A^{(3)}\!\left[\Delta_{\delta_2 \delta_3 \delta_4,}- h; \{-k_2, -k_3, -k_4 \}; \left\{\Delta_{\delta_2} -h+1, \Delta_{\delta_3} -h+1,\Delta_{\delta_4} -h+1 \right\}; 1,1,1 \right]\!.
}
The Lauricella functions appearing above can be simplified further into combinations of Pochhammer symbols  and generalized hypergeometric functions (see appendix~\ref{LAURICELLA}).
The final expression for the conformal block is given by~\eno{CBexpansion}-\eno{Feynman} with the cross-ratios, edge and vertex factors as determined above.

In section~\ref{7MIXPROOF} we will reproduce the seven-point mixed-channel conformal block of this section using holographic techniques which will involve the Mellin amplitude of a particular seven-point Witten diagram as the starting point.

\subsection{$n$-point comb channel block}
\label{NCOMB}

In this section we will illustrate how to apply the Feynman rules to reproduce the $n$-point comb channel conformal block (see figure~\ref{fig:CBexComb}) of ref.~\cite{Parikh:2019dvm}.
The first step involves picking the cross-ratios; in this section, we choose those from ref.~\cite{Parikh:2019dvm},\footnote{In ref.~\cite{Fortin:2019zkm} the authors obtained an alternate expression for the conformal block based on a different choice of cross-ratios than the one used in ref.~\cite{Parikh:2019dvm}. Nevertheless these different forms are expected to be equivalent. As noted previously, one of the places the dependence on the choice of cross-ratios shows up in the series expansion is in the precise post-Mellin parameters appearing in the Pochhammer symbol.  Moreover, in ref.~\cite{Rosenhaus:2018zqn} the author used a different set of cross-ratios for the five-point block. The proposed Feynman rules applied to this choice of cross-ratios readily reproduces the block obtained there.  }
\eqn{CombCR}{
u_i := {x_{1 (i+1)}^2 x_{(i+2)n}^2 \over x_{1(i+2)}^2  x_{(i+1)n}^2} \qquad 1\leq i \leq n-3\,, \qquad\quad v_{rs} := {x_{1n}^2 x_{rs}^2 \over x_{1s}^2 x_{rn}^2 } \qquad 2 \leq r < s \leq n-1\,.
}
It turns out, the associated canonical choice of $n$ {\it dependent} Mellin variables is given by the set
\eqn{DComb}{
\{ \gamma_{ij} : (ij) \in {\cal D}_{\rm comb} \} \quad {\rm where} \quad {\cal D}_{\rm comb} := \big\{ (jn) : 1 \leq j \leq n-1 \big\} \,\bigcup\, \big\{ (1(n-1)) \big\}\,.
}
Explicitly, the dependent variables take the form,\footnote{For a different choice of basis of cross-ratios, there will accordingly be a different canonical basis of independent Mellin variables.}
 \eqn{CombMellin}{
  \gamma_{1(n-1)} &= \Delta_{12\ldots (n-1),n} - \sum_{j=2}^{n-2} \gamma_{j(n-1)} - \sum_{j=2}^{n-2}\gamma_{1j} - \sum_{2 \leq i < j \leq n-2} \gamma_{ij} \cr 
 \gamma_{1n} &= \Delta_{1n, 23\ldots (n-1)} + \sum_{j=2}^{n-2}\gamma_{j(n-1)} + \sum_{2 \leq i < j \leq n-2} \gamma_{ij} \cr 
  \gamma_{in} &=  - \sum_{j=1}^{n-1} \gamma_{ij} \qquad (2 \leq i \leq n-2) \cr
 \gamma_{(n-1)n} &= \Delta_{(n-1)n,12\ldots (n-2)} + \sum_{j=2}^{n-2}\gamma_{1j} + \sum_{2 \leq i < j \leq n-2} \gamma_{ij} \,,
 }
which, along with $\gamma_{ij}=\gamma_{ji}$, and $\gamma_{ii}=-\Delta_i$ for all $i,j$ explicitly solves~\eno{MellinConstraints} as required.

After substituting in~\eno{CombMellin}, the Mellin product, expressed in terms of the cross-ratios~\eno{CombCR} becomes
\eqn{CombMellinProd}{
\prod_{1 \leq i < j \leq n} {1 \over (x_{ij}^2)^{\gamma_{ij}}} =: W_{n,{\rm comb}}^0(x_i) \left(\prod_{i=1}^{n-3} u_i^{s_{i}/2} \right) \left( \prod_{2 \leq r < s \leq n-1} v_{rs}^{-\gamma_{rs}} \right),
}
where the leg-factor $W_{n,{\rm comb}}^0$ is given by
\eqn{W0comb}{
W_{n,{\rm comb}}^0(x_i) = \left( { x_{2n}^2 \over x_{1n}^2 x_{12}^2} \right)^{\Delta_1 \over 2} \left( { x_{1(n-1)}^2 \over x_{1n}^2 x_{(n-1)n}^2} \right)^{\Delta_n \over 2} \prod_{i=2}^{n-1} \left( { x_{1n}^2 \over x_{1i}^2 x_{in}^2} \right)^{\Delta_i \over 2},
}
and the $s_{i}$ are expressible as,
\eqn{CombMandelstamExplicit}{
s_i = \sum_{j=1}^{i+1} \Delta_j - 2 \sum_{1 \leq r < s \leq i+1} \gamma_{rs} \qquad 1\leq i \leq n-3\,.
}
The set in~\eno{DComb} is the canonical choice of dependent Mellin variables precisely because it leads directly to~\eno{CombMellinProd}.

\begin{figure}
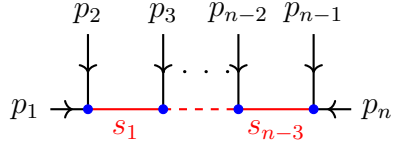

\centering
         \[  \musepic{\figNCombChannelMomentum} \]
    \caption{{\it Auxiliary momenta and Mandelstam variable assignments:}  Graphical representation of $n$-point conformal blocks in the  comb and channel from figure~\ref{fig:CBexComb} with auxiliary momenta and Mandelstam variables shown.}
    \label{fig:CBmomComb}
\end{figure}
As desired, the $s_i$ are the $n-3$ Mandelstam invariants associated with the $n-3$ internal edges. 
To see this, pass again to the auxiliary momentum space, 
and assign an incoming momentum to each external edge of the unrooted binary tree representation of the conformal block, such that the sum over all momenta is zero. 
Let $p_i$ be the momentum attached to the external edge labelled with conformal dimension $\Delta_i$, with the on-shell condition $p_i^2 = - \Delta_i$ and momentum conservation, with the identification $\gamma_{ij}:=p_i \cdot p_j$. 
Then the Mandelstam invariants associated to each internal leg are (see figure~\ref{fig:CBmomComb})
\eqn{CombMandelstam}{
s_i := -(p_1 + \cdots+ p_{i+1})^2 \qquad 1\leq i \leq n-3\,,
}
which precisely evaluates to~\eno{CombMandelstamExplicit}.
Additionally, as described in section~\ref{INTEGERS}, to each internal edge with Mandelstam invariant $s_i$, we also assign the single-trace parameter $k_i$.

Furthermore, from~\eno{CombMellinProd} we also identify the index set
\eqn{VComb}{
{\cal V}_{\rm comb} &:= \big\{ (rs) : 2 \leq r  < s \leq n-1 \big\} \,,
}
which leads directly to the set of Mellin parameters~\eno{MellinIntegers}. 
This, in turn, allows us to determine the post-Mellin parameters in terms of Mellin parameters.
First, let's work out the post-Mellin parameters associated to external dimensions/external edges~\eno{PostMellinExternal}:\footnote{For example, for $n\geq 8$
\eqn{}{
\begin{gathered} 
\ell_2 = \sum_{s=3}^{n-1} j_{2s} \qquad \qquad 
\ell_3 = j_{23} + \sum_{s=4}^{n-1} j_{3s} \qquad \qquad 
\ell_4 = \sum_{r=2}^3 j_{r4} + \sum_{s=5}^{n-1} j_{4s} \cr 
\ell_5 = \sum_{r=2}^4 j_{r5} + \sum_{s=6}^{n-1} j_{5s} \qquad \qquad 
\ell_6 = \sum_{r=2}^5 j_{r6} + \sum_{s=7}^{n-1} j_{6s} \qquad \qquad 
\ell_7 = \sum_{r=2}^6 j_{r7} + \sum_{s=8}^{n-1} j_{7s} \,.
\end{gathered} 
 } 
}
\eqn{}{
\begin{gathered}
\ell_1 = 0 \qquad \qquad  \ell_n = 0 \cr 
\ell_k = \sum_{r=2}^{k-1} j_{rk} + \sum_{s=k+1}^{n-1} j_{ks} \qquad \qquad (2\leq k \leq  n-1)\,. 
\end{gathered}
}
Next, applying~\eno{PostMellinInternal1} to the internal vertices at either extremes of the comb channel, we obtain the post-Mellin parameters $\ell_{\delta_1}$ and $\ell_{\delta_{n-3}}$:
\eqn{CombPMextremes}{
\DrawVertexPostMellin{\Delta_1}{}{\Delta_2}{}{\Delta_{\delta_1}}{1} &: \quad  \ell_{\delta_1} \stackrel{2{\cal J}}{=} \ell_1 + \ell_2 \qquad \Rightarrow \ell_{\delta_1} = \ell_{2} = \sum_{s=3}^{n-1} j_{2s} \cr 
\DrawVertexPostMellin{\Delta_{n-1}}{}{\Delta_n}{}{\Delta_{\delta_{n-3}}}{1} &: \quad  \ell_{\delta_{n-3}} \stackrel{2{\cal J}}{=} \ell_{n-1} + \ell_n \qquad \Rightarrow \ell_{\delta_{n-3}} = \ell_{n-1} =  \sum_{r=2}^{n-2} j_{r(n-1)}\,. 
}

Finally to determine the remaining post-Mellin parameters, we use~\eno{PostMellinInternal2} on vertices with two internal edges and one external edge attached. For example, one can start with the vertex:
\eqn{}{
\DrawVertexPostMellin{\Delta_{\delta_1}}{1}{\Delta_3}{}{\Delta_{\delta_2}}{1} &: \quad  \ell_{\delta_2} \stackrel{2{\cal J}}{=} \ell_{\delta_1} + \ell_3 \qquad \Rightarrow \ell_{\delta_2} = \sum_{s=4}^{n-1} j_{2s} + \sum_{s=4}^{n-1} j_{3s}\,,
}
and then proceed one vertex to the right:
\eqn{}{
\DrawVertexPostMellin{\Delta_{\delta_2}}{1}{\Delta_4}{}{\Delta_{\delta_3}}{1} &: \quad  \ell_{\delta_3} \stackrel{2{\cal J}}{=} \ell_{\delta_2} + \ell_4 \qquad \Rightarrow \ell_{\delta_3} = \sum_{s=5}^{n-1} j_{2s} + \sum_{s=5}^{n-1} j_{3s} + \sum_{s=5}^{n-1} j_{4s} \,,
}
and so on.
Proceeding iteratively, we find
\eqn{pMComb}{
\ell_{\delta_i} = \sum_{r=2}^{i+1} \sum_{s=i+2}^{n-1} j_{rs} \qquad \qquad (1\leq i\leq n-3)\,,
}
where the results from~\eno{CombPMextremes} have been included in the formula above.

This is the full extent of computations needed to write down the $n$-point comb channel conformal block.
The final step involves substituting the single-trace and post-Mellin parameters into the edge~\eno{EdgeDef} and vertex~\eno{VertexDef} factors, which immediately yields the explicit conformal block via~\eno{Feynman}.

For the sake of completeness, we provide the explicit edge and vertex factors below. The $(n-3)$ edge factors are:
\eqn{CombEdge}{
\DrawEdge{\Delta_{\delta_i}}{k_i} &: \quad E_i  = {(\Delta_{\delta_i} - h +1)_{k_i} \over  (\Delta_{\delta_i})_{2k_i + \sum_{r=2}^{i+1} \sum_{s=i+2}^{n-1} j_{rs} } }
 \qquad\qquad  (1\leq i\leq n-3)\,.
}
Similarly, for the vertex factors we simply substitute all the ingredients from above into~\eno{VertexDef}. To facilitate comparison with the result from ref.~\cite{Parikh:2019dvm} (as well as the new derivation in section~\ref{COMBPROOF}) we will simplify the linear combination of post-Mellin parameters appearing in the vertex factors.
Rewriting,
\begingroup\makeatletter\def\f@size{10}\check@mathfonts
\[ \ell_{\delta_{i}} =  \sum_{r=2}^{i+1} \sum_{s=i+3}^{n-1} j_{rs}  + \sum_{r=2}^{i+1} j_{r(i+2)} \quad
\ell_{\delta_{i+1}} =  \sum_{r=2}^{i+1} \sum_{s=i+3}^{n-1} j_{rs}  + \sum_{s=i+3}^{n-1} j_{(i+2)s} \quad
\ell_{i+2} = \sum_{r=2}^{i+1} j_{r(i+2)} + \sum_{s=i+3}^{n-1} j_{(i+2)s} \]
\endgroup
for $0 \leq i\leq n-3$ where we used the identifications $\Delta_{\delta_0} := \Delta_1$ and $\Delta_{\delta_{n-2}} := \Delta_n$, simple arithmetic leads to
\eqn{}{
{1\over 2} \ell_{(i+2)\delta_i ,\delta_{i+1}} = \sum_{r=2}^{i+1} j_{r(i+2)} \qquad 
{1\over 2} \ell_{\delta_i \delta_{i+1}, (i+2)} = \sum_{r=2}^{i+1} \sum_{s=i+3}^{n-1} j_{rs} \qquad 
{1\over 2} \ell_{(i+2)\delta_{i+1} ,\delta_i} = \sum_{s=i+3}^{n-1} j_{(i+2)s}
}
for $0\leq i\leq n-3$.
Then the $n-2$ internal vertices,
\eqn{CombVFig}{
V_1 : \DrawVertex{\Delta_1}{}{\Delta_2}{}{\Delta_{\delta_1}}{k_1}  \qquad 
V_{i+1} : \DrawVertex{\Delta_{\delta_i}}{k_i}{\Delta_{\delta_{i+1}}}{k_{i+1}}{\Delta_{i+2}}{} \qquad 
V_{n-2} : \DrawVertex{\Delta_{n-1}}{}{\Delta_n}{}{\Delta_{\delta_{n-3}}}{k_{n-3}}   \,,
}
for $1 \leq i \leq n-4$, are associated with the vertex factors
\eqn{CombVertex}{
V_1 &= (\Delta_{12,\delta_1})_{-k_1} (\Delta_{\delta_11,2})_{k_1} (\Delta_{\delta_12,1})_{k_1+ \sum_{s=3}^{n-1} j_{2s} }  \cr 
     & \quad  \times F_A^{(1)}\!\left[\Delta_{12\delta_1,}- h; \{-k_1 \}; \left\{\Delta_{\delta_1} -h+1\right\}; 1 \right]  
\cr 
V_{i+1} &:= (\Delta_{\delta_i\delta_{i+1},i+2})_{k_{i(i+1),}+ \sum_{r=2}^{i+1} \sum_{s=i+3}^{n-1} j_{rs} } 
(\Delta_{(i+2) \delta_i,\delta_{i+1}})_{k_{i,i+1}+ \sum_{r=2}^{i+1} j_{r(i+2)} } \cr 
   &\quad \times 
(\Delta_{(i+2) \delta_{i+1},\delta_i})_{k_{i+1,i}+ \sum_{s=i+3}^{n-1} j_{(i+2)s} } \cr 
     &\quad   \times F_A^{(2)}\!\left[\Delta_{\delta_i \delta_{i+1} (i+2),}- h; \{-k_i, -k_{i+1}\}; \left\{\Delta_{\delta_i} -h+1, \Delta_{\delta_{i+1}} -h+1 \right\}; 1,1 \right] \cr 
V_{n-2} &= (\Delta_{(n-1)n,\delta_{n-3}})_{-k_{n-3}} (\Delta_{\delta_{n-3} (n-1),n})_{k_{n-3}+ \sum_{r=2}^{n-2} j_{r(n-1)} } (\Delta_{\delta_{n-3}n,n-1})_{k_{n-3}} \cr 
     & \quad  \times F_A^{(1)}\!\left[\Delta_{(n-1)n\delta_{n-3},}- h; \{-k_{n-3} \}; \left\{\Delta_{\delta_{n-3}} -h+1 \right\}; 1 \right]  .
}
 One can re-express the Lauricella functions of one and two variables above in terms of Pochhammer symbols and the generalized hypergeometric function ${}_3F_2$, respectively (see~\eno{FA1}-\eno{FA2}).
Upon doing so, the final conformal block given by substituting the edge and vertex factors above into~\eno{CBexpansion}-\eno{Feynman} finds precise agreement with the result of ref.~\cite{Parikh:2019dvm}. 

Furthermore, one can check this reproduces the well-known four-point block upon setting $n=4$. 
Finally, ref.~\cite{Rosenhaus:2018zqn} worked out the $n=5$ block for a different set of cross-ratios. 
Starting with those cross-ratios as the input data, we checked that the Feynman rules reproduced precisely the block of ref.~\cite{Rosenhaus:2018zqn}. 
Generally,  blocks from different choices of cross-ratios, though perhaps not manifestly identical, are still equivalent in the shared domain of convergence. In particular it can be checked that the five-point blocks of ref.~\cite{Parikh:2019dvm}, ref.~\cite{Fortin:2019dnq} and ref.~\cite{Rosenhaus:2018zqn} are equivalent, even though they seem slightly different.

\subsection{$n$-point OPE channel block}
\label{NOPE}

In the OPE channel (see figure~\ref{fig:CBexOPE}), we choose to represent the $n$-point conformal block (for even $n \geq 6$) in terms of the following cross-ratios:\footnote{Note that for certain choices of $(ij)$, the cross-ratio $v_{ij}$ in~\eno{OPECR} identically evaluates to unity, in which case it is not to be included as an independent cross-ratio. The counting leading to $n(n-3)/2$ independent cross-ratios accounts for such occurrences.}
\eqn{OPECR}{
\begin{gathered}
u_1 := {x_{12}^2 x_{4n}^2 \over x_{1n}^2 x_{24}^2} \qquad \qquad \qquad u_{n-3} := { x_{(n-1)n}^2 x_{2(n-2)}^2 \over x_{(n-2)(n-1)}^2 x_{2n}^2}   \cr 
u_{2 \times j} := {x_{(2j+1)(2j+2)}^2 x_{2n}^2  \over  x_{(2j+1)n}^2 x_{2(2j+2)}^2} \quad (1\leq j\leq {n\over 2}-2) 
\qquad u_{2\times j-1} := {x_{2(2j)}^2 x_{(2j+2)n}^2 \over x_{(2j)n}^2 x_{2(2j+2)}^2 } \quad (2\leq j\leq {n\over 2}-2) \cr 
v_{i(n-1)} := {x_{i(n-1)}^2 x_{(n-2)n}^2 \over x_{in}^2 x_{(n-2)(n-1)}^2} \qquad (1 \leq i \leq n-3) \cr 
v_{ij} := {x_{ij}^2  x_{2n}^2 x_{(2t)n}^2 \over x_{in}^2 x_{jn}^2 x_{2(2t)}^2}  \qquad (2 \leq t \leq {n\over 2}-1\,,\;{\rm with\ } 1 \leq i \leq 2t-2\,,\; 2t-1 \leq j \leq 2t)\,.
\end{gathered}
}

A convenient choice of dependent Mellin variables associated with the choice of cross-ratios above is given by the following index set
\eqn{DOPEDef}{
{\cal D}_{\rm OPE} := \big\{ (in)\, \big|\, 1 \leq i \leq n-2 \big\} \bigcup \big\{ ((n-2)(n-1)), (2(n-2)) \big\}\,.
}
The dependent Mellin variables take the form
\eqn{OPEMellin}{
\gamma_{in} &=  -\sum_{j=1}^{n-1} \gamma_{ij} \qquad (i=1,3,4, 5, 6, \ldots, n-4, n-3) \cr 
\gamma_{2n} &= \Delta_{2(n-1)n,134\ldots (n-2)} -\gamma_{(n-1)n} - \gamma_{2(n-1)} + \sum_{j=3}^{n-2} \gamma_{1j} + \sum_{3\leq i<j\leq (n-2)} \gamma_{ij} \cr 
\gamma_{(n-2)n} &=  \Delta_{(n-2)n, 123\ldots(n-3)(n-1)} + \sum_{j=1}^{n-3} \gamma_{j(n-1)} + \sum_{1 \leq i < j \leq n-3} \gamma_{ij} \cr 
\gamma_{(n-2)(n-1)} &= -\sum_{\substack{j=1\\ j \neq n-2}}^{n} \gamma_{j(n-1)} \cr 
\gamma_{2(n-2)} &= \Delta_{12\ldots (n-2),(n-1)n} + \gamma_{(n-1)n} - \sum_{j=2}^{n-2} \gamma_{1j} - \sum_{j=3}^{n-3} \gamma_{2j}  - \sum_{3 \leq i<j\leq n-2} \gamma_{ij}  \,.
}
\begin{figure}
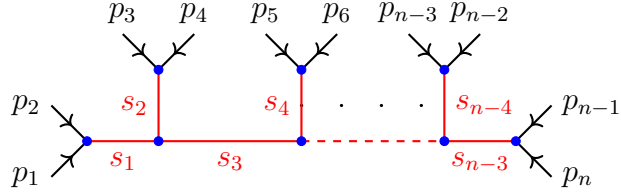

\centering
          \[  \musepic{\figNOPEChannelMomentum}  \]
    \caption{{\it Auxiliary momenta and Mandelstam variable assignments:}  Graphical representation of $n$-point conformal blocks in the  OPE channel from figure~\ref{fig:CBexOPE} with auxiliary momenta and Mandelstam variables shown.}
    \label{fig:CBmomOPE}
\end{figure}
Just like for the comb channel, it is useful to consider the auxiliary momentum space in the OPE channel.
We use the convention for momentum and Mandelstam invariant assignments as depicted in figure~\ref{fig:CBmomOPE}.
In this convention, the Mandelstam invariants associated with the internal legs take the following explicit form in terms of the independent Mellin variables:
\eqn{OPEMandelstam}{
s_1 &:= -(p_1 + p_2)^2 = \Delta_1 + \Delta_2 - 2\gamma_{12} \cr 
s_{n-3} &:= -(p_{n-1} + p_n)^2 = \Delta_{n-1} + \Delta_n - 2\gamma_{(n-1)n} \cr 
s_{2j} &:= -(p_{2j+1} + p_{2j+2})^2 = \Delta_{2j+1} + \Delta_{2j+2} - 2\gamma_{(2j+1)(2j+2)} \qquad (1 \leq j \leq {n\over 2}-2) \cr 
s_{2j-1} &:= -(p_1 + p_2 + p_3 + \cdots + p_{2j})^2 = \sum_{k=1}^{2j} \Delta_k - 2 \sum_{1 \leq r<s \leq 2j} \gamma_{rs} \qquad (2\leq j \leq {n\over 2}-2) \,.
}
With this in hand, it is straightforward to re-express the Mellin product~\eno{MellinProd} in terms of conformal cross-ratios~\eno{OPECR}:
\eqn{OPEMellinProd}{
\prod_{1 \leq i < j \leq n} {1 \over (x_{ij}^2)^{\gamma_{ij}}} = W_{n,{\rm OPE}}^0(x_i) \left(\prod_{i=1}^{n-3} u_i^{s_{i}/2} \right) \left( \prod_{(rs) \in {\cal V}_{\rm OPE}} v_{rs}^{-\gamma_{rs}} \right),
}
where the  leg-factor $W_{n,{\rm OPE}}^0$ is defined to be
\eqn{W0OPE}{
W_{n,{\rm OPE}}^0(x_i) &:= \left( { x_{(n-2)n}^2 \over x_{(n-2)(n-1)}^2 x_{(n-1)n}^2} \right)^{\Delta_{n-1} \over 2} \left( { x_{(n-2)(n-1)}^2 \over x_{(n-2)n}^2 x_{(n-1)n}^2} \right)^{\Delta_n \over 2} \cr 
 & \quad \times \left( \prod_{i=1}^{{n\over 2}-1} \left( { x_{(2i-1)n}^2 \over x_{(2i-1)(2i)}^2 x_{(2i)n}^2} \right)^{\Delta_{2i} \over 2} \left( { x_{(2i)n}^2 \over x_{(2i-1)(2i)}^2 x_{(2i-1)n}^2} \right)^{\Delta_{2i-1} \over 2} \right),
}
and the $(rs)$ index in the final product in~\eno{OPEMellinProd} runs over the index set 
\eqn{VOPEDef}{
{\cal V}_{\rm OPE} := \big\{ (ab) \; \big|\; 1 \leq a < b \leq n-1 \} \smallsetminus {\cal S}_{\rm OPE}\,,
}
where we have defined\footnote{Explicitly, ${\cal S}_{\rm OPE} = \big\{(12), (34), \ldots, ((n-3)(n-2)) , (24), (26), \ldots , (2(n-2)),  ((n-2)(n-1))\big\}$.}
\eqn{SOPEDef}{
{\cal S}_{\rm OPE} &:= \big\{ ((2j+1)(2j+2)) \; \big|\; 0 \leq j \leq {n\over 2}-2 \big\}\; \bigcup\; \big\{ (2(2j)) \; \big| \; 2 \leq j \leq {n\over 2}-1 \big\} \cr 
 &\quad \bigcup\; \big\{ ((n-2)(n-1)) \big\}\,.
}
It can be easily checked that the subscripts $(rs)$ take $|{\cal V}_{\rm OPE}| = \binom{n-2}{2}$ pairs of values.
The index set ${\cal V}_{\rm OPE}$ in turn allows us to compute the post-Mellin parameters for the external conformal dimensions via~\eno{PostMellinExternal} (it may be helpful here to consult figure~\ref{fig:OPESets} from section~\ref{OPEPROOF} for a visual representation of the set ${\cal V}_{\rm OPE}$):
 \eqn{opeExtpM}{
\begin{gathered}
\ell_{1}= \sum_{s=3}^{n-1} j_{1s} \qquad 
\ell_{2}= \sum_{i=1}^{{n \over 2}-1} j_{2(2i+1)} 
\qquad
\ell_{n}=0
\cr 
\ell_{n-1} = \sum_{r=1}^{n-3} j_{r(n-1)} \qquad 
\ell_{n-2}= j_{1(n-2)}+ \sum_{r=3}^{n-4} j_{r(n-2)} \qquad
\ell_{n-3}=j_{(n-3)(n-1)}+\sum_{r=1}^{n-4}j_{r(n-3)}
\cr 
\ell_{2i+2}=j_{1(2i+2)}+ \sum_{s=2i+3}^{n-1}j_{(2i+2)s} + \sum_{r=3}^{2i} j_{r(2i+2)} \qquad (1 \leq i \leq {n\over 2}-3)
\cr 
\ell_{2i+1}= \sum_{s=2i+3}^{n-1} j_{(2i+1)s}+\sum_{r=1}^{2i} j_{r(2i+1)} \qquad (1 \leq i \leq {n \over 2}-3)\,.
\end{gathered}
}
Now we can use~\eno{PostMellinInternal1} and~\eno{PostMellinInternal2} to determine the post-Mellin parameters for internal conformal dimensions iteratively. 
First, focusing on vertices with two incident external legs, we can determine the post-Mellin parameters from previously determined data, as shown:
 \eqn{opeIntpM1}{
\DrawVertexPostMellin{\Delta_1}{}{\Delta_2}{}{\Delta_{\delta_1}}{1} &:  \ell_{\delta_{1}} \stackrel{2{\cal J}}{=}  
\ell_{1}+\ell_{2} = \sum_{s=3}^{n-1} j_{1s}+\sum_{i=1}^{{n \over 2}-1} j_{2(2i+1)}
\cr 
\DrawVertexPostMellin{\Delta_{n-1}}{}{\Delta_n}{}{\Delta_{\delta_{n-3}}}{1} &: \ell_{\delta_{n-3}} \stackrel{2{\cal J}}{=} \ell_{n-1}+\ell_n  = \sum_{r=1}^{n-3} j_{r(n-1)}
\cr 
\DrawVertexPostMellin{\Delta_{n-3}\;\;}{}{\;\;\Delta_{n-2}}{}{\Delta_{\delta_{n-4}}}{1} &:   \ell_{\delta_{n-4}} \stackrel{2{\cal J}}{=} \ell_{n-3}+\ell_{n-2}  =  j_{(n-3)(n-1)}+\sum_{r=1}^{n-4}j_{r(n-3)}+j_{1(n-2)}+ \sum_{r=3}^{n-4} j_{r(n-2)}
\cr 
\DrawVertexPostMellin{ \Delta_{2i+1}\;\;}{}{\;\;\Delta_{2i+2}}{}{\Delta_{\delta_{2i}}}{1} &: \ell_{\delta_{2i}} \stackrel{2{\cal J}}{=} \ell_{2i+1} + \ell_{2i+2} = \sum_{s=2i+3}^{n-1} j_{(2i+1)s}+\sum_{r=1}^{2i} j_{r(2i+1)} + j_{1(2i+2)}+ \sum_{s=2i+3}^{n-1}j_{(2i+2)s} \cr 
 & \qquad \qquad \qquad \qquad \qquad + \sum_{r=3}^{2i} j_{r(2i+2)} 
}
for $1 \leq i \leq {n\over 2} -3 $. In fact, in all $\stackrel{2{\cal J}}{=}$ equalities in~\eno{opeIntpM1}, we can freely drop the $2{\cal J}$ as the post-Mellin parameters on the RHS do not share common Mellin parameters.

The post-Mellin parameters for the remaining internal conformal dimensions satisfy
\eqn{}{
\ell_{\delta_{2i+1}}\stackrel{2{\cal J}}{=} \ell_{\delta_{2i-1}}+\ell_{\delta_{2i}} \,,
}
for $1 \leq i \leq {n \over 2}-3 $, thus they need to be determined iteratively.
 Working out a few explicit cases such as $\ell_{\delta_3}, \ell_{\delta_5}$ and $\ell_{\delta_7}$ allows us to conjecture, and subsequently prove by induction in appendix~\ref{APP:ELLODDPROOF},  the general form
\eqn{ellOddGuess}{
\ell_{\delta_{2i+1}} &= \sum_{s=2i+3}^{n-1}j_{1s}+\sum_{r=3}^{2i+2}\sum_{s=2i+3}^{n-1}j_{rs}+\sum_{z=i+1}^{{n\over 2}-1} j_{2(2z+1)}\,,
}
for $1 \leq i \leq {n \over 2}-3$.

We now have all the ingredients to write down the full conformal block in the OPE channel. The $(n-3)$ edge factors are given by~\eno{EdgeDef} with the post-Mellin parameters as determined above.
The $(n-2)$ internal vertices of the binary graph, enumerated as follows
\eqn{OPEVFig}{
V_1^{(1)} : \DrawVertex{\Delta_1}{}{\Delta_2}{}{\Delta_{\delta_1}}{k_1}  \qquad 
V_{a}^{(1)} : \DrawVertex{\Delta_{2a-1}\;}{}{\;\Delta_{2a}}{}{\Delta_{\delta_{2a-2}}}{k_{2a-2}} \qquad  
V_{b}^{(3)} : \DrawVertex{\Delta_{\delta_{2b-1}}\;\;}{k_{2b-1}}{\;\;\Delta_{\delta_{2b}}}{k_{2b}}{\Delta_{\delta_{2b+1}}}{k_{2b+1}} \qquad 
V_{n/2}^{(1)} : \DrawVertex{\Delta_{n-1}}{}{\Delta_n}{}{\Delta_{\delta_{n-3}}}{k_{n-3}}   ,
}
for $2\leq a \leq n/2-1$ and $1 \leq b \leq n/2-2$, correspond to the vertex factors~\eno{VertexDef}. Explicitly\footnote{To write the vertex factors compactly, we made the additional stipulation that an ill-defined post-Mellin parameter, $j_{(n-2)(n-1)}$, which appears in the subscript of one of the Pochhammer symbols of $V_{n/2-1}^{(1)}$ as a consequence of our compact rewriting, should be set to zero.}
\eqn{OPEVertex}{
V_1^{(1)} &= (\Delta_{12,\delta_1})_{-k_1} (\Delta_{\delta_11,2})_{k_1+ \sum_{s=3}^{n-1} j_{1s}} (\Delta_{\delta_12,1})_{k_1+ \sum_{i=1}^{{n \over 2}-1} j_{2(2i+1)}}  \cr 
     & \quad  \times F_A^{(1)}\!\left[\Delta_{12\delta_1,}- h; \{-k_1 \}; \left\{\Delta_{\delta_1} -h+1\right\}; 1 \right]  
\cr 
V_{a}^{(1)} &:= (\Delta_{(2a-1)(2a),\delta_{2a-2}})_{-k_{2a-2} }    
(\Delta_{(2a-1) \delta_{2a-2},2a})_{k_{2a-2} + \sum_{s=2a+1}^{n-1} j_{(2a-1)s}+\sum_{r=1}^{2a-2} j_{r(2a-1)} } \cr 
   &\quad \times 
(\Delta_{(2a) \delta_{2a-2},2a-1})_{k_{2a-2} + j_{1(2a)}+ \sum_{s=2a+1}^{n-1}j_{(2a)s} + \sum_{r=3}^{2a-2} j_{r(2a)} } \cr 
     &\quad   \times F_A^{(1)}\!\left[\Delta_{(2a-1)(2a) \delta_{2a-2},}- h; \{-k_{2a-2}\}; \left\{\Delta_{\delta_{2a-2}} -h+1 \right\}; 1 \right] \cr 
V_{b}^{(3)} &:= (\Delta_{\delta_{2b-1}\delta_{2b},\delta_{2b+1}})_{k_{(2b-1)(2b),(2b+1)}+  \sum_{s=2b+1}^{2b+2}j_{1s} + \sum_{r=2}^{2b}j_{r(2b+1)} + \sum_{r=3}^{2b} j_{r(2b+2)} }  \cr 
&\quad \times 
(\Delta_{\delta_{2b+1} \delta_{2b-1},\delta_{2b}})_{k_{(2b+1)(2b-1),(2b)}+  \sum_{s=2b+3}^{n-1}j_{1s}+ \sum_{r=3}^{2b}\sum_{s=2b+3}^{n-1}j_{rs}+ \sum_{z=b+1}^{{n\over 2}-1} j_{2(2z+1)}} \cr 
   &\quad \times 
(\Delta_{\delta_{2b+1} \delta_{2b},\delta_{2b-1}})_{k_{(2b+1)(2b),(2b-1)}+ \sum_{r=2b+1}^{2b+2}\sum_{s=2b+3}^{n-1}j_{rs}  } \cr 
     &\quad   \times F_A^{(3)}\!\left[\Delta_{\delta_{2b-1} \delta_{2b} \delta_{2b+1},}- h; \{-k_{2b-1}, -k_{2b}, -k_{2b+1} \}; \right. \cr 
     & \qquad \qquad \left. \left\{\Delta_{\delta_{2b-1}} -h+1, \Delta_{\delta_{2b}} -h+1, \Delta_{\delta_{2b+1}} -h+1 \right\}; 1,1,1 \right] \cr 
V_{n-2}^{(1)} &= (\Delta_{(n-1)n,\delta_{n-3}})_{-k_{n-3}} (\Delta_{\delta_{n-3} (n-1),n})_{k_{n-3}+ \sum_{r=1}^{n-3} j_{r(n-1)} } (\Delta_{\delta_{n-3}n,n-1})_{k_{n-3}} \cr 
     & \quad  \times F_A^{(1)}\!\left[\Delta_{(n-1)n\delta_{n-3},}- h; \{-k_{n-3} \}; \left\{\Delta_{\delta_{n-3}} -h+1 \right\}; 1 \right]  .
}
where  to work out the linear combination of post-Mellin parameters in $V_a^{(1)}$ we used
\eqn{}{
{1\over 2} \ell_{(2a-1)(2a),\delta_{2a-2}} = 0 \qquad  
{1\over 2} \ell_{(2a-1)\delta_{2a-2},2a} = \ell_{2a-1} \qquad 
{1\over 2} \ell_{(2a)\delta_{2a-2},2a-1} = \ell_{2a} \,,
}
for $2\leq a \leq n/2-1$, and in $V_b^{(3)}$ we used
\eqn{}{
{1\over 2} \ell_{\delta_{2b-1}\delta_{2b},\delta_{2b+1}} &=  \sum_{s=2b+1}^{2b+2}j_{1s}
+ \sum_{r=2}^{2b}j_{r(2b+1)}
+ \sum_{r=3}^{2b} j_{r(2b+2)}   \cr
{1\over 2} \ell_{\delta_{2b+1}\delta_{2b-1},\delta_{2b}} &= \sum_{s=2b+3}^{n-1}j_{1s}+ \sum_{r=3}^{2b}\sum_{s=2b+3}^{n-1}j_{rs}+ \sum_{z=b+1}^{{n\over 2}-1} j_{2(2z+1)}   \cr 
{1\over 2} \ell_{\delta_{2b+1}\delta_{2b},\delta_{2b-1}} &= \sum_{r=2b+1}^{2b+2}\sum_{s=2b+3}^{n-1}j_{rs} \,,
}
for $1 \leq b \leq n/2-2$, which can both be verified easily.
Substituting all the edge and vertex factors above into~\eno{CBexpansion}-\eno{Feynman} gives the $n$-point OPE channel block.

In ref.~\cite{Fortin:2020yjz}, a power-series expansion was worked out for the special case of the $n=6$ block.\footnote{The authors of ref.~\cite{Fortin:2020yjz} referred to it as the ``snowflake channel,'' which is the same as the ``OPE channel'' above.} 
Notably, the same generalized hypergeometric function makes an appearance in both their paper and the result above. 
For $n=6$, it can be seen from~\eno{OPEVertex} that in all exactly one factor of the Lauricella function $F_A^{(3)}$ appears in the vertex factors, which is directly related to the Kamp\'{e} de F\'{e}riet function via~\eno{FA3}. 
Precisely the same Kamp\'{e} de F\'{e}riet function appeared in the result of ref.~\cite{Fortin:2020yjz}. 
The choice of cross-ratios in that paper differs from the general choice made above in~\eno{OPECR}, so to make a precise comparison, we can start with the cross-ratios of ref.~\cite{Fortin:2020yjz} and apply to them the Feynman rules of section~\ref{FEYNMAN}.
We confirmed that doing so exactly reproduces the conformal block of ref.~\cite{Fortin:2020yjz}.

This section generalizes this result to any even $n \geq 6$.\footnote{Recall that the ``OPE channel'' in this paper is well-defined only for even number of external operators.} At higher $n$, precisely $(n/2-2)$ factors of the Lauricella function of three variables $F_A^{(3)}$ (equivalently $(n/2-2)$ factors of the Kamp\'{e} de F\'{e}riet function) will appear in the power series expansion.\footnote{The remaining vertex factors in~\eno{OPEVertex} contribute one factor of the Lauricella function $F_A^{(1)}$ each, but this function can be trivially expressed in terms of Gamma functions or Pochhammer symbols; see~\eno{FA1}.}

\section{From Mellin amplitudes to conformal blocks}
\label{PROOF}

In this section we will prove the Feynman prescription for conformal blocks for the examples considered in the previous section. 
As outlined in section~\ref{INTRO}, our starting point will be certain canonical tree-level Witten diagrams in an effective $\phi^3$ scalar field theory in AdS. 
We will write down their Mellin amplitudes  using the Feynman rules for Mellin amplitude~\cite{Fitzpatrick:2011ia,Paulos:2011ie,Nandan:2011wc}, from which we will be able to extract the desired conformal blocks via single-trace projections. 
A canonical Witten diagram is a tree-level Witten diagram of the same topology as the conformal block we are interested in computing, with generic scalar dimensions running along each edge.
Since we will only be interested in the contribution coming from single-trace exchanges, and not the full amplitude, we would like to project onto the single-trace part of this Witten diagram.
This is convenient to do in large $N$ theories, since the poles of the meromorphic Mellin amplitude correspond precisely to the exchange of single-trace primaries. Evaluating the residue at these poles  furnishes the required projection.

Concretely, the Mellin amplitude ${\cal M}_n(\gamma_{ij})$ for an $n$-point Witten diagram, whose position space ampitude is denoted $A_n$, is defined via a multi-dimensional inverse Mellin transform,
\eqn{MellinAmpDef}{
A_n = {\cal N} \left( \prod_{(rs) \in {\cal U} \bigcup {\cal V}} \int {d\gamma_{rs} \over 2\pi i} \right) {\cal M}_n(\gamma_{ij})  \left(\prod_{(ij) \in {\cal U} \bigcup {\cal V} \bigcup {\cal D}} {\Gamma(\gamma_{ij}) \over (x_{ij}^2)^{\gamma_{ij}} } \right),
}
where $A_n$ is an AdS integral over products of bulk-to-bulk and bulk-to-boundary propagators which are normalized as follows: In Poincar\'{e} coordinates $z=(z_0,z^i) \in \mathbb{R}^{+} \times \mathbb{R}^d$,
\eqn{GKNorm}{
	\hat{G}_\Delta(w,z)  &= \left({\xi(w,z) \over 2}\right)^\Delta\: {}_2F_1\left[{\Delta\over 2},{\Delta+1 \over 2};\Delta-{d\over 2}+1;\xi(w,z)^2\right] \cr 
	\xi(w,z) &= {2 w_0 z_0 \over w_0^2 + z_0^2 + (w^i-z^i)^2 } \cr 
	\hat{K}_\Delta(x^i,z) &= {z_0^\Delta \over (z_0^2 + (z^i- x^i)^2)^{\Delta}}\,.
}
The contours of integration on the RHS of~\eno{MellinAmpDef} run parallel to the imaginary axis for $\Re \gamma_{rs} >0$ such that they separate out the semi-infinite sequence of poles running to the left or to the right.
 The overall normalization constant
 ${\cal N}$ will be fixed shortly.    
 The set ${\cal U} \bigcup {\cal V}$ is the index set of $n(n-3)/2$ independent Mellin variables $\gamma_{ij}$ and the set ${\cal D}$ is the index set of the $n$ dependent Mellin variables.
 We have chosen to decompose the independent variable index set into a union of two disjoint subsets ${\cal U}$ and ${\cal V}$.
 The precise prescription for this choice of sets was explained in section~\ref{INTEGERS} and illustrated in section~\ref{FEYNMANEXAMPLES}. 
 Briefly, this  choice will be dictated by the choice of cross-ratios and the channel (i.e.\ binary tree topology) for the precise conformal block we wish to extract from $A_n$.
 To stress this dependence, we will call the sets  ${\cal U}_{\rm chan}, {\cal V}_{\rm chan}$ and ${\cal D}_{\rm chan}$.
The choice of cross-ratios and the index sets served as the input in section~\ref{FEYNMANEXAMPLES} for writing down the conformal block using the proposed Feynman rules.
In this section, this choice will serve as the input for deriving the block from Mellin amplitudes.
 
 The union of all three sets ${\cal U}_{\rm chan}, {\cal V}_{\rm chan}$ and ${\cal D}_{\rm chan}$ gives the full range of indices~\eno{UnionAll} associated with the product over Gamma functions and powers of pairwise distances in~\eno{MellinAmpDef}. 
 This product over powers of pairwise distances was called the ``Mellin product''; see~\eno{MellinProd}. For a canonical choice of index sets, the Mellin product admits a convenient rewriting, to be substituted in~\eno{MellinAmpDef}, in terms of the chosen cross-ratios as shown in~\eno{ArbMellinProd} and repeated below:
 \eqn{ArbMellinProdAgain}{ 
 \left(\prod_{(ij) \in {\cal U}_{\rm chan}  \bigcup {\cal V}_{\rm chan} \bigcup {\cal D}_{\rm chan} } {1 \over (x_{ij}^2)^{\gamma_{ij}} } \right) =  W_{n, {\rm chan}}^0(x_i) \left(\prod_{i=1}^{n-3} u_i^{s_{i}/2} \right) \left( \prod_{(rs) \in {\cal V}_{\rm chan}} v_{rs}^{-\gamma_{rs}} \right) ,
 }
where $s_i$ are the Mandelstam invariants associated with the internal legs (and expressible in terms of Mellin variables drawn from the index sets ${\cal U}_{\rm chan}$ and ${\cal V}_{\rm chan}$), while $W_{n, {\rm chan}}^0(x_i)$ is the leg-factor which depends solely on external conformal dimensions and position coordinates.
We note that we have indexed the Mandelstam invariants and the cross-ratios in a manner that allows us to write~\eno{ArbMellinProdAgain} as displayed.

To evaluate the single-trace contribution to $A_n$, one needs the Mellin amplitude ${\cal M}(\gamma_{ij})$ for the Witten diagram.
For tree-level scalar Witten diagrams, the Mellin amplitude is readily available via the ``Feynman rules for Mellin amplitudes''~\cite{Fitzpatrick:2011ia,Paulos:2011ie,Nandan:2011wc}.
According to these rules, in the normalization conventions we are following, the Mellin amplitude of a scalar $n$-point tree-level Witten diagram in a $\phi^3$ theory is constructed as follows:
\begin{itemize}
    \item Label the internal lines of the Witten diagram with an index $i$ running from $1$ to $n-3$, and to it associate an integer parameter $k_i$ (which will double as single-trace parameters) and a factor of
    \eqn{EdgeMellin}{
    E_i^{\rm Mellin} := {1\over k_i!} {(\Delta_{\delta_i} - h+1)_{k_i} \over    {\Delta_{\delta_i}-s_i\over 2} +k_i} \qquad (1\leq i \leq n-3)\,,
    }
    where $s_i$ is the Mandelstam invariant associated to that leg, and $\Delta_{\delta_i}$ is the conformal dimension of the dual operator running along the line.
    \item Label each internal vertex of the diagram with an index $j$ running from $1$ to $n-2$ and assign a factor of
    \eqn{VertexMellin}{
    V_j^{\rm Mellin} &:= {1 \over 2} \Gamma(\Delta_{abc,}-h)  \cr 
    & \hspace{-1em} \times F_A^{(3)}\!\left[\Delta_{abc,}- h; \{-k_a, -k_b, -k_c \}; \left\{\Delta_a -h+1, \Delta_b -h+1,\Delta_c -h+1 \right\}; 1,1,1 \right],
    }
    where $\Delta_a, \Delta_b$ and $\Delta_c$ are the conformal dimensions incident at the vertex, and $k_a, k_b$ and $k_c$ are the respective integer parameters (or single-trace parameters) associated with the {\it internal} exchanged dimensions. Set the integer parameter to zero if the conformal dimension associated to it is an external dimension.
\end{itemize}
Then for the choice of normalization constant,
\eqn{NDef}{
{\cal N} := \pi^{(n-2)h} \left( \prod_{i=1}^{n-3} {1 \over \Gamma(\Delta_{\delta_i}) } \right) \left( \prod_{i=1}^n {1\over \Gamma(\Delta_i)} \right),
}
where $\Delta_i$ are the external conformal dimensions and $\Delta_{\delta_i}$ are the internal exchanged dimensions, the Mellin amplitude is given by
\eqn{MellinFeynman}{
{\cal M}_n(\gamma_{ij}) = \left( \prod_{i=1}^{n-3} \sum_{k_i=0}^{\infty} \right) \left( \prod_{i=1}^{n-3} E_i^{\rm Mellin} \right)  \left( \prod_{i=1}^{n-2} V_i^{\rm Mellin} \right) .
}

With the Mellin amplitude in hand, we can proceed to evaluate the single-trace projection of the  position space amplitude $A_{n,}$, by performing the contour integrals in~\eno{MellinAmpDef}.
The single-trace contribution comes from the poles of the Mellin amplitude.
From the point of view of the Feynman prescription for Mellin amplitudes, these arise from simple poles occurring in the  denominator of the Mellin amplitude edge factors~\eno{EdgeMellin}.
In terms of Mandelstam invariants, these poles occur at
\eqn{MandelstamPoles}{
s_i = \Delta_{\delta_i} + 2k_i \qquad (1 \leq i \leq n-3)\,,
}
which correspond to putting the internal legs on-shell in the auxiliary momentum space.
These poles should be viewed as lying in the complex $\gamma_{rs}$ planes for $(rs) \in {\cal U}_{\rm chan}$, and our task is to evaluate the residue at these poles.

Before we do so, we point out that precisely the same Lauricella functions as those in~\eno{VertexMellin} appeared in~\eno{VertexDef} in the Feynman rules for {\it conformal blocks}. 
This is expected for the simple reason that the vertex factors~\eno{VertexMellin} are independent of Mellin variables $\gamma_{ij}$, so they remain unaffected through the following computation of Mellin integrals.

Turning to evaluating the residue at the ``single-trace poles''~\eno{MandelstamPoles} 
 in the ``${\cal U}_{\rm chan}$-plane,'' we obtain the following single-trace projection of the AdS diagram, denoted $A_n^{\rm s.t.}$,
\eqn{Ast}{
A_n^{\rm s.t.} &:= {\cal N} W_{n, {\rm chan}}^0(x_i) \left(\prod_{i=1}^{n-3} \sum_{k_i=0}^\infty \right)  
 \left(\prod_{i=1}^{n-3} { u_i^{{\Delta_{\delta_i} \over 2} + k_i} \over k_i!} (\Delta_{\delta_i} -h+1)_{k_i}  \right)  \left( \prod_{i=1}^{n-2} V_i^{\rm Mellin} \right)
  \cr 
 & \times \left(\prod_{(rs) \in {\cal V}_{\rm chan}} \int {d\gamma_{rs} \over 2\pi i} \right)  \!
 \left( \prod_{(ij) \in {\cal V}_{\rm chan}} \Gamma(\gamma_{ij}) \right) \!
 \left(\prod_{ (ij) \in {\cal U}_{\rm chan} \bigcup {\cal D}_{\rm chan} } \Gamma(\gamma_{ij})\Big|_{\rm s.t.}  \right) \!
 \left( \prod_{(rs) \in {\cal V}_{\rm chan}} v_{rs}^{-\gamma_{rs}} \right) \! ,
}
where $ \Gamma(\gamma_{ij})\Big|_{\rm s.t.}$ stands for Gamma functions with arguments from the index sets ${\cal U}_{\rm chan}$ and ${\cal D}_{\rm chan}$ evaluated at the poles~\eno{MandelstamPoles}.\footnote{We note that the dependent Mellin variables $\gamma_{ij}$ with $(ij) \in {\cal D}_{\rm chan}$ in the Gamma functions are assumed to have already been expressed in terms of the independent Mellin variables from the sets ${\cal U}_{\rm chan}$ and ${\cal V}_{\rm chan}$. Still, for brevity, we prefer to use the notation in the second line of~\eno{Ast}. We also emphasize the obvious fact that the Gamma functions with arguments from the index set ${\cal V}_{\rm chan}$ remain unaffected after taking the single-trace residues.}
Equation~\eno{Ast} is proportional to the desired conformal block. More precisely, 
\begin{empheq}[box=\narrowtcboxmath]{align} 
& W_{n,{\rm chan}}(x_i) \!= \!{ {\cal N}\, W_{n, {\rm chan}}^0(x_i) \over \left(\prod_{i=1}^{n-2} f_i \right)} \left(\prod_{i=1}^{n-3} \sum_{k_i=0}^\infty \right)  \!\!
 \left(\prod_{i=1}^{n-3} {u_i^{{\Delta_{\delta_i} \over 2} + k_i} \over k_i!} (\Delta_{\delta_i} -h+1)_{k_i} \!\right) \!\! \left( \prod_{i=1}^{n-2} V_i^{\rm Mellin} \!\right)
  \cr 
 & \; \times \left(\prod_{(rs) \in {\cal V}_{\rm chan}} \int {d\gamma_{rs} \over 2\pi i} \right)  \!\!
 \left( \!\prod_{(ij) \in {\cal V}_{\rm chan}}\!\! \Gamma(\gamma_{ij})\! \right) \!\!
 \left(\prod_{ (ij) \in {\cal U}_{\rm chan} \bigcup {\cal D}_{\rm chan} } \!\!\! \Gamma(\gamma_{ij})\Big|_{\rm s.t.}  \right) \!\!
 \left( \prod_{(rs) \in {\cal V}_{\rm chan}} \!\! v_{rs}^{-\gamma_{rs}}\! \right) 
\label{CBSerInt}
\end{empheq}
furnishes a mixed series-integral representation of the desired conformal block, where $f_i= C_{\Delta_{a_i}\Delta_{b_i}\Delta_{c_i}}$ are the known $(n-2)$ MFT OPE coefficients, one for each internal vertex of the binary unrooted tree representing the AdS diagram with scalars of conformal dimensions $\Delta_{a_i}, \Delta_{b_i}$ and $\Delta_{c_i}$ on the incident edges.

To obtain an integral-free representation of the block, one must evaluate the residual $\binom{n-2}{2}$-dimensional contour integrals in the second line of~\eno{CBSerInt}. 
This will introduce $\binom{n-2}{2}$ new summations. 
Evaluating these integrals in general for an an arbitrary $n$-point conformal block in an arbitrary channel is not clear to us.
Instead, in the remainder of this section, we will focus on evaluating these integrals explicitly for the three classes of examples from section~\ref{FEYNMANEXAMPLES}. We will reproduce the blocks as prescribed by the Feynman rules of section~\ref{FEYNMAN}, which serves as a  highly non-trivial check of the proposed Feynman rules.

Before specializing to specific examples, we can do further general manipulations. First, isolating the integral in the second line of~\eno{CBSerInt},
\eqn{IDef}{
I := \left(\prod_{(rs) \in {\cal V}_{\rm chan}} \int {d\gamma_{rs} \over 2\pi i} \right)  
\left(\prod_{(rs) \in  {\cal V}_{\rm chan}} \Gamma(\gamma_{rs}) \right)
\left(\prod_{(ij) \in {\cal U}_{\rm chan}  \bigcup {\cal D}_{\rm chan}} \Gamma(\gamma_{ij})\Big|_{\rm s.t.}     \right)
  \left( \prod_{(rs) \in {\cal V}_{\rm chan}} v_{rs}^{-\gamma_{rs}} \right) \!,
}
we rewrite each factor of $v_{rs}^{-\gamma_{rs}}$ above by introducing an additional contour integral, as follows\footnote{Use the Mellin-Barnes representation, 
\eqn{MBR}{ 
\frac{1}{ (x+y)^a} = \frac{ 1}{\Gamma{\left(a\right)}} 
\int{ \frac{ds}{ 2\pi i} \Gamma{(-s)} \Gamma{(s+a)} x^s y^{-a-s}}
}
with $x=v_{rs}-1, y=1$ and $a=\gamma_{rs}$. The contour of integration is chosen such that it separates the poles of $\Gamma(s+a)$ from those of $\Gamma(-s)$; see e.g. the discussion around ~\cite[eq.~(B.11)]{Rosenhaus:2018zqn}.}. 
\eqn{vRewrite}{
v_{rs}^{-\gamma_{rs}} = {1 \over \Gamma(\gamma_{rs})} \int {d \widetilde{\gamma}_{rs} \over 2\pi i} \, \Gamma(-\widetilde{\gamma}_{rs}) \Gamma(\widetilde{\gamma}_{rs} + \gamma_{rs}) (v_{rs}-1)^{\widetilde{\gamma}_{rs}} \,,
}
where the $\widetilde{\gamma}_{rs}$ contour runs vertically such that it separates the semi-infinite sequence of poles running to the left and to right of origin. 
Then we get
\eqn{Ifinal}{
I =  \left(\prod_{(rs) \in {\cal V}_{\rm chan}} \int {d\widetilde{\gamma}_{rs} \over 2\pi i}  \int {d \gamma_{rs} \over 2\pi i}\, \Gamma(-\widetilde{\gamma}_{rs}) \Gamma(\widetilde{\gamma}_{rs} + \gamma_{rs}) (v_{rs}-1)^{\widetilde{\gamma}_{rs}} \right) \!\!
 \left(\prod_{(ij) \in {\cal D}_{\rm OPE}\bigcup {\cal U}_{\rm OPE}} \Gamma(\gamma_{ij})\Big|_{\rm s.t.}     \right)\!,
}
where we switched the order of integrals.
The rewriting in~\eno{vRewrite} makes it easier to obtain a convergent series expansion of the conformal block in powers of $(1-v_{rs})$ as desired (see~\eno{CBexpansion}-\eno{Feynman}).
The overall strategy now will be to repeatedly use the first Barnes lemma~\cite{Barnes1908}, 
\eqn{FirstBarnes}{ 
\int_{-i \infty}^{+i \infty} \!\!\frac{ds}{2 \pi i} \Gamma{(a_1+s)}\Gamma{(a_2+s)}\Gamma{(b_1-s)}\Gamma{(b_2-s)} = \frac{\Gamma{(a_1+b_1)}\Gamma{(a_1+b_2)}\Gamma{(a_2+b_1)}\Gamma{(a_2+b_2)}}{\Gamma{(a_1+a_2+b_1+b_2)}} ,
}
 to evaluate all $\gamma_{rs}$ integrals, which it turns out will leave us with trivial-to-evaluate $\widetilde{\gamma}_{rs}$ contour integrals.

\subsection{Seven-point mixed channel}
\label{7MIXPROOF}

To obtain the $7$-point ``mixed-channel'' conformal block (topology shown in figure~\ref{fig:CBexMix}), we start with the following ``canonical'' tree-level AdS diagram:
\eqn{7CanonDiag}{
A_{\rm 7, mix} := \musepic{\figSevenWitten}\,.
}
Here we have labeled the external scalar operators with their conformal dimensions but suppressed the insertion coordinates (e.g.\ $\Delta_1$ should be understood as a scalar operator of conformal dimension $\Delta_1$ inserted at boundary coordinate $x_1$). 
By design, the operator insertions and exchanged operator labels in the canonical Witten diagram above match with the corresponding binary graph of the 7-point conformal block shown in figure~\ref{fig:CBexMix}.
Finally, the green disks represent cubic interaction vertices to be integrated over all of AdS$_{d+1}$.

As previously noted, to express the conformal block, one must start by making a choice of independent conformal cross-ratios. 
To reproduce the block obtained via the Feynman rules of section~\ref{FEYNMAN}, we will utilize the same choice of cross-ratios as in section~\ref{7MIXED}, which comes with associated index sets for the dependent and (a subset of) independent Mellin variables,  ${\cal D}_{\rm 7,mix}$ and ${\cal V}_{\rm 7,mix}$ respectively (see~\eno{Dmix} and~\eno{Vmix}). 
Using~\eno{UnionAll} we can determine the remaining independent Mellin variables, associated with the index set ${\cal U}_{\rm 7,mix}$,
\eqn{Umix}{
{\cal U}_{\rm 7,mix} = \{(12),(23),(45),(67)\}\,.
}
With the help of a color-coded upper-triangular matrix, we can present these sets visually as shown in figure~\ref{fig:7mixSets}.
\begin{figure}
\centering
  \[
  \begin{array}{|c|c|c|c|c|c|}
    \cline{1-6}
      \Uchan 12  & \Vchan 13 & \Vchan 14 & \Vchan 15 & \Vchan 16 & \Dchan 17 \\
    \cline{1-6}
        \multicolumn{1}{c|}{}    &  \Uchan 23 & \Vchan 24 &\Dchan 25 &\Vchan 26 &  \Dchan 27 \\
    \cline{2-6}
        \multicolumn{2}{c|}{}   &  \Vchan 34 &\Vchan  35 & \Vchan 36 &  \Dchan 37 \\
    \cline{3-6}
        \multicolumn{1}{c}{\Uchan {\cal U}} & \multicolumn{2}{c|}{}   & \Uchan  45 & \Vchan 46 &  \Dchan 47  \\
    \cline{4-6}
        \multicolumn{1}{c}{\Vchan {\cal V}} & \multicolumn{3}{c|}{} & \Dchan 56 & \Dchan 57  \\
    \cline{5-6}
        \multicolumn{1}{c}{\Dchan {\cal D}} & \multicolumn{4}{c|}{} & \Uchan 67  \\
    \cline{6-6}
  \end{array}
  \]
  \caption{Color-coded matrix displaying the choice of Mellin variable index subsets ${\cal U}_{\rm 7,mix}, {\cal V}_{\rm 7,mix}$ and ${\cal D}_{\rm 7,mix}$.}
\label{fig:7mixSets}
 \end{figure}
With these choices in place, the Mellin product admits a rewriting in terms of the cross-ratios as shown in~\eno{7mixMellinProd}, which we substitute into~\eno{MellinAmpDef}.

The other ingredient which goes into~\eno{MellinAmpDef} is the Mellin amplitude ${\cal M}_7(\gamma_{ij})$ for the seven-point AdS diagram above. According to the Feynman rules for {\it Mellin amplitudes}, the Mellin amplitude in this case is given by~\eno{MellinFeynman} for $n=7$ with 
the edge factors given in~\eno{EdgeMellin}  and the vertex factors taking the explicit form:
\eqn{7VertexMellin}{ 
V_1^{\rm Mellin} &= {1 \over 2} \Gamma(\Delta_{12\delta_1,}-h)  \,
F_A^{(1)}\!\left[\Delta_{12\delta_1,}- h; \{-k_1 \}; \left\{\Delta_{\delta_1} -h+1 \right\}; 1\right] \cr 
V_2^{\rm Mellin} &= {1 \over 2} \Gamma(\Delta_{3\delta_1\delta_2,}-h)  \, F_A^{(2)}\!\left[\Delta_{3\delta_1\delta_2,}- h; \{-k_1, -k_2 \}; \left\{\Delta_{\delta_1} -h+1, \Delta_{\delta_2} -h+1 \right\}; 1,1 \right] \cr 
V_3^{\rm Mellin} &= {1 \over 2} \Gamma(\Delta_{\delta_2\delta_3\delta_4,}-h)  \cr 
 & \times F_A^{(3)}\!\left[\Delta_{\delta_2\delta_3\delta_4}- h; \{-k_2, -k_3, -k_4 \}; \left\{\Delta_{\delta_2} -h+1, \Delta_{\delta_3} -h+1,\Delta_{\delta_4} -h+1 \right\}; 1,1,1 \right] \cr 
V_4^{\rm Mellin} &= {1 \over 2} \Gamma(\Delta_{45\delta_3,}-h)  \,
F_A^{(1)}\!\left[\Delta_{45\delta_3,}- h; \{-k_3 \}; \left\{\Delta_{\delta_3} -h+1 \right\}; 1 \right] \cr 
V_5^{\rm Mellin} &= {1 \over 2} \Gamma(\Delta_{67\delta_4,}-h)  \,
F_A^{(1)}\!\left[\Delta_{67\delta_4,}- h; \{-k_4 \}; \left\{\Delta_{\delta_4} -h+1 \right\}; 1 \right].
 }
 
The single-trace projection of the Witten diagram~\eno{7CanonDiag}, described in general terms in the discussion preceding this example, then leads to~\eno{Ast} which is proportional to the desired conformal block. 
This projection involves evaluating the residue at the ``single-trace poles,'' occuring at
\eqn{7UPoles}{
\gamma_{12} = \Delta_{12,\delta_1}-k_1 \qquad \!\!
\gamma_{23}= \Delta_{3\delta_1,\delta_2}+k_{1,2}-\gamma_{13}
\qquad \!\!
\gamma_{45}= \Delta_{45,\delta_3}-k_{3}
\qquad\!\!
\gamma_{67}= \Delta_{67,\delta_4}-k_{4} .
}
The block itself is given by~\eno{CBSerInt} by projecting out the theory dependent OPE coefficients:
\eqn{7OPE}{
f_1 = C_{\Delta_1 \Delta_2 \Delta_{\delta_1}} \qquad  \!\!
f_2 = C_{\Delta_3 \Delta_{\delta_1} \Delta_{\delta_2}} \qquad \!\!
f_3 = C_{\Delta_{\delta_2} \Delta_{\delta_3} \Delta_{\delta_4}}  \qquad \!\!
f_4 = C_{\Delta_4 \Delta_5 \Delta_{\delta_3}} \qquad \!\!
f_5 = C_{\Delta_6 \Delta_7 \Delta_{\delta_4}}.
}
The MFT OPE coefficients above take the well known form,
\eqn{MFTOPE}{
C_{\Delta_a \Delta_b \Delta_c} = \frac{\pi^h}{2} \, \Gamma\left(\Delta_{abc,}-h\right)
\frac{\Gamma(\Delta_{ab,c})
\Gamma(\Delta_{bc,a})
\Gamma(\Delta_{ca,b})}{\Gamma(\Delta_a)\Gamma(\Delta_b)\Gamma(\Delta_c)}\,.
}
The precise form of these OPE coefficients will be utilized at the end of the computation.

The non-trivial computation which needs to be done is the contour integral on the second line of~\eno{CBSerInt}, or equivalently the integral~\eno{Ifinal}. 
The arguments of Gamma functions in the product $\left( \prod_{(ij) \in {\cal D}_{\rm 7,mix} {\cal U}_{\rm 7,mix}} \Gamma(\gamma_{ij}) \Big|_{\rm s.t.} \right)$ are expressed entirely in terms of Mellin variables from the index set ${\cal V}_{\rm 7,mix}$. Four of them were shown in~\eno{7UPoles} which correspond to the index set ${\cal U}_{\rm 7,mix}$; the other seven, corresponding to the index set ${\cal D}_{\rm 7,mix}$ take the form
\eqn{7DArgumentsAgain}{
\gamma_{17}&= \Delta_{1\delta_1,2}+k_1-\gamma_{13}-\gamma_{14}-\gamma_{15}-\gamma_{16}
\cr 
\gamma_{25}
&=\Delta_{\delta_2\delta_3,\delta_4}+k_{23,4}-\gamma_{14}-\gamma_{15}-\gamma_{24}-\gamma_{34}-\gamma_{35}
\cr 
\gamma_{27}
&= \Delta_{2\delta_4,13\delta_3}+k_{4,3}+\gamma_{13}+\gamma_{14}+\gamma_{15}-\gamma_{26}+\gamma_{34}+\gamma_{35}
\cr 
\gamma_{37}
&= \Delta_{3\delta_2,\delta_1}+k_{2,1}-\gamma_{34}-\gamma_{35}-\gamma_{36}
\cr 
\gamma_{47}
&= \Delta_{4\delta_3,5}+k_3 - \gamma_{14}- \gamma_{24}- \gamma_{34}- \gamma_{46}
\cr 
\gamma_{56}&= \Delta_{6\delta_4,7}+k_4 - \gamma_{16}- \gamma_{26}- \gamma_{36}- \gamma_{46}
\cr 
\gamma_{57}&= \Delta_{57,46\delta_2}-k_2 +\gamma_{14}+\gamma_{16}+\gamma_{24}+\gamma_{26}+\gamma_{34}+\gamma_{36}+\gamma_{46}\,.
}
These were obtained by substituting~\eno{7UPoles} into~\eno{DMixVars}.

Explicitly, $I$ in~\eno{Ifinal} then takes the form 
\eqn{}{
I &=  \left(\prod_{(rs) \in {\cal V}_{\rm 7,mix}} {d\widetilde{\gamma}_{rs} \over 2\pi i} \int {d\gamma_{rs} \over 2\pi i}  \int \, \Gamma(-\widetilde{\gamma}_{rs}) \Gamma(\widetilde{\gamma}_{rs} + \gamma_{rs}) (v_{rs}-1)^{\widetilde{\gamma}_{rs}} \right)
\Gamma\left(  \Delta_{12,\delta_1}-k_1 \right)
 \cr
&\times \Gamma\left( \Delta_{3\delta_1,\delta_2}+k_{1,2}-\gamma_{13} \right)
 \Gamma\left(  \Delta_{45,\delta_3}-k_{3} \right) 
 \Gamma\left(  \Delta_{67,\delta_4}-k_{4} \right)
 \cr 
 &\times
 \Gamma \left( \Delta_{1\delta_1,2}+k_1-\gamma_{13}-\gamma_{14}-\gamma_{15}-\gamma_{16} \right)
 \cr 
 &\times
 \Gamma \left( \Delta_{\delta_2\delta_3,\delta_4}+k_{23,4}-\gamma_{14}-\gamma_{15}-\gamma_{24}-\gamma_{34}-\gamma_{35} \right)
 \cr 
 &\times 
  \Gamma \left( \Delta_{2\delta_4,13\delta_3}+k_{4,3}+\gamma_{13}+\gamma_{14}+\gamma_{15}-\gamma_{26}+\gamma_{34}+\gamma_{35} \right)
  \cr 
  &\times 
  \Gamma \left( \Delta_{3\delta_2,\delta_1}+k_{2,1}-\gamma_{34}-\gamma_{35}-\gamma_{36} \right)
  \cr 
  &\times \Gamma \left( \Delta_{4\delta_3,5}+k_3 - \gamma_{14}- \gamma_{24}- \gamma_{34}- \gamma_{46} \right)
  \cr 
  &\times \Gamma \left(  \Delta_{6\delta_4,7}+k_4 - \gamma_{16}- \gamma_{26}- \gamma_{36}- \gamma_{46} \right)
  \cr 
  &\times \Gamma \left( \Delta_{57,46\delta_2}-k_2 +\gamma_{14}+\gamma_{16}+\gamma_{24}+\gamma_{26}+\gamma_{34}+\gamma_{36}+\gamma_{46} \right).
}
 We first evaluate the 10-dimensional contour integral over the $\gamma_{rs}$ variables for $(rs) \in {\cal V}_{\rm 7,mix}$.  
We  perform the contour integrations one at a time, in the following order:
\[ \gamma_{16},\gamma_{36},\gamma_{26},\gamma_{46},\gamma_{35},\gamma_{15},\gamma_{24},\gamma_{34},\gamma_{13},\gamma_{14}.\] 
With this choice of ordering, we are able to make direct use of the first Barnes lemma~\eno{FirstBarnes} at every step.\footnote{Notably, after the $\gamma_{15}$ integral, we need to do a linear change of variables $\gamma_{14} \rightarrow \gamma_{14}-\gamma_{24}-\gamma_{34}$. }
To keep the manuscript to a reasonable length, we refrain from including the lengthy but straightforward computational details, and merely present the final result of this 10-fold contour integral:
\eqn{7Igamma}{
I &=  \left(\prod_{(rs) \in {\cal V}_{\rm 7,mix} }   \int {d\widetilde{\gamma}_{rs} \over 2\pi i}\, \Gamma(-\widetilde{\gamma}_{rs})  (v_{rs}-1)^{\widetilde{\gamma}_{rs}} \right) \cr 
&\times 
\Big( \Gamma(  \Delta_{12,\delta_1}-k_1)
\Gamma( \widetilde{\gamma}_{24} +\widetilde{\gamma}_{26} +\Delta_{2\delta_1,1}+k_1 )
\Gamma(\widetilde{\gamma}_{13}+ \widetilde{\gamma}_{14} +\widetilde{\gamma}_{15}  + \widetilde{\gamma}_{16}  + \Delta_{1\delta_1,2}+k_1)
\Big) \cr 
&\times 
\Big(   
\Gamma(\widetilde{\gamma}_{14} + \widetilde{\gamma}_{15}  + \widetilde{\gamma}_{16} +\widetilde{\gamma}_{24}  +\widetilde{\gamma}_{26}+ \Delta_{\delta_1\delta_2,3}+k_{12,})
\Gamma( \widetilde{\gamma}_{13}  +\Delta_{3\delta_1,\delta_2}+k_{1,2} )\cr 
&\quad \times 
\Gamma( \widetilde{\gamma}_{34} +\widetilde{\gamma}_{35} +\widetilde{\gamma}_{36}+\Delta_{3\delta_2,\delta_1}+k_{2,1} ) 
\Big)
\cr 
&\times 
\Big( \Gamma( \widetilde{\gamma}_{14}+ \widetilde{\gamma}_{15} +\widetilde{\gamma}_{24} + \widetilde{\gamma}_{34}  +\widetilde{\gamma}_{35}  +\Delta_{\delta_2\delta_3,\delta_4}+k_{23,4} )
\Gamma(\widetilde{\gamma}_{46} + \Delta_{\delta_3\delta_4,\delta_2}+k_{34,2}) 
\cr 
&\quad \times 
\Gamma( \widetilde{\gamma}_{16}+\widetilde{\gamma}_{26}+\widetilde{\gamma}_{36} +\Delta_{\delta_2\delta_4,\delta_3}+k_{24,3})
\Big)
\cr 
&\times 
\Big(  \Gamma(  \Delta_{45,\delta_3}-k_{3}) 
\Gamma(\widetilde{\gamma}_{14}+\widetilde{\gamma}_{24}  + \widetilde{\gamma}_{34}  +\widetilde{\gamma}_{46} +\Delta_{4\delta_3,5}+k_3) 
\Gamma(\widetilde{\gamma}_{15} +\widetilde{\gamma}_{35} +\Delta_{5\delta_3,4}+k_{3} )
\Big)
 \cr
 &\times 
\Big( 
\Gamma(  \Delta_{67,\delta_4}-k_{4} )
 \Gamma(\Delta_{7\delta_4,6}+k_4)
 \Gamma( \widetilde{\gamma}_{16}+ \widetilde{\gamma}_{26} +\widetilde{\gamma}_{36} +\widetilde{\gamma}_{46} +\Delta_{6\delta_4,7}+k_4)
 \Big) 
 \cr 
&\times 
{1\over \Gamma(\widetilde{\gamma}_{13}+\widetilde{\gamma}_{14} +\widetilde{\gamma}_{15}  + \widetilde{\gamma}_{16}  +\widetilde{\gamma}_{24} +\widetilde{\gamma}_{26} +\Delta_{\delta_1}+2k_1)}
\cr 
&\times 
{1\over \Gamma(\widetilde{\gamma}_{14}+\widetilde{\gamma}_{15}  + \widetilde{\gamma}_{16}  +\widetilde{\gamma}_{24}  +\widetilde{\gamma}_{26} +\widetilde{\gamma}_{34} +\widetilde{\gamma}_{35} +\widetilde{\gamma}_{36}+\Delta_{\delta_2}+2k_2)}
\cr 
&\times 
{1\over \Gamma(\widetilde{\gamma}_{14}+ \widetilde{\gamma}_{15} + \widetilde{\gamma}_{24}  +\widetilde{\gamma}_{34} +\widetilde{\gamma}_{35} +\widetilde{\gamma}_{46}+\Delta_{\delta_3}+2k_3) }
\cr 
&\times {
 1\over 
\Gamma(\widetilde{\gamma}_{16} + \widetilde{\gamma}_{26} +\widetilde{\gamma}_{36}+ \widetilde{\gamma}_{46}  + \Delta_{\delta_4}+2k_4) }\,.
}
Next, we evaluate the remaining integrals via the Cauchy residue theorem. We close all $\widetilde{\gamma}_{rs}$ contours to the right to be able to drop the contribution from the arc at infinity, picking the lone semi-infinite sequence of poles starting at the origin, at $\widetilde{\gamma}_{rs} = j_{rs}$ for $j_{rs} \in \mathbb{Z}^{\geq 0}$ for each $(rs) \in {\cal V}_{\rm 7,mix}$.\footnote{We recall that the contour for the $\widetilde{\gamma}_{rs}$ integrals was chosen such that it separates the semi-infinite series of poles running to the left from those running to the right; see~\eno{MBR}.} These poles come from the poles of $\Gamma(-\widetilde{\gamma}_{rs})$ in the first line of~\eno{7Igamma}, and the residues, which are elementary to compute, introduce $\binom{n-2}{2}$ additional infinite sums:
\eqn{7Igamma2}{
I &=  \left(\prod_{(rs) \in {\cal V}_{\rm 7,mix} }    \sum_{j_{rs}=0}^{\infty} { (1-v_{rs})^{j_{rs}} \over j_{rs}!} \right) \cr 
&\times 
\Big( \Gamma(  \Delta_{12,\delta_1}-k_1)
\Gamma( j_{24} +j_{26} +\Delta_{2\delta_1,1}+k_1 )
\Gamma(j_{13}+ j_{14} +j_{15}  + j_{16}  + \Delta_{1\delta_1,2}+k_1)
\Big) \cr 
&\times 
\Big(   
\Gamma(j_{14} + j_{15}  + j_{16} +j_{24}  +j_{26}+ \Delta_{\delta_1\delta_2,3}+k_{12,})
\Gamma( j_{13}  +\Delta_{3\delta_1,\delta_2}+k_{1,2} )\cr 
&\quad \times 
\Gamma( j_{34} +j_{35} +j_{36}+\Delta_{3\delta_2,\delta_1}+k_{2,1} ) 
\Big)
\cr 
&\times 
\Big( \Gamma( j_{14}+ j_{15} +j_{24} + j_{34}  +j_{35}  +\Delta_{\delta_2\delta_3,\delta_4}+k_{23,4} )
\Gamma(j_{46} + \Delta_{\delta_3\delta_4,\delta_2}+k_{34,2}) 
\cr 
&\quad \times 
\Gamma( j_{16}+j_{26}+j_{36} +\Delta_{\delta_2\delta_4,\delta_3}+k_{24,3})
\Big)
\cr 
&\times 
\Big(  \Gamma(  \Delta_{45,\delta_3}-k_{3}) 
\Gamma(j_{14}+j_{24}  + j_{34}  +\widetilde{\gamma}_{46} +\Delta_{4\delta_3,5}+k_3) 
\Gamma(\widetilde{\gamma}_{15} +\widetilde{\gamma}_{35} +\Delta_{5\delta_3,4}+k_{3} )
\Big)
 \cr
 &\times 
\Big( 
\Gamma(  \Delta_{67,\delta_4}-k_{4} )
 \Gamma(\Delta_{7\delta_4,6}+k_4)
 \Gamma( j_{16}+ j_{26} +j_{36} +j_{46} +\Delta_{6\delta_4,7}+k_4)
 \Big) 
 \cr 
&\times 
{1\over \Gamma(j_{13}+j_{14} +j_{15}  + j_{16}  +j_{24} +j_{26} +\Delta_{\delta_1}+2k_1)}
\cr 
&\times 
{1\over \Gamma(j_{14}+j_{15}  + j_{16}  +j_{24}  +j_{26} +j_{34} +j_{35} +j_{36}+\Delta_{\delta_2}+2k_2)}
\cr 
&\times 
{1\over \Gamma(j_{14}+ j_{15} + j_{24}  +j_{34} +j_{35} +j_{46}+\Delta_{\delta_3}+2k_3) }
\cr 
&\times {
 1\over 
\Gamma(j_{16} + j_{26} +j_{36}+ j_{46}  + \Delta_{\delta_4}+2k_4) }\,.
}

Now we are ready to put everything together into~\eno{CBSerInt} to obtain the conformal block $W_{7, {\rm mix}}$. 
There will be a host of simplifying cancellations between factors of Gamma functions.
For example, the factors of one-half times a Gamma function in the five Mellin vertex factors~\eno{7VertexMellin} cancel against a factor of one-half times a Gamma function appearing in each of the five OPE coefficients in~\eno{7OPE} (after employing the explicit form of the OPE coefficients~\eno{MFTOPE}).
The remaining triplet of factors of Gamma functions of the form $\Gamma(\Delta_{ab,c})\Gamma(\Delta_{bc,a})\Gamma(\Delta_{ca,b})$,  in the numerators of each of the five OPE coefficients combine with the five groups of a triplet of Gamma functions in the numerator of~\eno{7Igamma2} separated by parentheses, of the form $\Gamma(\Delta_{ab,c}+p)\Gamma(\Delta_{bc,a}+q)\Gamma(\Delta_{ca,b}+r)$, to give rise to five groups of triplets of Pochhammer symbols of the form $(\Delta_{ab,c})_p (\Delta_{bc,a})_q (\Delta_{ca,b})_r$.
These combine with the Lauricella functions in the Mellin vertex factors to give what we call the ``conformal block vertex factors,'' one associated to each vertex of the canonical AdS diagram.
The Gamma functions in the denominator of the five OPE coefficients with {\it external} dimensions in the argument cancel the same factors appearing in the normalization constant ${\cal N}$ in~\eno{NDef}.
For Gamma functions in the denominator of the OPE coefficients with {\it internal} dimensions in their arguments, there are precisely two copies for each internal dimension, while there is only one such factor for each internal dimension in the normalization constant ${\cal N}$. 
Thus after cancellations, we are left with a product  in the numerator of $W_{7, {\rm mix}}$ over factors of Gamma function, one each for every internal dimension. 
These combine with the Gamma functions in the denominator of~\eno{7Igamma2} of the form $\Gamma(\Delta_{\delta_i} + t_i)$ to give rise to four Pochhammer symbols, $(\Delta_{\delta_i})_{t_i}$ in the denominator.
This together with the Pochhamer symbols originally appearing in~\eno{CBSerInt} combine to give what we call the ``conformal block edge factor,'' one for each internal leg of the canonical AdS diagram.
Finally, the factor of $\pi^{(n-2)h}$ in ${\cal N}$ for $n=7$ cancels with $(n-2)$ factors of $\pi^h$ distributed over $(n-2)$ OPE coefficients~\eno{7OPE}. Thus all explicit factors of $\pi$ cancel.

This leads to the following final expression for the conformal block:
\eqn{}{
 W_{7,{\rm mix}}(x_i) =  W_{7, {\rm mix}}^0(x_i)  \! \left(\! \prod_{i=1}^{4} \sum_{k_i=0}^\infty \right)  \!\!
 \left(\! \prod_{i=1}^{4} {u_i^{{\Delta_{\delta_i} \over 2} + k_i} \over k_i!} \!\! \right) \!\!
 \left(\! \prod_{(rs) \in {\cal V}_{\rm 7,mix}} \!\sum_{j_{rs}=0}^\infty \!\! { (1-v_{rs})^{j_{rs}} \over j_{rs}!} \!\! \right) \!\!
 \left(\! \prod_{i=1}^{4} E_i \right) \!\! 
 \left(\! \prod_{i=1}^{5} V_i \right)\! ,
 }
where $E_i$ are precisely the conformal block edge factors~\eno{7mixEfac} and $V_i$ are the conformal block vertex factors~\eno{7mixVfac} prescribed by the Feynman rules, thus confirming the Feynman rules in this particular example.

\subsection{Comb channel}
\label{COMBPROOF}

To derive the comb channel conformal block obtained in section~\ref{NCOMB} via Feynman rules, we start with the Mellin amplitude of the following tree-level $n$-point AdS diagram ($n \geq 4$): 
\eqn{CombCanonDiag}{
A_{n, \rm comb} := \musepic{\figNCombWitten}
}
where the external scalar operators of dimensions $\Delta_i$ are inserted at coordinates $x_i$. In total $(n-3)$ single-particle bulk fields are exchanged in the interior, which are dual to single-trace operators with conformal dimensions $\Delta_{\delta_i}$ for $1\leq i \leq n-3$.  The ellipses in the middle indicate a repeating pattern.
This is the canonical AdS diagram for the comb channel block shown in figure~\ref{fig:CBexComb}.

To reproduce the block from section~\ref{NCOMB}, we will utilize the same input data as before.
This data comprises of a set of $n(n-3)/2$ independent cross-ratios~\eno{CombCR}.  
Correspondingly, we choose the dependent Mellin index set ${\cal D}_{\rm comb}$ to be~\eno{DComb} which allows us to rewrite the Mellin product~\eno{MellinAmpDef} in terms of the cross-ratios and the Mandelstam invariants for internal legs~\eno{CombMandelstamExplicit}, as shown in~\eno{CombMellinProd}. 
The choice of cross-ratios also determines the index set ${\cal V}_{\rm comb}$~\eno{VComb}, which represents a subset of independent Mellin variables.
With ${\cal D}_{\rm comb}$ and ${\cal V}_{\rm comb}$ in hand, one can use~\eno{UnionAll} to obtain the index set associated with the remaining $(n-3)$ independent Mellin parameters,
\eqn{UComb}{ 
{\cal U}_{\rm comb} &:= \big\{ (1j) \;\big|\; 2 \leq j \leq n-2 \big\} \,.
}
For reference, we have collected all comb channel index sets together into a color-coded upper-triangular matrix in figure~\ref{fig:combSets}. 
\begin{figure}
\centering
  \[ 
  \begin{array}{|c|c|c|c|c|c|c|c|c|>{\columncolor{red!50}}c|}
    \cline{1-10}
      \rowcolor{brown!30} 12  & 13 & 14 & 15 & 16 & \ldots & 1(n-3) & 1(n-2) & \Dchan 1(n-1) &  \Dchan  1n \\
    \cline{1-10}
        \multicolumn{1}{c|}{}    &  \Vchan 23 & \Vchan  24 &\Vchan 25 &\Vchan 26 & \Vchan \ldots & \Vchan 2(n-3) &\Vchan 2(n-2) &\Vchan 2(n-1) &   2n \\
    \cline{2-10}
        \multicolumn{2}{c|}{}   &  \Vchan 34 & \Vchan 35 & \Vchan 36 & \Vchan \ldots & \Vchan 3(n-3) &\Vchan  3(n-2) & \Vchan 3(n-1) &   3n \\
    \cline{3-10}
        \multicolumn{3}{c|}{}   &   \Vchan \ddots & \Vchan &  \Vchan &  \Vchan & \Vchan \vdots  & \Vchan & \vdots  \\
    \cline{4-10}
        \multicolumn{4}{c|}{} & \Vchan \ddots &  \Vchan &  \Vchan & \Vchan \vdots & \Vchan & \vdots  \\
    \cline{5-10}
        \multicolumn{5}{c|}{} & \Vchan \ddots &  \Vchan & \Vchan \vdots  & \Vchan & \vdots  \\
    \cline{6-10}
        \multicolumn{6}{c|}{} & \Vchan (n-4)(n-3) &\Vchan (n-4)(n-2) &\Vchan (n-4)(n-1) &  (n-4)n \\
    \cline{7-10}
        \multicolumn{1}{c}{\Uchan {\cal U}}   &  \multicolumn{6}{c|}{}   & \Vchan (n-3)(n-2) &\Vchan (n-3)(n-1) &  (n-3)n\\
    \cline{8-10}
        \multicolumn{1}{c}{\Vchan {\cal V}} &   \multicolumn{7}{c|}{}   & \Vchan (n-2)(n-1) &  (n-2)n\\
    \cline{9-10}
        \multicolumn{1}{c}{\Dchan {\cal D}}  & \multicolumn{8}{c|}{}  & (n-1)n \\
    \cline{10-10}
  \end{array}
  \]
 \caption{Color-coded matrix displaying the canonical choice of Mellin variable index subsets ${\cal U}_{\rm comb}, {\cal V}_{\rm comb}$ and ${\cal D}_{\rm comb}$.}
\label{fig:combSets}
\end{figure}

Using the Feynman rules for Mellin amplitudes, it is trivial to write down Mellin amplitude for the diagram~\eno{CombCanonDiag},
\eqn{}{
{\cal M}_{n,\rm comb}(\gamma_{ij}) = \left(\prod_{i=1}^{n-3} \sum_{k_i=0}^\infty \right) \left( \prod_{i=1}^{n-3} E_i^{\rm Mellin} \right) \left(  V_1^{\rm Mellin}  \left( \prod_{i=1}^{n-4} V_{i+1}^{\rm Mellin} \right) V_{n-2}^{\rm Mellin} \right) ,
}
where each of the $(n-3)$ edge factors is given  in terms of the Mandelstam invariants $s_i$ and the associated single-trace parameters $k_i$ by~\eno{EdgeMellin}, and the $(n-2)$ vertex factors  follow directly from~\eno{VertexMellin}:
\eqn{MellinCombVertex}{
V_1^M &= {1 \over 2} \Gamma(\Delta_{12\delta_1,}-h) \,  F_A^{(1)}\!\left[\Delta_{12\delta_1,}- h; \{-k_1 \}; \left\{\Delta_{\delta_1} -h+1 \right\}; 1 \right] \cr
V_{i+1}^M &= {1 \over 2} \Gamma(\Delta_{\delta_{i} \delta_{i+1} (i+2),}-h)   \cr 
 &\times F_A^{(2)}\!\left[\Delta_{\delta_{i} \delta_{i+1} (i+2),}- h; \{-k_{i}, -k_{i+1} \}; \left\{\Delta_{\delta_{i}} -h+1, \Delta_{\delta_{i+1}} -h+1\right\}; 1,1 \right] \cr 
V_{n-2}^M &= {1 \over 2} \Gamma(\Delta_{(n-1)n\delta_{n-3},}-h) \,  F_A^{(1)}\!\left[\Delta_{(n-1)n\delta_{n-3},}- h; \{-k_{n-3} \}; \left\{\Delta_{\delta_{n-3}} -h+1 \right\}; 1 \right] ,
}
for $1 \leq i \leq n-4$. 

One can now obtain the single-trace projection of the AdS diagram~\eno{CombCanonDiag} by evaluating the residue at the ``single-trace poles'' in the ``${\cal U}_{\rm comb}$-plane.'' 
The poles are situated at~\eno{MandelstamPoles}, which in the ${\cal U}_{\rm comb}$-plane corresponds to
\eqn{gamma1jExp}{
\gamma_{1j} = \Delta_{j\delta_{j-2},\delta_{j-1}} + k_{j-2,j-1} - \sum_{i=2}^{j-1}\gamma_{ij} \qquad (2 \leq j \leq n-2)\,,
}
which we obtained by inverting~\eno{CombMandelstamExplicit} and where made the identifications $\Delta_{\delta_0} := \Delta_1$ and $k_0 :=0$ to write~\eno{gamma1jExp} compactly. 

The residue at the poles~\eno{gamma1jExp} is given by~\eno{Ast} 
where the dependent variables inside Gamma functions, written in~\eno{CombMellin}, transform to
\eqn{DCombSt}{
\gamma_{1(n-1)} &=   \Delta_{\delta_{n-3} (n-1),n}   + k_{n-3} - \sum_{j=2}^{n-2} \gamma_{j(n-1)}  \cr 
\gamma_{1n} &= \Delta_{1n, 23\ldots (n-1)} + \sum_{j=2}^{n-2}\gamma_{j(n-1)} + \sum_{2 \leq i < j \leq n-2} \gamma_{ij}  \cr 
\gamma_{in}  &=  \Delta_{i\delta_{i-1},\delta_{i-2}} + k_{i-1,i-2} -\sum_{j=i+1}^{n-1} \gamma_{ij} \qquad\qquad (2 \leq i\leq n-2) \cr 
\gamma_{(n-1)n} &=  \Delta_{(n-1)n,\delta_{n-3}} -k_{n-3} \,.
 }

The conformal block~\eno{CBSerInt} is then obtained  by projecting out the following OPE coefficients:  
\eqn{fiComb}{
f_1 = C_{\Delta_1 \Delta_2 \Delta_{\delta_1}} \qquad f_{i+1} = C_{\Delta_{\delta_i} \Delta_{\delta_{i+1}} \Delta_{i+2}} \qquad f_{n-2} = C_{\Delta_{n-1} \Delta_n \Delta_{\delta_{n-3}}} \,,
}
for $1 \leq i \leq n-4$, whose general form was given in~\eno{MFTOPE}.

We will now evaluate the second line of~\eno{CBSerInt}, or more precisely, the equivalent form in~\eno{Ifinal}. Substituting~\eno{gamma1jExp} and~\eno{DCombSt} in~\eno{Ifinal}, we get
\eqn{IfinalComb}{
I &= \left( \prod_{(rs) \in {\cal V}_{\rm comb}} \int { d\widetilde{\gamma}_{rs} \over 2\pi i} \int { d\gamma_{rs} \over 2\pi i} \right)
 \left( \prod_{2 \leq r < s \leq n-1}  \Gamma(-\widetilde{\gamma}_{rs}) \Gamma(\widetilde{\gamma}_{rs} + \gamma_{rs}) (v_{rs}-1)^{\widetilde{\gamma}_{rs}} \right) 
 \cr 
 &\times 
   \Gamma\left(\Delta_{n1, 23\ldots (n-1)} + \sum_{2 \leq r < s \leq n-1} \gamma_{rs} \right)  
 \cr 
 &\times \left(\prod_{i=2}^{n-1} \Gamma\left(\Delta_{i\delta_{i-2},\delta_{i-1}} + k_{i-2,i-1} - \sum_{j=2}^{i-1}\gamma_{ji}\right) 
  \Gamma\left(\Delta_{i\delta_{i-1},\delta_{i-2}} + k_{i-1,i-2} -\sum_{j=i+1}^{n-1} \gamma_{ij} \right) \right),
 }
where we employed the additional identifications $\Delta_{\delta_{n-2}}:=\Delta_n$ and $k_{n-2}:=0$ to write $I$ compactly.

To evaluate the $\binom{n-2}{2}$-dimensional contour integral over $\gamma_{rs}$ variables, we will employ a multi-dimensional variant of the first Barnes lemma
 \eqn{MultiFirstBarnes}{
&\left( \prod_{r=1}^K \int {d s_{r} \over 2 \pi i} \, \Gamma(A_r + s_r) \Gamma(B_r-s_r) \right) \Gamma\left(C+ \sum_{r=1}^K s_r\right) \Gamma\left(D-\sum_{r=1}^K s_r\right) 
\cr 
&= { \left(\prod_{r=1}^K \Gamma(A_r + B_r) \right) \Gamma\left(C+D\right) \Gamma\left(\sum_{r=1}^K A_r + D\right) \Gamma\left(\sum_{r=1}^K B_r + C\right) \over \Gamma\left(\sum_{r=1}^K (A_r + B_r) + C +D \right) } \,,
}
which can be easily proven by a repeated application of the first Barnes lemma~\eno{FirstBarnes}. 
 
Looking forward, our strategy will be to evaluate the contour integrals in the order shown below (see figure~\ref{fig:combSets} for color key):
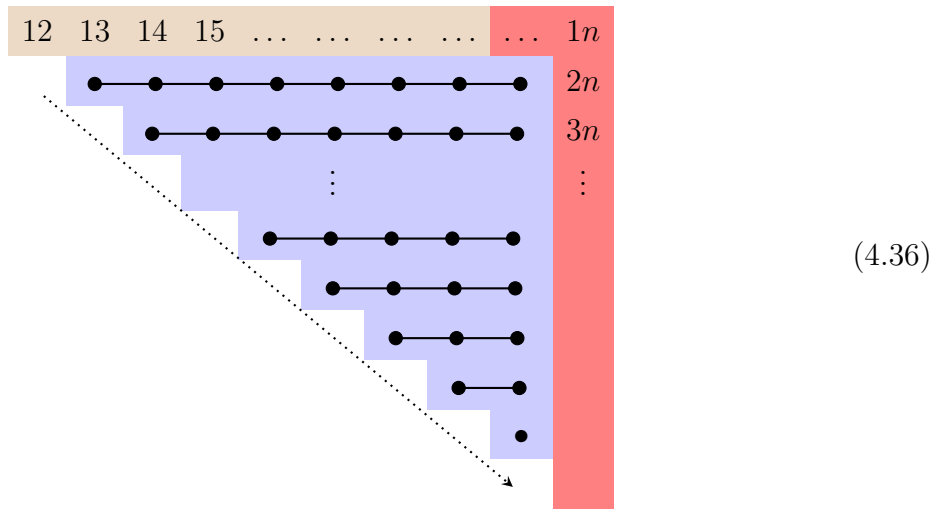
\eqn{combOrder}{
  \begin{array}{ccccccccc>{\columncolor{red!50}}c}
      \rowcolor{brown!30}  12 & 13  & 14 & 15 & \ldots & \ldots &  \ldots & \ldots  & \Dchan \ldots &  \Dchan 1n \\
         \tikzmarkAlt{20}{}   & \tikzmarkAlt{21}{} \Vchan  & \Vchan   &\Vchan  &\Vchan  & \Vchan  & \Vchan  &\Vchan  &\Vchan \tikzmarkAlt{2n}{} &   2n \\
             &   &  \Vchan \tikzmarkAlt{31}{} & \Vchan & \Vchan  & \Vchan  & \Vchan  &\Vchan   & \Vchan \tikzmarkAlt{3n}{} &    3n\\
      & &   &   \Vchan & \Vchan &  \Vchan \vdots &  \Vchan & \Vchan   & \Vchan &   \vdots\\
      & & & & \Vchan \tikzmarkAlt{41}{}  &  \Vchan &  \Vchan & \Vchan  & \Vchan \tikzmarkAlt{4n}{} &   \\
        & & & & & \Vchan \tikzmarkAlt{51}{}  &  \Vchan & \Vchan    & \Vchan \tikzmarkAlt{5n}{}  &   \\
            &    &   &     &   &  & \Vchan  \tikzmarkAlt{61}{}  &\Vchan   &\Vchan \tikzmarkAlt{6n}{}  &   \\
           &   &    &     &   &   &   & \Vchan \tikzmarkAlt{71}{}  &\Vchan \tikzmarkAlt{7n}{}    &  \\
          &   &    &     &   &  &   &   & \Vchan \bullet &  \\
         & &  &     &   &    & &   & \tikzmarkAlt{8n}{}  & 
  \end{array}
 \linkhor{21}{2n}{black}
 \linkhor{31}{3n}{black}
 \linkhor{41}{4n}{black}
 \linkhor{51}{5n}{black}
 \linkhor{61}{6n}{black}
 \linkhor{71}{7n}{black}
 \linkarrow{20}{8n}{black}
}
Here, integrals over elements of the index ${\cal V}_{\rm comb}$ (shown in blue) are indicated with black dots. Integrals over all black dots connected via links are performed simultaneously by an application of~\eno{MultiFirstBarnes}, while the dotted arrow shows the order in which the contour integrals over the disconnected chains of dots are performed.

Concretely, for $2 \leq m \leq n-1$, define
\eqn{JmDef}{
J_m &:=   
 \left( \prod_{\substack{m \leq r< s \leq n-1 } }   \Gamma(\widetilde{\gamma}_{rs} + \gamma_{rs})  \right)
 \Gamma\left( \Delta_{n\delta_{m-2}, m(m+1)\ldots (n-1)}  + k_{m-2} + \sum_{\substack{m \leq a < b \leq n-1 }} \gamma_{ab}    \right) 
 \cr 
 &\times
 \left( \prod_{\substack{i=m}}^{n-1} \Gamma\left( \sum_{j=2}^{m-1}  \widetilde{\gamma}_{ji} + \Delta_{i\delta_{i-2},\delta_{i-1}} + k_{i-2,i-1} - \sum_{j=m}^{i-1} \gamma_{ji} \right)    \right) \cr 
 &\times 
\left(\prod_{i=m}^{n-1}  \Gamma\left(\Delta_{i\delta_{i-1},\delta_{i-2}} + k_{i-1,i-2} -\sum_{j=i+1}^{n-1} \gamma_{ij} \right) \right). 
}
Then for  $2 \leq m \leq n-2$ integrating $J_m$ over the elements in the $m$-th row of the index set ${\cal V}_{\rm comb}$ in~\eno{combOrder} gives
\eqn{CombInduction}{
 \left(\prod_{m+1 \leq  j \leq n-1} \int {d\gamma_{mj} \over 2\pi i}  \right) J_m = \lambda_m J_{m+1}\,,
 }
where
\eqn{lambdaDef}{
\lambda_m &:=  
{ \Gamma\left(\Delta_{m\delta_{m-2},\delta_{m-1}}\right) 
 \Gamma\left(  \Delta_{m\delta_{m-1},\delta_{m-2}} \right)
  \Gamma\left(\Delta_{\delta_{m-2} \delta_{m-1}, m} \right)
  \over 
  \Gamma\left(  \Delta_{\delta_{m-1}} \right)
  }
 {1 \over 
  \left(  \Delta_{\delta_{m-1}}   \right)_{2k_{m-1}  + \sum_{i=2}^{m}\sum_{j=m+1}^{n-1} \widetilde{\gamma}_{ij}}
  }
 \cr 
&\quad \times 
 \left(\Delta_{m\delta_{m-2},\delta_{m-1}} \right)_{ k_{m-2,m-1}  +  \sum_{i=2}^{m-1} \widetilde{\gamma}_{im} }
\left(  \Delta_{m\delta_{m-1},\delta_{m-2}}   \right)_{ k_{m-1,m-2} +  \sum_{j=m+1}^{n-1} \widetilde{\gamma}_{mj} } 
\cr 
&\quad \times 
 \left(\Delta_{\delta_{m-2} \delta_{m-1}, m}   \right)_{k_{(m-2)(m-1),}  +  \sum_{i=2}^{m-1}\sum_{j=m+1}^{n-1} \widetilde{\gamma}_{ij}}\,,
 }
which is proven in appendix~\ref{APP:COMBPROOF}. 
Moreover, at $m=n-1$, $J_{m}$ reduces to a $\gamma_{ij}$-independent expression,
\eqn{JnMinus1}{
J_{n-1} &= 
 \Gamma\left(   \Delta_{(n-1)\delta_{n-3},n}  \right) 
  \Gamma\left(\Delta_{(n-1)n,\delta_{n-3}}  \right)
   \Gamma\left(\Delta_{n\delta_{n-3}, (n-1)}  \right)  
   \cr 
  &\quad \times 
 \left(   \Delta_{(n-1)\delta_{n-3},n}   \right)_{ k_{n-3} + \sum_{i=2}^{n-2} \widetilde{\gamma}_{i(n-1)} } 
  \left(\Delta_{(n-1)n,\delta_{n-3}}  \right)_{- k_{n-3} }
   \left(\Delta_{n\delta_{n-3}, (n-1)}  \right)_{  k_{n-3}} \,.
 }
The careful reader may notice that both~\eno{lambdaDef} and~\eno{JnMinus1} can be written more compactly purely in terms of Gamma functions. We have chosen to express them in terms of Pochhammer symbols to facilitate matching with the Feynman rules of section~\ref{FEYNMAN} at the end of this section.

In terms of $J_m$, the original contour integral~\eno{IfinalComb} can be written as
\eqn{Icomb}{
I  &=  \left( \prod_{(rs) \in {\cal V}_{\rm comb}} \int { d\widetilde{\gamma}_{rs} \over 2\pi i} 
\Gamma(-\widetilde{\gamma}_{rs}) (v_{rs}-1)^{\widetilde{\gamma}_{rs}} \right) 
\left(\prod_{2 \leq i < j \leq n-1} \int {d\gamma_{ij}  \over 2\pi i}  \right) J_2 \cr 
&= \left( \prod_{(rs) \in {\cal V}_{\rm comb}} \int { d\widetilde{\gamma}_{rs} \over 2\pi i}  \Gamma(-\widetilde{\gamma}_{rs}) (v_{rs}-1)^{\widetilde{\gamma}_{rs}} \right) \lambda_2
\left(\prod_{3 \leq i < j \leq n-1} \int {d\gamma_{ij}  \over 2\pi i}  \right)  J_3 \cr 
&= \left( \prod_{(rs) \in {\cal V}_{\rm comb}} \int { d\widetilde{\gamma}_{rs} \over 2\pi i}   \Gamma(-\widetilde{\gamma}_{rs}) (v_{rs}-1)^{\widetilde{\gamma}_{rs}} \right) \lambda_2 \lambda_3
\left(\prod_{4 \leq i < j \leq n-1} \int {d\gamma_{ij}  \over 2\pi i}    \right)  J_4 \cr 
&= \left( \prod_{(rs) \in {\cal V}_{\rm comb}} \int { d\widetilde{\gamma}_{rs} \over 2\pi i}   \Gamma(-\widetilde{\gamma}_{rs}) (v_{rs}-1)^{\widetilde{\gamma}_{rs}} \right)   \lambda_2 \lambda_3 \ldots \lambda_{n-3}
\left(\prod_{n-2 \leq i < j \leq n-1} \int {d\gamma_{ij}  \over 2\pi i}  \right) J_{n-2} \cr
&= \left( \prod_{(rs) \in {\cal V}_{\rm comb}} \int { d\widetilde{\gamma}_{rs} \over 2\pi i}   \Gamma(-\widetilde{\gamma}_{rs}) (v_{rs}-1)^{\widetilde{\gamma}_{rs}} \right) \left(\prod_{m=2}^{n-2} \lambda_m \right) J_{n-1},
  } 
where in the second step onward, we made repeated use of~\eno{CombInduction} to perform all $\gamma_{rs}$ integrals in the manner indicated in~\eno{combOrder}.

Now we turn to the $\widetilde{\gamma}_{rs}$ integrals. For carrying out the contour integrals, it is convenient to rewrite $I$ in terms of Gamma functions as
\eqn{Icomb2}{
I &= \left(\! \prod_{(rs) \in {\cal V}_{\rm comb}} \!\!\!\!\!\! \int { d\widetilde{\gamma}_{rs} \over 2\pi i}   \Gamma(-\widetilde{\gamma}_{rs}) (v_{rs}-1)^{\widetilde{\gamma}_{rs}} \!\!\right) \!\! \Bigg( \prod_{m=2}^{n-2} \Gamma\!\left( \! \Delta_{\delta_{m-2}\delta_{m-1},m} +{k_{(m-2)(m-1),}  + \sum_{i=2}^{m-1}\sum_{j=m+1}^{n-1} \widetilde{\gamma}_{ij}} \!\right)
\cr
&\times 
{\Gamma(  \Delta_{m\delta_{m-2},\delta_{m-1}} +{k_{m-2,m-1}  + \sum_{i=2}^{m-1} \widetilde{\gamma}_{im}}) 
\Gamma(  \Delta_{m\delta_{m-1},\delta_{m-2}} +{k_{m-1,m-2}  + \sum_{j=m+1}^{n-1} \widetilde{\gamma}_{mj}}) 
 \over 
 \Gamma(  \Delta_{\delta_{m-1}} +{2k_{m-1}  + \sum_{i=2}^{m}\sum_{j=m+1}^{n-1} \widetilde{\gamma}_{ij}})} \! \Bigg)
 \cr
 &\times \Gamma\!\left(\!   \Delta_{(n-1)\delta_{n-3},n} + k_{n-3}+ \sum_{i=2}^{n-2} \widetilde{\gamma}_{i(n-1)}  \!\! \right) \!
 \Gamma(\Delta_{(n-1)n,\delta_{n-3}}-k_{n-3}  )
   \Gamma(\Delta_{n\delta_{n-3}, (n-1)} +k_{n-3} ) \,.
 }
Examining the pole structure of the integrand~\eno{Icomb2}, we notice that just like in the example of the seven-point block in the previous subsection, we can evaluate the remaining $\widetilde{\gamma}_{rs}$ contour integrals by closing the contours to the right. 
In the process, each integral picks up a semi-infinite sequence of poles originating from $\Gamma(-\widetilde{\gamma}_{rs})$ at $\widetilde{\gamma}_{rs} = j_{rs}$ for non-negative integers $j_{rs}$, for each $(rs) \in {\cal V}_{\rm comb}$.
All other poles lie to the left of the contour and the contribution from the arc at infinity vanishes.
This immediately leads to
\eqn{CombFinalSum}{
I &=   
\left(\prod_{(rs) \in {\cal V}_{\rm Comb} } \sum_{j_{rs}=0}^{\infty}  {(1-v_{rs})^{j_{rs}} \over j_{rs}!} \! \right) \!\!
\Bigg( \prod_{m=2}^{n-2} \Gamma\!\left(\!  \Delta_{\delta_{m-2}\delta_{m-1},m} +{k_{(m-2)(m-1),}  + \sum_{r=2}^{m-1}\sum_{s=m+1}^{n-1} j_{rs}} \!\right) 
\cr
&\times 
 {\Gamma(  \Delta_{m\delta_{m-2},\delta_{m-1}} +{k_{m-2,m-1}  + \sum_{r=2}^{m-1} j_{rm}} ) 
\Gamma(  \Delta_{m\delta_{m-1},\delta_{m-2}} +{k_{m-1,m-2}  + \sum_{s=m+1}^{n-1} j_{ms}}) 
 \over 
 \Gamma(  \Delta_{\delta_{m-1}} +{2k_{m-1}  + \sum_{r=2}^{m}\sum_{s=m+1}^{n-1} j_{rs}})} \Bigg)
 \cr
 &\times \Gamma\!\left(\!   \Delta_{(n-1)\delta_{n-3},n} + k_{n-3}+ \sum_{r=2}^{n-2} j_{r(n-1)} \!\! \right) \!
 \Gamma(\Delta_{(n-1)n,\delta_{n-3}}-k_{n-3}  )
  \Gamma(\Delta_{n\delta_{n-3}, (n-1)} +k_{n-3} ) \, .
  }
 
It is now suggestive to re-express $I$ in terms of Pochhammer symbols, as shown here:
\eqn{CombFinalPoch}{
I &=   
\left(\prod_{(rs) \in {\cal V}_{\rm Comb} } \sum_{j_{rs}=0}^{\infty}  {(1-v_{rs})^{j_{rs}} \over j_{rs}!} \! \right) \!\!
\Bigg( \prod_{m=2}^{n-2} (\Delta_{\delta_{m-2}\delta_{m-1},m})_{k_{(m-2)(m-1),}  + \sum_{r=2}^{m-1}\sum_{s=m+1}^{n-1} j_{rs}}
\cr
&\times 
 { (\Delta_{m\delta_{m-2},\delta_{m-1}} )_{k_{m-2,m-1}  + \sum_{r=2}^{m-1} j_{rm}}  
(\Delta_{m\delta_{m-1},\delta_{m-2}})_{k_{m-1,m-2}  + \sum_{s=m+1}^{n-1} j_{ms}} 
 \over 
 (  \Delta_{\delta_{m-1}})_{2k_{m-1}  + \sum_{r=2}^{m}\sum_{s=m+1}^{n-1} j_{rs}} } \Bigg)
 \cr
 &\times ( \Delta_{(n-1)\delta_{n-3},n})_{k_{n-3}+ \sum_{r=2}^{n-2} j_{r(n-1)}}
 (\Delta_{(n-1)n,\delta_{n-3}})_{-k_{n-3}}
  (\Delta_{n\delta_{n-3}, n-1})_{k_{n-3}} \cr 
  &\times 
  \left(\prod_{m=2}^{n-2} {\Gamma(\Delta_{\delta_{m-2}\delta_{m-1},m}) \Gamma(  \Delta_{m\delta_{m-2},\delta_{m-1}}) \Gamma(  \Delta_{m\delta_{m-1},\delta_{m-2}})  \over \Gamma(\Delta_{\delta_{m-1}}) }\right) \cr 
  &\times 
  \Gamma( \Delta_{(n-1)\delta_{n-3},n} ) \Gamma(\Delta_{(n-1)n,\delta_{n-3}}) \Gamma(\Delta_{n\delta_{n-3}, n-1})\,.
  }
  Putting this $I$  back into the expression for the full conformal block~\eno{CBSerInt}, 
it is clear that, just like for the seven-point example above, all explicitly shown $(n-2)$ triplets of Gamma functions in the numerator of~\eno{CombFinalPoch} cancel against the same triplets appearing in the $(n-2)$ OPE coefficients~\eno{fiComb}. 
The remaining factor of the Gamma function in the numerator of each OPE coefficient cancels against the Gamma function in each of the Mellin vertex factors~\eno{MellinCombVertex}.
The explicit factor of Gamma function in the denominator in~\eno{CombFinalPoch} cancels against one of two such identical factors in the normalization constant ${\cal N}$ written in~\eno{NDef}, while the factors of Gamma functions in the denominators of the OPE coefficients cancel out all factors of Gamma functions in ${\cal N}$ which carry external conformal dimensions in their argument.
All explicit factors of $\pi$ cancel out too, leaving only the Pochhammer symbols in~\eno{CombFinalPoch}, and powers of cross-ratios (as well as the expected factors of  factorials).

It is now straightforward to check that the $(n-2)$ triplets of Pochhammer symbols in~\eno{CombFinalPoch} reproduce precisely the $(n-2)$ triplets of Pochhammer symbols appearing in the vertex factors of the Feynman rules~\eno{CombVertex}. 
Likewise the $(n-2)$ factors of Lauricella functions in the Mellin space vertex factors~\eno{MellinCombVertex} appearing inside~\eno{CBSerInt} find precise term by term agreement with the Lauricella functions in the conformal block vertex factors~\eno{CombVertex}.
The remaining $(n-3)$ ratios of Pochhamer symbols, of the form $(\Delta_{\delta_i}-h+1)_{k_i} / (\Delta_{\delta_i})_{2k_i + \ell_i}$ for an appropriately defined $\ell_{\delta_i}$ agree perfectly with the  edge factors in the conformal block Feynman rules~\eno{EdgeDef}, where $\ell_{\delta_i}$ are identified as the post-Mellin parameters~\eno{pMComb} associated with the internal legs labeled with conformal dimensions $\Delta_{\delta_i}$.

Thus, starting from first principles ({\it viz.}\ using the Feynman rules for Mellin amplitudes), we have reproduced the conformal block of section~\ref{NCOMB} which was obtained from an application of the proposed Feynman rules (and also previously obtained in ref.~\cite{Parikh:2019dvm} using geodesic bulk diagram techniques).

\subsection{OPE channel}
\label{OPEPROOF}

To reproduce the OPE channel conformal block obtained previously via Feynman rules, we start with the following canonical AdS diagram: 
\eqn{OPECanonDiag}{
A_{n, \rm OPE} := \musepic{\figNOPEWitten}\,,
}
where the ellipses represent a repeating pattern of ``upright Y-shaped'' interacting legs attached to the central horizontal line. The vertical internal exchanges are labeled with even-indexed scaling dimensions $\Delta_{\delta_2},\Delta_{\delta_4},\Delta_{\delta_6},\ldots,\Delta_{\delta_{n-4}}$, while the horizontal internal exchanges are labeled with odd-indexed scaling dimensions $\Delta_{\delta_1},\Delta_{\delta_3},\Delta_{\delta_5},\ldots,\Delta_{\delta_{n-3}}$. 

Just like in the previous two examples, we will use the same cross-ratios~\eno{OPECR} as input data as used for Feynman rules.
Let us recall the associated index sets which will be important in the computations to follow.
The associated choice of dependent and independent Mellin index sets will also be identical. The dependent set ${\cal D}_{\rm OPE}$ was given in~\eno{DOPEDef} which allowed a rewriting of the Mellin product as shown in~\eno{OPEMellinProd}, with the Mandelstam invariants for each internal leg as defined in~\eno{OPEMandelstam}, and also determined the set ${\cal V}_{\rm OPE}$ as shown in~\eno{VOPEDef}.
 The remaining independent Mellin variables are associated with the set ${\cal U}_{\rm OPE}$ which can be found using~\eno{UnionAll}:\footnote{Explicitly, ${\cal U}_{\rm OPE} = \{ (12),(34),(56),(78), \ldots, ((n-1)n)\} \,\bigcup\, \{(24), (26), (28), \ldots, (2(n-4))\}$.}  
\eqn{UOPEDef}{
{\cal U}_{\rm OPE} := \big\{ \left((2j+1)(2j+2)\right) \, \big| \, 0 \leq j \leq n/2-1 \big\} 
\; \bigcup \;
\big\{ \left(2(2j)\right) \, \big|\, 2\leq j\leq n/2-2 \big\} \,.
}
It is useful to represent the index sets visually as shown in figure~\ref{fig:OPESets}.
 \begin{figure}
 \centering
 \[
  \begin{array}{|c|c|c|c|c|c|c|c|c|>{\columncolor{red!50}}c|}
    \cline{1-10}
      \rowcolor{blue!20} \Uchan  12  & 13 & 14 & 15 & 16 & \ldots & 1(n-3) & 1(n-2) &  1(n-1) &  \Dchan  1n \\
    \cline{1-10}
         \multicolumn{1}{c|}{}   &  \Vchan 23 & \Uchan  24 &\Vchan 25 &\Uchan 26 &  \cellBG{double color fill={blue!20}{brown!30}, shading angle=90, opacity=1}{c}{\ldots}   & \Vchan 2(n-3) &\Dchan 2(n-2) &\Vchan 2(n-1) &   2n \\
    \cline{2-10}
        \multicolumn{2}{c|}{}    &  \Uchan 34 & \Vchan 35 & \Vchan 36 & \Vchan \ldots & \Vchan 3(n-3) &\Vchan  3(n-2) & \Vchan 3(n-1) &   3n \\
    \cline{3-10}
      \multicolumn{3}{c|}{}    &   \Vchan 45 & \Vchan 46 &  \Vchan \ddots &  \Vchan & \Vchan \vdots  & \Vchan & \vdots  \\
    \cline{4-10}
     \multicolumn{4}{c|}{}  & \Uchan 56 &  \Vchan \ddots &  \Vchan & \Vchan \vdots & \Vchan & \vdots  \\
    \cline{5-10}
       \multicolumn{5}{c|}{}  &  \cellBG{double color fill={blue!20}{brown!30}, shading angle=45, opacity=1}{c}{\ddots}  &  \Vchan & \Vchan \vdots  & \Vchan & \vdots  \\
    \cline{6-10}
               \multicolumn{6}{c|}{}   & \Vchan (n-4)(n-3) &\Vchan (n-4)(n-2) &\Vchan (n-4)(n-1) &  (n-4)n \\
    \cline{7-10}
          \multicolumn{1}{c}{\Uchan {\cal U}}   &  \multicolumn{6}{c|}{}    & \Uchan (n-3)(n-2) &\Vchan (n-3)(n-1) &  (n-3)n\\
    \cline{8-10}
          \multicolumn{1}{c}{\Vchan {\cal V}} &  \multicolumn{7}{c|}{}   & \Dchan (n-2)(n-1) &  (n-2)n\\
    \cline{9-10}
        \multicolumn{1}{c}{\Dchan {\cal D}}  &  \multicolumn{8}{c|}{}    & \Uchan (n-1)n \\
    \cline{10-10}
  \end{array}
 \]
  \caption{Color-coded matrix displaying the canonical choice of Mellin variable index subsets ${\cal U}_{\rm OPE}, {\cal V}_{\rm OPE}$ and ${\cal D}_{\rm OPE}$.}
\label{fig:OPESets}
\end{figure}

According to the Feynman rules for Mellin amplitudes, the AdS diagram~\eno{OPECanonDiag} has the Mellin amplitude,
\eqn{MellinAmpOPE}{
{\cal M}_{n,\rm OPE}(\gamma_{ij}) = \left(\prod_{i=1}^{n-3} \sum_{k_i=0}^\infty \right) \left( \prod_{i=1}^{n-3} E_i^{\rm Mellin} \right)  \left(\prod_{i=1}^{n/2} V^{(1){\rm Mellin}}_i\right)  \left( \prod_{i=1}^{n/2-2} V_i^{(3){\rm Mellin}} \right).
}
Here each of the edge factors is given by~\eno{EdgeMellin} with the Mandelstam invariants $s_i$ given in~\eno{OPEMandelstam} and the associated legs assigned single-trace parameters $k_i$, and the vertex factors come directly from~\eno{VertexMellin}:
\eqn{MellinOPEVertex}{
V_1^{(1){\rm Mellin}} &= {1 \over 2} \Gamma(\Delta_{12\delta_1,}-h) \,  F_A^{(1)}\!\left[\Delta_{12\delta_1,}- h; \{-k_1 \}; \left\{\Delta_{\delta_1} -h+1 \right\}; 1 \right] \cr 
V_a^{(1){\rm Mellin}} &= {1 \over 2} \Gamma(\Delta_{(2a-1)(2a)\delta_{2a-2},}-h) \,  F_A^{(1)}\!\left[\Delta_{(2a-1)(2a)\delta_{2a-2},}- h; \{-k_{2a-2} \};\! \left\{\!\Delta_{\delta_{2a-2}} -h+1 \!\right\}\!;\! 1  \right] \cr 
V_{n\over 2}^{(1){\rm Mellin}} &= {1 \over 2} \Gamma(\Delta_{(n-1)n\delta_{n-3},}-h) \,  F_A^{(1)}\!\left[\Delta_{(n-1)n\delta_{n-3},}- h; \{-k_{n-3} \}; \left\{\Delta_{\delta_{n-3}} -h+1 \right\}; 1 \right] \cr 
V_{b}^{(3){\rm Mellin}} &= {1 \over 2} \Gamma(\Delta_{\delta_{2b-1} \delta_{2b} \delta_{2b+1},}-h)   \,
 F_A^{(3)}\!\left[\Delta_{\delta_{2b-1} \delta_{2b} \delta_{2b+1},}- h; \{-k_{2b-1}, -k_{2b},-k_{2b+1} \}; \right. \cr 
&\qquad\qquad\qquad\qquad\qquad\qquad\quad        \left. \left\{\Delta_{\delta_{2b-1}} -h+1, \Delta_{\delta_{2b}} -h+1, \Delta_{\delta_{2b+1}} \right\}; 1,1,1 \right],
}
for $2 \leq a \leq n/2-1$ and $1 \leq b \leq n/2-2$. 

Substituting~\eno{MellinAmpOPE} into~\eno{MellinAmpDef}, we proceed to obtain the single-trace projection of the AdS diagram as described around~\eno{MandelstamPoles}. 
This leads to~\eno{Ast} which is the desired conformal block times a set of known OPE coefficients,
\eqn{fiOPE}{
f_1^{(1)} = C_{\Delta_1\Delta_2\Delta_{\delta_1}} \quad f_a^{(1)} = C_{\Delta_{2a-1}\Delta_{2a}\Delta_{\delta_{2a-2}}} \quad f_{n\over 2}^{(1)} = C_{\Delta_{n-1}\Delta_n\Delta_{\delta_{n-3}}} \quad f_b^{(3)} = C_{\Delta_{\delta_{2b-1}}\Delta_{\delta_{2b}}\Delta_{\delta_{2b+1}}}\,,
} 
for $2 \leq a \leq n/2-1$ and $1 \leq b \leq n/2-2$,
which can be factored out to obtain the block~\eno{CBSerInt}.
This single-trace projection is obtained by evaluating the residue at the poles~\eno{MandelstamPoles}, which in the ${\cal U}_{\rm OPE}$-plane occur at
\eqn{stPolesOPE}{
\begin{gathered}
\gamma_{12} = \Delta_{12,\delta_1} - k_1 \qquad 
\gamma_{(n-1)n} = \Delta_{(n-1)n,\delta_{n-3}} - k_{n-3}  \qquad 
\gamma_{(2i+1)(2i+2)} = \Delta_{(2i+1)(2i+2),\delta_{2i}} - k_{2i} \cr 
\gamma_{2(2j)} = \Delta_{\delta_{2j-3}\delta_{2j-2},\delta_{2j-1}} + k_{(2j-3)(2j-2),(2j-1)} - \sum_{\substack{(ab) \in {\cal V}_{\rm OPE} \\ a<b, b=2j-1 {\rm \ or\ }2j}} \gamma_{ab} \,,
\end{gathered}
}
for $1 \leq i \leq n/2-2$, and $2 \leq j \leq n/2-2$.
Substituting these in the dependent Mellin variables $\gamma_{ab}$~\eno{OPEMellin} for $(ab) \in {\cal D}_{\rm OPE}$, we get\footnote{
A notational remark: If $(ab) \in {\cal V}_{\rm OPE}$, then so is $(ba) \in {\cal V}_{\rm OPE}$. So if $\{(14),(24),(34),(45),(46),(47)\} \subseteq {\cal V}_{\rm OPE}$, whenever there is a restriction of the form $(a4) \in {\cal V}_{\rm OPE}$ with $a< 4$, it only admits elements from the set $\{(14),(24),(34)\}$ and not elements from the set $\{(45),(46),(47),\ldots\}$. Without the restriction $a<4$, all elements above will be admitted upon selecting $(a4) \in {\cal V}_{\rm OPE}$.}
\eqn{DOPEst}{
\gamma_{1n} &= \Delta_{1\delta_1,2} + k_1  - \sum_{j=3}^{n-1} \gamma_{1j}
\cr 
\gamma_{2n} &=  \Delta_{2\delta_{n-3},1\delta_2 \delta_4 \ldots \delta_{n-4}}  + k_{(n-3),24\ldots (n-4)}  - \gamma_{2(n-1)} + \sum_{j=3}^{n-2} \gamma_{1j} + \sum_{\substack{3\leq i<j\leq (n-2)\\ (ij) \in {\cal V}_{\rm OPE}}} \gamma_{ij} 
\cr 
 \gamma_{(2i+1)n}  &=  \Delta_{(2i+1)\delta_{2i},(2i+2)} + k_{2i}   -\sum_{((2i+1)b) \in {\cal V}_{\rm OPE}} \gamma_{(2i+1)b} 
\cr 
 \gamma_{(2j+2)n} &= 
   \Delta_{(2j+2)\delta_{2j+1},(2j+1)\delta_{2j-1}} + k_{2j+1,2j-1}  + \! \sum_{\substack{(a(2j+1)) \in {\cal V}_{\rm OPE} \\ a<2j+1}} \!\!\!\! \gamma_{a(2j+1)}  -\! \sum_{\substack{((2j+2)b) \in {\cal V}_{\rm OPE}\\ b> 2j+2}} \!\!\!\! \gamma_{(2j+2)b} 
\cr 
\gamma_{(n-2)n} &=  \Delta_{(n-2)n, (n-3)(n-1)\delta_{n-5}} - k_{n-5}  + \sum_{j=1}^{n-3} \gamma_{j(n-1)} + \sum_{j=1}^{n-4} \gamma_{j(n-3)}    
\cr 
\gamma_{(n-2)(n-1)} &=   \Delta_{(n-1)\delta_{n-3},n} + k_{n-3}  - \sum_{\substack{j=1}}^{n-3} \gamma_{j(n-1)} 
\cr 
\gamma_{2(n-2)} &=  \Delta_{\delta_{n-5}\delta_{n-4},\delta_{n-3}} + k_{(n-5)(n-4),(n-3)} - \sum_{\substack{(ab) \in {\cal V}_{\rm OPE} \\ a<b, b=n-3 {\rm \ or\ }n-2}} \gamma_{ab}\,,
}
for $1\leq i\leq n/2-2$ and $1 \leq j \leq n/2-3$,
where we made use of 
 \eqn{}{
\sum_{j=2}^{{n\over 2}-2} \gamma_{2(2j)} = \Delta_{\delta_1 \delta_2 \delta_4 \delta_6 \ldots \delta_{n-6}, \delta_{n-5}} + k_{1246\ldots (n-6),(n-5)} - \sum_{\substack{(ij) \in {\cal V}_{\rm OPE}\\ i<j,j=3,4,\ldots,n-4}} \gamma_{ij}\,,
 }
and 
 \eqn{}{
 \sum_{j=1}^{{n\over 2}-2} \gamma_{(2j+1)(2j+2)} = \Delta_{3456\ldots(n-3)(n-2), \delta_2 \delta_4 \ldots \delta_{n-4}} - k_{24\ldots (n-4),} \,.
 }

Substituting~\eno{stPolesOPE} and~\eno{DOPEst} back in~\eno{Ifinal}, we turn to evaluating the remaining contour integrals.
Like in the seven-point and comb channel examples, we will first integrate over the $\gamma_{rs}$ variables for $(rs) \in {\cal V}_{\rm OPE}$.
The order in which we will integrate is shown below (consult figure~\ref{fig:OPESets} for reference):
 \eqn{OPEOrder}
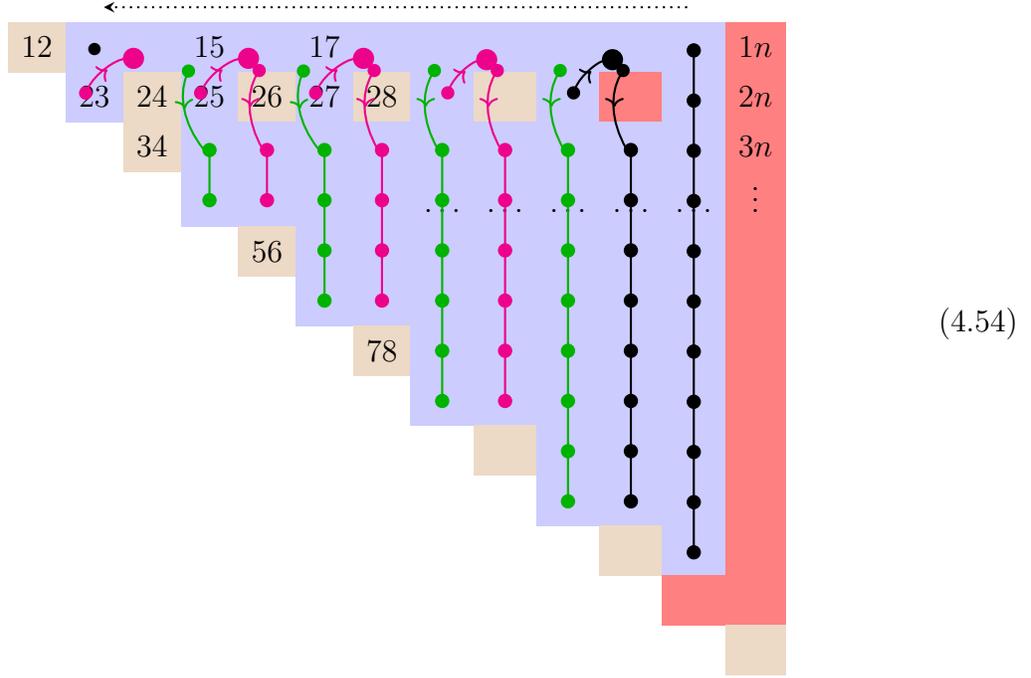
{  
\begin{array}{cccccccccccc>{\columncolor{red!50}}c}
   &  \tikzmarkAlt{03}{} & &  &  &  & &  & &  &   & \tikzmarkAlt{013}{} &  {\cellcolor{red!0}} \\
      \rowcolor{blue!20}  \Uchan  12 &  \bullet & \tikzmarkAlt{14}{} & 
      \tikzmarkAlt{15}{} 15 & \tikzmarkAlt{16}{} & \tikzmarkAlt{17}{} 17 & \tikzmarkAlt{18}{} & \tikzmarkAlt{19}{} & \tikzmarkAlt{110}{} & \tikzmarkAlt{111}{}  &  \tikzmarkAlt{112}{}  & \tikzmarkAlt{113}{} &  \Dchan 1n \\
           &  \Vchan \tikzmarkAlt{23}{} 23 & \Uchan 24  &\Vchan \tikzmarkAlt{25}{} 25 &\Uchan 26 & \Vchan \tikzmarkAlt{27}{} 27 & \Uchan 28 & \Vchan \tikzmarkAlt{29}{} & \Uchan & \Vchan \tikzmarkAlt{211}{} & \Dchan &\Vchan   &   2n \\
             &   &  \Uchan 34 & \Vchan  \tikzmarkAlt{35}{} & \Vchan \tikzmarkAlt{36}{} & \Vchan  \tikzmarkAlt{37}{} & \Vchan \tikzmarkAlt{38}{} &\Vchan \tikzmarkAlt{39}{} & \Vchan \tikzmarkAlt{310}{} & \Vchan \tikzmarkAlt{311}{} &\Vchan \tikzmarkAlt{312}{}  & \Vchan \tikzmarkAlt{3n}{} &    3n\\
      & &   &   \Vchan \tikzmarkAlt{45}{} & \Vchan \tikzmarkAlt{46}{} &  \Vchan  &  \Vchan  & \Vchan \ldots & \Vchan \ldots  & \Vchan\ldots  &\Vchan \ldots  & \Vchan \ldots &   \vdots\\
      & & & & \Uchan 56 &  \Vchan &  \Vchan & \Vchan & \Vchan  & \Vchan  &\Vchan  & \Vchan \tikzmarkAlt{4n}{} &   \\
        & & & & & \Vchan \tikzmarkAlt{67}{} &  \Vchan \tikzmarkAlt{68}{} & \Vchan   & \Vchan  & \Vchan  &\Vchan   & \Vchan   &   \\
            &    &   &     &   &  & \Uchan 78 & \Vchan  & \Vchan  &\Vchan  &\Vchan   &\Vchan   &   \\
           &   &    &     &   &   &   & \Vchan \tikzmarkAlt{89}{} & \Vchan \tikzmarkAlt{810}{} & \Vchan  &\Vchan   &\Vchan     &  \\
          &   &    &     &   &  &   &  & \Uchan  & \Vchan  &\Vchan  & \Vchan  &  \\
         & &  &     &   &    & &   &   & \Vchan \tikzmarkAlt{1011}{} & \Vchan \tikzmarkAlt{1012}{}  & \Vchan  &  \\
         & &  &     &   &    & &   &   &  & \Uchan  & \Vchan \tikzmarkAlt{1113}{} &  \\
         & &  &     &   &    & &   &   & &   & \Dchan  &  \\
         & &  &     &   &    & &   &   &  &   &   &  \Uchan
  \end{array}
 \link{113}{1113}{black}
 \linkarcCW{211}{112}{black}
 \linkarcACW{112}{312}{black}
 \link{312}{1012}{black}
 \linkarcACW{111}{311}{green!70!black}
 \link{311}{1011}{green!70!black}
 \linkarcCW{29}{110}{magenta}
 \linkarcACW{110}{310}{magenta}
 \link{310}{810}{magenta}
 \linkarcACW{19}{39}{green!70!black}
 \link{39}{89}{green!70!black}
 \linkarcCW{27}{18}{magenta}
 \linkarcACW{18}{38}{magenta}
 \link{38}{68}{magenta}
 \linkarcACW{17}{37}{green!70!black}
 \link{37}{67}{green!70!black}
 \linkarcCW{25}{16}{magenta}
 \linkarcACW{16}{36}{magenta}
 \link{36}{46}{magenta}
 \linkarcACW{15}{35}{green!70!black}
 \link{35}{45}{green!70!black}
 \linkarcCWterminal{23}{14}{magenta}
 \linkarrow{013}{03}{black}
 }
The direction of the dotted arrow  (right to left) indicates the order in which we integrate over the connected elements of the set ${\cal V}_{\rm OPE}$. 
Each connected chain corresponds to a subset of contour integrals that will be evaluated with the help of the inductive first Barnes lemma~\eno{MultiFirstBarnes}. 
While any ordering works, the precise ordering chosen here makes it possible to set up an inductive step.
The strategy will be as follows:
We will first evaluate integrals~\eno{Ifinal} associated with the two (right-most) black-colored chains in~\eno{OPEOrder}.
Using the resulting expression from the black-colored chain integrals, we will establish a two-step induction; the green and magenta colored chains above suggest how the induction will work.
 
 In appendix~\ref{APP:OPEFIRST}, we present the computation of the integrals marked as  black-colored chains above. 
 The end result of this computation is given in~\eno{IOPEAfterTwo}. 
 To set up induction, we define a new contour integral $\widehat{I}_{n-2K-1}$ such that 
\eqn{BaseCase}{
\widehat{I}_{n-2K-1}\Big|_{K=1} = I 
}
where $I$ is given by~\eno{IOPEAfterTwo}, and $\widehat{I}_{n-2K-1}$ is defined to be
 \eqn{IhatDef}{
\widehat{I}_{n-2K-1} &:=  %
\left(\prod_{(rs) \in {\cal V}_{\rm OPE}} \int  {d\widetilde{\gamma}_{rs} \over 2\pi i} \, \Gamma(-\widetilde{\gamma}_{rs}) (v_{rs}-1)^{\widetilde{\gamma}_{rs}} \right) 
\left( \prod_{j=n-2K-2}^{n-3} W_j  \right)
\cr 
&\times
\left(\prod_{\substack{(rs) \in {\cal V}_{\rm OPE}\\s\neq n-2K-1, \ldots ,n-1}} \int {d\gamma_{rs} \over 2\pi i} \,  \right)
  M_{n-2K-1}
\left(\prod_{r=1,r\neq 2}^{n-2K-2} \int {d\gamma_{r(n-2K-1)} \over 2\pi i}  \right)   L_{n-2K-1}\,,
}
for $1 \leq K \leq n/2-2$. Here $W_{n-3}$ which was defined in~\eno{WnMinus3}, is repeated below,
 \begingroup\makeatletter\def\f@size{10}\check@mathfonts
\eqn{WnMinus3Again}{
& W_{n-3} \cr 
&:=   {\Gamma\left(\Delta_{(n-1)n,\delta_{n-3}}-k_{n-3}\right)  
   \Gamma\left(\Delta_{n  \delta_{n-3},  (n-1) } + k_{n-3}  
 \right)  
 \Gamma\left( \displaystyle{\sum_{(i(n-1)) \in {\cal V}_{\rm OPE}}} \widetilde{\gamma}_{i(n-1)}  + \Delta_{(n-1) \delta_{n-3},n} + k_{n-3}\right) 
\over 
 \Gamma\left( \displaystyle{\sum_{(i(n-1)) \in {\cal V}_{\rm OPE}}} \widetilde{\gamma}_{i(n-1)}    + \Delta_{ \delta_{n-3}} + 2k_{n-3}  
 \right) },
}
\endgroup
and
\eqn{Wj}{
W_{j} &:= \begin{cases} \quad \Gamma\left(\sum_{i=1, i \neq 2}^{j+1}  \widetilde{\gamma}_{i(j+3)} + \sum_{i=1}^{j+1}  \widetilde{\gamma}_{i(j+2)}   +  \Delta_{\delta_{j}\delta_{j+1},\delta_{j+2}} + k_{(j)(j+1),(j+2)}  \right) 
\cr 
\times 
\Gamma\left( \sum_{i=1,i\neq 2}^{j+1} \sum_{b=j+4}^{n-1} \widetilde{\gamma}_{ib} + \sum_{b={j+3 \over 2}}^{{n\over 2}-1} \widetilde{\gamma}_{2(2b+1)}  +  \Delta_{\delta_{j}\delta_{j+2},\delta_{j+1}}  + k_{(j)(j+2),(j+1)}  \right) 
\cr 
\times
{
\Gamma\left(\displaystyle{\sum_{a=j+4}^{n-1}} \widetilde{\gamma}_{(j+3)a} +   \displaystyle{\sum_{a=j+4}^{n-1}} \widetilde{\gamma}_{(j+2
)a} + \Delta_{\delta_{j+2}\delta_{j+1},\delta_{j}} + k_{(j+2)(j+1),(j)}     \right) 
 \over 
 \Gamma\left( \sum_{i=1,i\neq 2}^{j+1} \sum_{b=j+2}^{n-1} \widetilde{\gamma}_{ib}  + \sum_{b={j+1 \over 2}}^{{n\over 2}-1} \widetilde{\gamma}_{2(2b+1)} + \Delta_{\delta_{j}} + 2k_{j} \right) }
 & j {\rm \ odd}
\cr \cr 
  \quad     { \Gamma(\Delta_{(j+1)(j+2),\delta_{j}}-k_{j}) \,
 \Gamma\left(\sum_{(r(j+2)) \in {\cal V}_{\rm OPE} } \widetilde{\gamma}_{r(j+2)}  + \Delta_{(j+2)\delta_{j},(j+1)} + k_{j}   \right)
  \over 
 \Gamma\left( \displaystyle{\sum_{(r(j+1)) \in {\cal V}_{\rm OPE} }} \widetilde{\gamma}_{r(j+1)}  + \sum_{(r(j+2)) \in {\cal V}_{\rm OPE} } \widetilde{\gamma}_{r(j+2)} + \Delta_{\delta_{j}} + 2k_{j}    \right) } 
\cr 
\times \,
 \Gamma\left(\sum_{(r(j+1)) \in {\cal V}_{\rm OPE} } \widetilde{\gamma}_{r(j+1)}  + \Delta_{(j+1)\delta_{j},(j+2)}  + k_{j} \right)   & j {\rm \ even }
 \end{cases}
}
for $n-2K-2 \leq j \leq n-4$.\footnote{We note that $W_j$ at $j=n-4$ coincides with~\eno{WnMinus4}.} Additionally, we define
\eqn{MnMinus2KMinus1}{
& M_{n-2K-1} \cr 
&:= \Gamma(\Delta_{12,\delta_1}-k_1) 
 \left(\prod_{j=1}^{{n\over 2}-K-2} \Gamma(\Delta_{(2j+1)(2j+2),\delta_{2j}}-k_{2j}) \right) 
 \left(\prod_{\substack{(rs) \in {\cal V}_{\rm OPE}\\s\neq n-2K-1, \ldots ,n-1}}  \Gamma(\widetilde{\gamma}_{rs} + \gamma_{rs}) \right)
 \cr 
 &\times 
\left(\prod_{j=1}^{{n\over 2}-K-2} \Gamma\left( \Delta_{\delta_{2j-1}\delta_{2j},\delta_{2j+1}} + k_{(2j-1)(2j),(2j+1)} - \sum_{\substack{(ab) \in {\cal V}_{\rm OPE} \\ a<b, b=2j+1 {\rm \ or\ }2j+2}} \gamma_{ab}\right) \right)
\cr 
&\times 
\Gamma\left(\sum_{\substack{((n-2K)a) \in {\cal V}_{\rm OPE}\\ a>n-2K}} \!\!\!\!\!\! \widetilde{\gamma}_{(n-2K)a} +  \!\!\! \sum_{\substack{((n-2K-1)a) \in {\cal V}_{\rm OPE}\\ a>n-2K}} \!\!\!\!\!\!\! \widetilde{\gamma}_{(n-2K-1)a} + \Delta_{\delta_{n-2K-1}\delta_{n-2K-2},\delta_{n-2K-3}}  \right. \cr 
 &\qquad + k_{(n-2K-1)(n-2K-2),(n-2K-3)}     \Bigg) \,,
}
and
\eqn{LnMinus2KMinus1}{
& L_{n-2K-1} \cr 
&:= \left(\prod_{r=1,r\neq 2}^{n-2K-2}  \Gamma(\widetilde{\gamma}_{r(n-2K)} + \widetilde{\gamma}_{r(n-2K-1)} + \gamma_{r(n-2K-1)}) \right) 
\cr 
&\times
 \Gamma\left(\sum_{j=n-2K+1}^{n-1} \widetilde{\gamma}_{1j}  +  \Delta_{1\delta_1,2} + k_1 -\sum_{j=3}^{n-2K-1} \gamma_{1j}\right)
\cr 
&\times 
 \Gamma\!\left(\!\widetilde{\gamma}_{2(n-2K-1)} +  \Delta_{\delta_{n-2K-3}\delta_{n-2K-2},\delta_{n-2K-1}} + k_{(n-2K-3)(n-2K-2),(n-2K-1)} - \!\! \sum_{a=1,a\neq 2}^{n-2K-2}\!\! \gamma_{a(n-2K-1)} \!\right)
\cr 
&\times 
\left(\prod_{j=1}^{{n\over 2}-K-2} \Gamma\left( \sum_{a=n-2K+1}^{n-1} \widetilde{\gamma}_{(2j+1)a} + \Delta_{(2j+1)\delta_{2j},(2j+2)} + k_{2j} - \sum_{\substack{((2j+1)b) \in {\cal V}_{\rm OPE}\\ b\neq n-2K,\ldots, n-1}}  \gamma_{(2j+1)b} \right) \right)
\cr 
&\times 
\left(\prod_{j=1}^{{n\over 2}-K-2} \Gamma\left( \sum_{a=n-2K+1}^{n-1} \widetilde{\gamma}_{(2j+2)a} + \Delta_{(2j+2)\delta_{2j+1},(2j+1)\delta_{2j-1}} + k_{2j+1,2j-1} + \sum_{\substack{(b(2j+1)) \in {\cal V}_{\rm OPE}\\ b<2j+1}} \gamma_{b(2j+1)} \right. \right. 
\cr 
&\quad 
 \left. \left. - \sum_{\substack{((2j+2)b) \in {\cal V}_{\rm OPE}\\ b>2j+2, b\neq n-2K,\ldots, n-1}}  \gamma_{(2j+2)b}  \right)  \right)
   \cr
&\times 
\Gamma\left(\sum_{i=0}^{K-1} \widetilde{\gamma}_{2(n-2i-1)}  +  \Delta_{2\delta_{n-2K-1},1\delta_2\delta_4\ldots \delta_{n-2K-2}} + k_{(n-2K-1),24\ldots(n-2K-2)} + \sum_{j=3}^{n-2K-1} \gamma_{1j} \right. 
\cr 
&\quad \left. + \sum_{\substack{3\leq i<j\leq n-2K-1\\ (ij) \in {\cal V}_{\rm OPE} }} \gamma_{ij} \right) .
}
It can be checked that~\eno{BaseCase} holds. This will serve as the base case for an inductive argument which we develop next.

We would like to integrate $L_{n-2K-1}$ over $\gamma_{1(n-2K-1)}, \gamma_{3(n-2K-1)},\gamma_{4(n-2K-1)},\ldots, \gamma_{(n-2K-2)(n-2K-1)}$.
This will turn out to be associated with integrating out a green-colored chain in~\eno{OPEOrder} in an intermediate step,
 where this  computation is described in appendix~\ref{APP:OPESECOND}.
Here we rewrite the result of this computation, given in~\eno{IhatAgainOne}  as follows:
\begingroup\makeatletter\def\f@size{11}\check@mathfonts
\eqn{IhatTwo}{
& \widehat{I}_{n-2K-1} \cr 
&=  \Gamma(\Delta_{12,\delta_1}-k_1) 
 \left(\prod_{j=1}^{{n\over 2}-K-2} \Gamma(\Delta_{(2j+1)(2j+2),\delta_{2j}}-k_{2j}) \right) 
\left(\prod_{(rs) \in {\cal V}_{\rm OPE}} \int  {d\widetilde{\gamma}_{rs} \over 2\pi i} \, \Gamma(-\widetilde{\gamma}_{rs}) (v_{rs}-1)^{\widetilde{\gamma}_{rs}} \right) 
\cr 
&\times
\left( \prod_{j=n-2K-3}^{n-3} W_j  \right)
\left(\prod_{\substack{(rs) \in {\cal V}_{\rm OPE}\\s\neq n-2K-3, \ldots ,n-1}} \int {d\gamma_{rs} \over 2\pi i} \,  \Gamma(\widetilde{\gamma}_{rs} + \gamma_{rs}) \right)
\left(\prod_{r=1, r\neq 2}^{n-2K-4} \int {d\gamma_{r(n-2K-3)} \over 2\pi i}  \right) 
\cr &\times 
\int {d \gamma_{2(n-2K-3)} \over 2\pi i}  
\left(\prod_{r=1,r\neq 2}^{n-2K-4} \int {d\gamma_{r(n-2K-2)} \over 2\pi i}   \right)   L_{n-2K-2}\,,
}
\endgroup
where
\begingroup\makeatletter\def\f@size{11}\check@mathfonts
\eqn{LnMinus2KMinus2}{
& L_{n-2K-2}  := \left(\prod_{\substack{(rs) \in {\cal V}_{\rm OPE} \\ r<s, s= n-2K-3, n-2K-2}}   \Gamma(\widetilde{\gamma}_{rs} + \gamma_{rs}) \right)
\cr 
&  \times 
\left(\prod_{j=1}^{{n\over 2}-K-2} \Gamma\left( \Delta_{\delta_{2j-1}\delta_{2j},\delta_{2j+1}} + k_{(2j-1)(2j),(2j+1)} - \sum_{\substack{(ab) \in {\cal V}_{\rm OPE} \\ a<b, b=2j+1 {\rm \ or\ }2j+2}} \gamma_{ab}\right) \right)
\cr 
&\times 
\Gamma\left(\widetilde{\gamma}_{1(n-2K)} + \widetilde{\gamma}_{1(n-2K-1)} + \sum_{j=n-2K+1}^{n-1} \widetilde{\gamma}_{1j}  +  \Delta_{1\delta_1,2} + k_1 -\sum_{j=3}^{n-2K-2} \gamma_{1j}\right) 
\cr
&\times 
\Gamma\left(\sum_{i=0}^{K-1} \widetilde{\gamma}_{2(n-2i-1)} + \widetilde{\gamma}_{2(n-2K-1)} +  \Delta_{2\delta_{n-2K-3},1\delta_2\delta_4\ldots \delta_{n-2K-4}} + k_{(n-2K-3),24\ldots(n-2K-4)} \right. 
\cr 
&\quad \left.  + \sum_{j=3}^{n-2K-2} \gamma_{1j} + \sum_{\substack{3\leq i<j\leq n-2K-2\\ (ij) \in {\cal V}_{\rm OPE} }} \gamma_{ij}     \right) 
\cr 
&\times 
 \left(\prod_{j=1}^{{n\over 2}-K-2} \Gamma\left(  \displaystyle{\sum_{a=n-2K-1}^{n-1}} \widetilde{\gamma}_{(2j+1)a} + \Delta_{(2j+1)\delta_{2j},(2j+2)} + k_{2j}  - \sum_{\substack{((2j+1)b) \in {\cal V}_{\rm OPE}\\ b\neq n-2K-1,\ldots, n-1}}  \gamma_{(2j+1)b} \right) \right) 
 \cr 
&\times 
 \left(\prod_{j=1}^{{n\over 2}-K-2} \Gamma( \displaystyle{\sum_{a=n-2K-1}^{n-1}} \widetilde{\gamma}_{(2j+2)a} + \Delta_{(2j+2)\delta_{2j+1},(2j+1)\delta_{2j-1}} + k_{2j+1,2j-1}  \right. \cr 
         &\quad \left.+ \sum_{\substack{(b(2j+1)) \in {\cal V}_{\rm OPE}\\ b<2j+1}} \gamma_{b(2j+1)}  - \displaystyle{\sum_{\substack{((2j+2)b) \in {\cal V}_{\rm OPE}\\ b>2j+2, b\neq n-2K-1,\ldots, n-1}}}  \gamma_{(2j+2)b}) \right).
 }
\endgroup
Now, we would like to integrate over $\gamma_{2(n-2K-3)}, \gamma_{1(n-2K-2)}, \gamma_{3(n-2K-2)},\gamma_{4(n-2K-2)},\ldots,\gamma_{(n-2K-4)(n-2K-2)}$. 
This will be associated with integrating over the  magenta-colored chain in~\eno{OPEOrder} immediately to the left of the green-colored chain we previously integrated out.
This computation is described in appendix~\ref{APP:OPETHIRD} and the final result is presented in~\eno{IhatThree}.
In fact, it is straightforward to show that~\eno{IhatThree} can be written as~\eno{IhatDef} upon sending $K \to K+1$. That is, comparing~\eno{IhatThree} with~\eno{IhatDef}, we conclude,
\eqn{Induction}{
\widehat{I}_{n-2K-1} = \widehat{I}_{n-2K-1}\Big|_{K\to K+1} = \widehat{I}_{n-2K-3} \,,
}
where on the RHS, $M_{n-2K-3}$ and $L_{n-2K-3}$ are given by~\eno{MnMinus2KMinus1} and~\eno{LnMinus2KMinus1} respectively, with $K \to K+1$, and  the range of validity for $W_j$ in~\eno{Wj} now becomes $n-2K-4 \leq j \leq n-4$.
This establishes the inductive step, and together with the base case~\eno{BaseCase} furnishes the following chain of equalities:
\eqn{Ichain}{
I = \widehat{I}_{n-3} = \widehat{I}_{n-5} = \cdots = \widehat{I}_{n-2K-1} = \cdots = \widehat{I}_{3}\,.
}
As we move progressively to the right down the chain of equalities above, we account for evaluations of more and more contour integrals from the set ${\cal V}_{\rm OPE}$, until we are left with just one integral.
At the right-most equality at $K={n\over 2}-2$, the original contour integral $I$ (see~\eno{IOPE}) reduces to
\eqn{ILast}{
I = \widehat{I}_3 &= 
 \left(\prod_{(rs) \in {\cal V}_{\rm OPE}} \int  {d\widetilde{\gamma}_{rs} \over 2\pi i} \, \Gamma(-\widetilde{\gamma}_{rs}) (v_{rs}-1)^{\widetilde{\gamma}_{rs}} \right) 
\left( \prod_{j=2}^{n-3} W_j  \right)
  M_{3}
\int {d\gamma_{13} \over 2\pi i}    L_{3}\,,
}
where the $W_j$ for $2 \leq j \leq n-3$ are given in~\eno{WnMinus3Again}-\eno{Wj}, $M_3$ is given by~\eno{MnMinus2KMinus1} which at $K={n\over 2}-2$ simplifies to
\eqn{M3}{
 M_{3} &= \Gamma(\Delta_{12,\delta_1}-k_1) \,
\Gamma\left(\sum_{\substack{(4a) \in {\cal V}_{\rm OPE}\\ a>4}} \!\!\!\!\!\! \widetilde{\gamma}_{4a} +  \!\!\! \sum_{\substack{(3a) \in {\cal V}_{\rm OPE}\\ a>4}} \!\!\!\!\!\!\! \widetilde{\gamma}_{3a} + \Delta_{\delta_{3}\delta_{2},\delta_{1}}   + k_{32,1}     \right) ,
}
and $L_3$ is obtained by setting $K={n\over 2}-2$ in~\eno{LnMinus2KMinus1},
\eqn{L3}{
 L_{3} &=   \Gamma(\widetilde{\gamma}_{14} + \widetilde{\gamma}_{13} + \gamma_{13})   \,
 \Gamma\left(\sum_{j=5}^{n-1} \widetilde{\gamma}_{1j}  +  \Delta_{1\delta_1,2} + k_1 - \gamma_{13}\right)
\,
 \Gamma\left(\widetilde{\gamma}_{23} +  \Delta_{\delta_{1}\delta_{2},\delta_{3}} + k_{12,3} -  \gamma_{13} \right)
   \cr
&\times 
\Gamma\left(\sum_{i=0}^{{n\over 2}-3} \widetilde{\gamma}_{2(n-2i-1)}  +  \Delta_{2\delta_{3},1\delta_{2}} + k_{3,2} +  \gamma_{13}   \right) .
}
The contour integral in~\eno{ILast}, which corresponds to the lone black dot in~\eno{OPEOrder} at the left-most extreme, can be  evaluated using the first Barnes lemma~\eno{FirstBarnes}, to give
\eqn{IBlock}{
I &=
 \left(\prod_{(rs) \in {\cal V}_{\rm OPE}} \int  {d\widetilde{\gamma}_{rs} \over 2\pi i} \, \Gamma(-\widetilde{\gamma}_{rs}) (v_{rs}-1)^{\widetilde{\gamma}_{rs}} \right) 
\left( \prod_{j=2}^{n-3} W_j  \right)
  M_{3} \cr 
  &\quad \times 
{ \Gamma\left(  \sum_{j=3}^{n-1} \widetilde{\gamma}_{1j}  +  \Delta_{1\delta_1,2} + k_1 \right) 
\Gamma(\widetilde{\gamma}_{13} + \widetilde{\gamma}_{14} + \widetilde{\gamma}_{23} +  \Delta_{\delta_{1}\delta_{2},\delta_{3}} + k_{12,3})  \over 
\Gamma\left( \sum_{j=3}^{n-1} \widetilde{\gamma}_{1j} + \sum_{i=0}^{{n\over 2}-2} \widetilde{\gamma}_{2(n-2i-1)} +  \Delta_{\delta_1} + 2k_1  \right) }
\cr
&\quad \times 
\Gamma\left(  \sum_{j=5}^{n-1} \widetilde{\gamma}_{1j} + \sum_{i=0}^{{n\over 2}-3} \widetilde{\gamma}_{2(n-2i-1)}  +  \Delta_{\delta_1\delta_{3},\delta_{2}} + k_{13,2}    \right) 
\Gamma\left(\sum_{i=0}^{{n\over 2}-2} \widetilde{\gamma}_{2(n-2i-1)}  +  \Delta_{2\delta_{1},1} + k_{1} \right) \cr 
\Rightarrow I &=  \left(\prod_{(rs) \in {\cal V}_{\rm OPE}} \int  {d\widetilde{\gamma}_{rs} \over 2\pi i} \, \Gamma(-\widetilde{\gamma}_{rs}) (v_{rs}-1)^{\widetilde{\gamma}_{rs}} \right) 
\left( \prod_{j=1}^{n-3} W_j  \right)
 \cr 
 & \quad \times 
  \Gamma(\Delta_{12,\delta_1}-k_1)\, \Gamma\left(  \sum_{j=3}^{n-1} \widetilde{\gamma}_{1j}  +  \Delta_{1\delta_1,2} + k_1 \right) 
\Gamma\left(\sum_{i=0}^{{n\over 2}-2} \widetilde{\gamma}_{2(n-2i-1)}  +  \Delta_{2\delta_{1},1} + k_{1} \right),
}
where we identified a factor of $W_1$ above by comparing with~\eno{Wj} at $j=1$, thus extending the regime of validity of the $W_j$ coefficients in~\eno{WnMinus3Again}-\eno{Wj} to $1\leq j\leq n-3$.

The $\binom{n-2}{2}$-dimensional contour integral over the $\widetilde{\gamma}_{rs}$ variables is significantly less complicated to evaluate. 
Just like in the seven-point and $n$-point comb channel examples, we close the contour to the right, and using the fact that all Gamma functions in the integrand contain positive linear combinations of $\widetilde{\gamma}_{rs}$ variables in their arguments except for the factors of $\Gamma(-\widetilde{\gamma}_{rs})$, the only poles picked are the ones at origin and the semi-infinite sequence of poles at positive integral values of $\widetilde{\gamma}_{rs}$ for each $(rs) \in {\cal V}_{\rm OPE}$.
This introduces $\binom{n-2}{2}$ new sums over the Mellin parameters $j_{rs}$, and furnishes a series expansion representation of the conformal block where,
effectively, all positive linear combinations of the $\widetilde{\gamma}_{rs}$ Mellin variables in the Gamma functions get replaced with positive linear combinations of the corresponding Mellin parameters $j_{rs}$.

Let us mention some salient points of comparison between the blocks of this section and section~\ref{NOPE}.
As shown in the previous two subsections, one can re-express all series coefficients in terms of Pochhammer symbols, such that all explicit factors of Gamma functions cancel out.
The Lauricella functions in~\eno{MellinOPEVertex} and~\eno{OPEVertex} are identical, and the positive linear combinations of Mellin parameters appearing in the Pochhammer symbols find perfect agreement as well. 
For instance, the arguments of the Gamma functions (or equivalently the Pochhammer symbols) in the numerators of $W_j$ for even $j=2a-2$  match with those of the Pochhammer symbols in $V_a^{(1)}$ in~\eno{OPEVertex} for $2\leq a\leq n/2-1$, and those of $W_j$ for odd $j=2b-1$  match with those of $V_b^{(3)}$ in~\eno{OPEVertex} for $1\leq b \leq n/2-2$.
Moreover, the numerators of $W_{n-3}$ are identified with the Pochhammer symbols in $V_{n-2}^{(1)}$ in~\eno{OPEVertex}, while the triplet of factors in the final line of~\eno{IBlock} are matched with the vertex factor $V_1^{(1)}$ in~\eno{OPEVertex}.
This accounts for $(n-2)$ triplet of Pochhammer symbol combinations, one for each internal vertex of the binary graph.
There are, additionally, $(n-3)$ factors of Pochhammer symbols in the denominators of the $(n-3)$ $W_j$ coefficients, and each of these is in one-to-one correspondence with the denominators of the $(n-3)$ internal edge factors~\eno{EdgeDef}.

A careful comparison  between the block found using the Feynman rules for conformal blocks, and the one found using the Mellin formalism in this section confirms that there is full agreement between the $n$-point conformal block of this section and section~\ref{NOPE}, thus confirming the Feynman rules for $n$-point blocks in the OPE channel.

\section{Discussion}
\label{DISCUSS}

In this paper we proposed a simple set of rules for constructing any scalar conformal block with scalar exchanges, given the appropriate cross-ratios as input data. 
The rules are summarized in~\eno{CBexpansion}-\eno{VertexDef}, and the prescription for obtaining the post-Mellin parameters appearing in the Pochhammer symbols and summations is described in section~\ref{INTEGERS}.\footnote{In section~\ref{PROOF}, we also obtained a mixed series-integral representation~\eno{CBSerInt}.}
These rules help bypass lengthy, often impossibly hard computations needed to obtain conformal blocks.
They  are very similar and closely related to the Feynman rules for Mellin amplitudes, as  in both methods we assign  a factor for each edge and vertex  appearing in the unrooted binary graph representation of the conformal block or Witten diagram.
In fact, the same Lauricella functions appear in the Feynman rules for both conformal blocks and Mellin amplitudes.
This was  exhibited to be a consequence of the derivation of the Feynman rules in specific examples in section~\ref{PROOF}, where our starting point was the Mellin amplitude of certain canonical Witten diagrams.
One important difference was that the type of Lauricella function which appears in a Mellin amplitude depends on the degree of the interaction vertex appearing in the AdS diagram; one associates the Lauricella function $F_A^{(\ell)}$ for a bulk interaction vertex with $\ell$ incident internal edges. 
For conformal blocks $\ell \leq 3$ because of the OPE structure of blocks; a related fact is that we needed to consider Mellin amplitudes for canonical Witten diagrams in an effective $\phi^3$ scalar field theory in AdS.
The Lauricella functions were also expected from the point of view of previous work on the holographic duals of higher-point conformal blocks~\cite{Parikh:2019ygo,Jepsen:2019svc}, where the same functions appeared in the context of three-propagator identities which were used in the derivation of the geodesic diagram representation of blocks.

The Feynman rules provide an interesting, explicit and analytical representation for arbitrary conformal blocks, which may help investigate hidden mathematical structure and properties of conformal blocks, such as higher-point recursion relations dimensional relations, and possible closed-form expressions.
The symmetric Lauricella functions in the power series expansion of the blocks also facilitate the symmetry analysis of conformal blocks; for instance permutation symmetries become manifest when the block is expressed in terms of the Lauricella functions. 
It would be interesting to undertake a detailed symmetry analysis, along the lines of ref.~\cite{Fortin:2019dnq}, of arbitrary conformal blocks as prescribed by the proposed Feynman rules.
Recent work~\cite{Pal:2020dqf} on expressing higher-point functions in two and four spacetime dimensions in terms of generalizations of Lauricella systems in the configuration space of $n$ points, generalized to complex and quaternionic settings respectively, also provides an interesting mathematical connection and avenue of exploration.

The proposed Feynman rules were conjectured based on known results in the literature.
We applied the rules to obtain the $n$-point block in the OPE channel which was not previously known, and verified  it independently via the Mellin space formalism.
Likewise we worked out a previously unknown seven-point block, both via Feynman rules and via Mellin amplitudes, and obtained a precise equivalence. 
These checks serve as non-trivial evidence in support of the proposed rules.
The methodology in section~\ref{PROOF} of proving the Feynman rules in particular examples, is also expected to work in exactly the same manner for  {\it any particular} choice of conformal block beyond those considered in this paper. 
However, proving it for an {\it arbitrary} choice of a conformal block will presumably require more work. Nevertheless, it would be useful to prove these rules in generality for arbitrary blocks.

Finally, it should also be possible to generalize these rules to arbitrary-point conformal blocks for external and/or exchanged operators in arbitrary representations of the Lorentz group. This would be especially useful from the point of view of setting up an $n$-point conformal bootstrap for external scalars where internal exchanges can still involve spinning operators.
Weight-shifting operators~\cite{Karateev:2017jgd} and differential operators~\cite{Sleight:2017fpc,Costa:2018mcg} may be helpful in determining such generalizations. 
In fact, Mellin amplitudes for representations other than scalars (see e.g.\ refs.~\cite{Paulos:2011ie,Kharel:2013mka,Goncalves:2014rfa,Faller:2017hyt,Chen:2017xdz,Sleight:2018epi}) may also inform the discussion on generalization of the conformal block Feynman rules beyond scalars.
Turning the logic around, it would be interesting to investigate whether generalizations of the block Feynman rules to other representations benefit the study of higher-point spinning (Mellin) amplitudes.

\section*{Acknowledgements}
The work of S.~H.\ was partially supported by the SCS Summer Research Grant, by the Pomona RAISE Grant, and by Caltech's Visiting Undergraduate Research Program (VURP).

\appendix 

\section{Lauricella functions}
\label{LAURICELLA}

The Lauricella function $F_A$ of $\ell$ variables is a generalized hypergeometric sum of $\ell$ variables~\cite{Lauricella1893,Srivastava1985book,AartsMathWorld} (see also ref.~\cite{Paulos:2011ie}) defined as
    \eqn{LauricellaDef}{
    F_A^{(\ell)}\Big[g;\{a_1,\ldots,a_\ell\};\{b_1,\ldots,b_\ell\};x_1,\ldots,x_\ell\Big]
    := \left[\prod_{i=1}^\ell\sum_{n_i=0}^\infty \right](g)_{\sum_{i=1}^\ell n_i}\prod_{i=1}^\ell \frac{(a_i)_{n_i}}{(b_i)_{n_i}}\frac{x_i^{n_i}}{n_i!}\,.
    }
    
One can always perform one of the sums in the~\eno{LauricellaDef} to re-express $F_A^{(\ell)}$ in terms of functions involving $\ell-1$ summations. For example, we present some identities for $\ell \leq 3$: 
\eqn{FA1}{ 
F_A^{(1)}\left[\Delta_{abc,}- h; \{ -k_c \}; \left\{ \Delta_c -h+1 \right\}; 1 \right] 
= {(1-\Delta_{ab,c})_{k_c} \over   (\Delta_c - h +1)_{k_c}} \,,
}
\eqn{FA2}{
 & F_A^{(2)}\left[\Delta_{abc,}- h; \{ -k_b, -k_c \}; \left\{ \Delta_b -h+1,\Delta_c -h+1 \right\}; 1,1 \right] \cr 
 & ={  (1-\Delta_{ac,b})_{k_b}  (1-\Delta_{ab,c})_{k_c} \over   (\Delta_b - h +1)_{k_b} (\Delta_c - h +1)_{k_c} } 
 {}_3F_2\left[ \{ -k_b, -k_c, \Delta_{abc,}-h\}; \{\Delta_{ac,b}-k_b, \Delta_{ab,c}-k_c \}; 1 \right] ,
 }
 and
\eqn{FA3}{
 F_A^{(3)} [\Delta_{abc,}- h; \{-k_a, -k_b, -k_c \}; & \left\{\Delta_a -h+1, \Delta_b -h+1,\Delta_c -h+1 \right\}; 1,1,1 ] \cr 
&= {(-1)^{k_a+k_b+k_c} \over (\Delta_a-h+1)_{k_a} (\Delta_b - h +1)_{k_b} (\Delta_c - h +1)_{k_c}}  \cr 
&  \times F^{1,3,2}_{2,1,0}\left[ \begin{array}{c} \Delta_{abc,}- h\,;  -k_b\,, \Delta_{ab,c}\,, \Delta_{bc,a}\,;  -k_a\,, -k_c \\ \Delta_{ab,c}-k_c\,,\Delta_{bc,a}-k_a\,; \Delta_b-h+1\,;- 
\end{array} \Bigg| 1,1 \right]  ,
 }
where $F_{q,s,v}^{p,r,u}$ is the Kamp\'{e} de F\'{e}riet function~\cite{exton1976multiple,Srivastava1985book} (see also ref.~\cite{Fortin:2020yjz}), defined by the following hypergeometric series,
\eqn{KdF}{
F_{q,s,v}^{p,r,u}\left[\left.\begin{array}{c}\boldsymbol{a};\boldsymbol{c};\boldsymbol{f}\\\boldsymbol{b};\boldsymbol{d};\boldsymbol{g}\end{array}\right|x,y\right] := \sum_{m,n = 0}^\infty {(\boldsymbol{a})_{m+n}(\boldsymbol{c})_m(\boldsymbol{f})_n \over (\boldsymbol{b})_{m+n}(\boldsymbol{d})_m(\boldsymbol{g})_n} {x^my^n \over m!n!} \,,
}
where
\eqn{}{
\begin{gathered}
(\boldsymbol{a})_{m+n} := (a_1)_{m+n}\cdots(a_p)_{m+n} \qquad(\boldsymbol{b})_{m+n} := (b_1)_{m+n}\cdots(b_q)_{m+n} \cr
(\boldsymbol{c})_m := (c_1)_m\cdots(c_r)_m \qquad(\boldsymbol{d})_m := (d_1)_m\cdots(d_s)_m \cr
(\boldsymbol{f})_n := (f_1)_n\cdots(f_u)_n \qquad(\boldsymbol{g})_n := (g_1)_n\cdots(g_v)_n \,.
\end{gathered}
}

\section{Technical details}
\label{TECHNICAL}

\subsection{Proof of~\eno{ellOddGuess}}
\label{APP:ELLODDPROOF}

In this appendix we will compute $\ell_{\delta_{2i+1}}$ which satisfies the recursion relation
\eqn{}{
\ell_{\delta_{2i+3}}\stackrel{2{\cal J}}{=} \ell_{\delta_{2i+1}}+\ell_{\delta_{2i+2}}  \qquad(0 \leq i \leq {n \over 2}-4 )\,,
}
where the even-indexed $\ell_{\delta_{2i+2}}$ and the smallest odd-indexed $\ell_{\delta_1}$ are  known from~\eno{opeIntpM1}.

We will now prove by induction that $\ell_{\delta_{2i+1}}$ takes the form
\eqn{oddEllConj}{
\ell_{\delta_{2i+1}} 
 &= \sum_{s=2i+3}^{n-1}j_{1s}+\sum_{r=3}^{2i+2}\sum_{s=2i+3}^{n-1}j_{rs}+\sum_{z=i+1}^{{n\over 2}-1} j_{2(2z+1)}\,,
} 
for $0 \leq i \leq n/2-3$.

Let's first establish the base case. For $i=1$,
\eqn{}{
\ell_{\delta_3} 
& \stackrel{2{\cal J}}{=} \ell_{\delta_1}+\ell_{\delta_2} \cr 
&= 
 \sum_{s=5 }^{n-1} j_{1s} + \sum_{r=3}^4 \sum_{s=5}^{n-1} j_{rs} 
 +\sum_{i=5,7,9,\ldots}^{n-1} j_{2i}
}
agrees with~\eno{oddEllConj}.

For the inductive step, assume~\eno{oddEllConj} is true for $i=K$, where $0 \leq K < n/2-3$.
We will now show that~\eno{oddEllConj} holds for $i=K+1$.
That is, assuming the following:
\eqn{oddEllAss}{
\ell_{\delta_{2K+1}}
 &= \sum_{s=2K+3}^{n-1}j_{1s}+\sum_{r=3}^{2K+2}\sum_{s=2K+3}^{n-1}j_{rs}+\sum_{z=K+1}^{{n\over 2}-1} j_{2(2z+1)}
} 
we can now compute
\eqn{oddEllComp}{
\ell_{\delta_{2K+3}}\stackrel{2{\cal J}}{=} \ell_{\delta_{2K+1}}+\ell_{\delta_{2K+2}}  \,,
}
where we can use
\eqn{}{
\ell_{\delta_{2K+2}} = \sum_{s=2K+5}^{n-1} j_{(2K+3)s}+\sum_{r=1}^{2K+2} j_{r(2K+3)} + j_{1(2K+4)}+ \sum_{s=2K+5}^{n-1}j_{(2K+4)s} + \sum_{r=3}^{2K+2} j_{r(2K+4)} 
}
and~\eno{oddEllAss} to compute the LHS of~\eno{oddEllComp}:
\eqn{}{
\ell_{\delta_{2K+3}} &\stackrel{2{\cal J}}{=}
\sum_{s=2K+5}^{n-1} j_{(2K+3)s}+\sum_{r=1}^{2K+2} j_{r(2K+3)} + j_{1(2K+4)}+ \sum_{s=2K+5}^{n-1}j_{(2K+4)s} + \sum_{r=3}^{2K+2}
j_{r(2K+4)}
\cr 
& \quad +\sum_{s=2K+3}^{n-1}j_{1s}+\sum_{r=3}^{2K+2}\sum_{s=2K+3}^{n-1}j_{rs}+\sum_{z=K+1}^{{n\over 2}-1} j_{2(2z+1)}
\cr 
&= \sum_{s=2K+3}^{n-1}j_{1s}+\sum_{r=3}^{2K+2}\sum_{s=2K+3}^{n-1}j_{rs}+\sum_{z=K+1}^{{n\over 2}-1} j_{2(2z+1)}-\sum_{2K+3}^{2K+4}j_{1s}+\sum_{2K+3}^{2K+4}\sum_{2K+5}^{n-1}j_{rs}
\cr 
&\quad -\sum_{r=3}^{2K+2}\sum_{s=2K+3}^{2K+4}j_{rs}-j_{2(2K+3)}
\cr 
&=  \sum_{s=2K+5}^{n-1}j_{1s}+\sum_{r=3}^{2K+4}\sum_{s=2K+5}^{n-1}j_{rs}+\sum_{z=K+2}^{{n\over 2}-1} j_{2(2z+1)}\,,
}
which is the same as~\eno{oddEllConj} for $i=K+1$, as needed.

\subsection{Proof of~\eno{CombInduction}} 
 \label{APP:COMBPROOF}

We notice that the LHS of~\eno{CombInduction} is of the form:
 \eqn{LHStoProve}{
  K_m \left( \prod_{j=m+1}^{n-1} \int {d s_{j} \over 2 \pi i} \, \Gamma(A_j + s_j) \Gamma(B_j-s_j) \right) \Gamma\left(C+ \sum_{j=m+1}^{n-1} s_j\right) \Gamma\left(D-\sum_{j=m+1}^{n-1} s_j\right) ,
}
where the integration variables are $s_i = \gamma_{mi}$, and the coefficients are
\eqn{}{ 
A_i &= \widetilde{\gamma}_{mi} \cr 
B_i &= \sum_{j=2}^{m-1}  \widetilde{\gamma}_{ji} + \Delta_{i\delta_{i-2},\delta_{i-1}} + k_{i-2,i-1} - \sum_{j=m+1}^{i-1} \gamma_{ji} \cr 
C &= \Delta_{n\delta_{m-2}, m(m+1)\ldots (n-1)}  + k_{m-2} +\sum_{\substack{m+1 \leq a < b \leq n-1 }} \gamma_{ab}  \cr 
D &= \Delta_{m \delta_{m-1},\delta_{m-2}}+k_{(m-1),(m-2)}\;,
}
with the overall factor
\eqn{nonintegrableconst}{
K_m &= \left( \prod_{m+1\leq r \leq s \leq n-1}
\Gamma\left( \widetilde{\gamma}_{rs}+\gamma_{rs} \right) \right)
\Gamma\left(\sum_{i=2}^{m-1} \widetilde{\gamma}_{im} + \Delta_{m \delta_{m-2},\delta_{m-1}}+k_{m-2,m-1}
\right)
\cr 
& \quad \times
\left( \prod_{i=m+1}^{n-1} \Gamma \left( \Delta_{i \delta_{i-1},\delta_{i-2}} + k_{i-1,i-2} - \sum_{j=i+1}^{n-1} \gamma_{ij} \right) \right).
}
Using the inductive Mellin Barnes lemma~\eno{MultiFirstBarnes}, we can easily evaluate~\eno{LHStoProve} to obtain 
 \eqn{LHSAgain}{
 & K_m \left(\prod_{i=m+1}^{n-1} \Gamma\left(\widetilde{\gamma}_{mi} + \sum_{j=2}^{m-1}  \widetilde{\gamma}_{ji} + \Delta_{i\delta_{i-2},\delta_{i-1}} + k_{i-2,i-1} - \sum_{j=m+1}^{i-1} \gamma_{ji} \right) \right)
\cr 
&\times 
\Gamma\left(\Delta_{n\delta_{m-1}, (m+1)\ldots (n-1)}  + k_{m-1} +\sum_{\substack{m+1 \leq a < b \leq n-1 }} \gamma_{ab} \right) 
\cr &\times 
\Gamma\left(\sum_{i=m+1}^{n-1} \widetilde{\gamma}_{mi} + \Delta_{m \delta_{m-1},\delta_{m-2}}+k_{(m-1),(m-2)}\right)
\cr 
&\times 
{\Gamma\left(\sum_{i=m+1}^{n-1} \sum_{j=2}^{m-1}  \widetilde{\gamma}_{ji} + \Delta_{\delta_{m-1}\delta_{m-2},m} + k_{m-2(m-1)} \right)
\over  \Gamma\left(\sum_{i=m+1}^{n-1} \sum_{j=2}^{m}  \widetilde{\gamma}_{ji}
+\Delta_{\delta_{m-1}}+2k_{m-1}\right) } .
} 
Substituting in the explicit form for $K_m$ and using the definitions~\eno{JmDef} and~\eno{lambdaDef}, we recognize~\eno{LHSAgain} to be precisely the RHS of~\eno{CombInduction}.

\subsection{OPE channel calculations}
\label{APP:OPE}

\subsubsection{Base case}
\label{APP:OPEFIRST}

Our starting point is the contour integral~\eno{Ifinal}, which takes the explicit form
\begingroup\makeatletter\def\f@size{10.5}\check@mathfonts
\eqn{IOPE}{
I &=   \Gamma(\Delta_{12,\delta_1}-k_1) 
\Gamma(\Delta_{(n-1)n,\delta_{n-3}}-k_{n-3}) 
 \left(\prod_{j=1}^{{n\over 2}-2} \Gamma(\Delta_{(2j+1)(2j+2),\delta_{2j}}-k_{2j}) \right) 
\cr 
&\times
\left(\prod_{(rs) \in {\cal V}_{\rm OPE}} \int  {d\widetilde{\gamma}_{rs} \over 2\pi i} \, \Gamma(-\widetilde{\gamma}_{rs}) (v_{rs}-1)^{\widetilde{\gamma}_{rs}} \right) 
\left(\prod_{(rs) \in {\cal V}_{\rm OPE}} \int {d\gamma_{rs} \over 2\pi i} \,  \Gamma(\widetilde{\gamma}_{rs} + \gamma_{rs}) \right)
 \cr 
 &\times 
\left(\prod_{j=1}^{{n\over 2}-2} \Gamma\left( \Delta_{\delta_{2j-1}\delta_{2j},\delta_{2j+1}} + k_{(2j-1)(2j),(2j+1)} - \sum_{\substack{(ab) \in {\cal V}_{\rm OPE} \\ a<b, b=2j+1 {\rm \ or\ }2j+2}} \gamma_{ab}\right) \right)
\cr 
&\times 
\Gamma\left( \Delta_{1\delta_1,2} + k_1 -\sum_{j=3}^{n-1} \gamma_{1j} \right) 
\Gamma\left(\Delta_{(n-1) \delta_{n-3},n} + k_{n-3} -\sum_{\substack{j=1}}^{n-3} \gamma_{j(n-1)} \right)
\cr 
&\times 
\left(\prod_{j=1}^{{n\over 2}-2} \Gamma\left(\Delta_{(2j+1)\delta_{2j},(2j+2)} + k_{2j}   -\sum_{((2j+1)b) \in {\cal V}_{\rm OPE}} \gamma_{(2j+1)b} \right)  \right)
  \cr 
 & \times 
{  \left(\prod_{j=1}^{{n\over 2}-3} \Gamma\left( \Delta_{(2j+2)\delta_{2j+1},(2j+1)\delta_{2j-1}} + k_{2j+1,2j-1}  + \sum_{\substack{(a(2j+1)) \in {\cal V}_{\rm OPE} \\ a<2j+1}} \gamma_{a(2j+1)}  -\sum_{\substack{((2j+2)b) \in {\cal V}_{\rm OPE}\\ b> 2j+2}} \gamma_{(2j+2)b} \right) \right)
}
\cr 
&\times
\Gamma\left( \Delta_{2\delta_{n-3},1\delta_2\delta_4\ldots \delta_{n-4}} + k_{(n-3),24\ldots(n-4)} - \gamma_{2(n-1)} + \sum_{j=3}^{n-2} \gamma_{1j} + \sum_{\substack{3\leq i<j\leq n-2\\ (ij) \in {\cal V}_{\rm OPE}}} \gamma_{ij} \right)
\cr 
&\times 
\Gamma\left( \Delta_{(n-2)n, (n-3)(n-1)\delta_{n-5}} - k_{n-5}  + \sum_{j=1}^{n-3} \gamma_{j(n-1)} + \sum_{j=1}^{n-4} \gamma_{j(n-3)}   \right).
}
\endgroup
Two straightforward applications of the first Barnes lemma~\eno{FirstBarnes} allow us to perform the $\gamma_{1(n-1)}$ and $\gamma_{2(n-1)}$ integrals. 
The result is
\begingroup\makeatletter\def\f@size{11}\check@mathfonts
\eqn{}{
I &=   \Gamma(\Delta_{12,\delta_1}-k_1) 
\Gamma(\Delta_{(n-1)n,\delta_{n-3}}-k_{n-3}) 
 \left(\prod_{j=1}^{{n\over 2}-2} \Gamma(\Delta_{(2j+1)(2j+2),\delta_{2j}}-k_{2j}) \right) 
\cr 
&\times
\left(\prod_{(rs) \in {\cal V}_{\rm OPE}} \int  {d\widetilde{\gamma}_{rs} \over 2\pi i} \, \Gamma(-\widetilde{\gamma}_{rs}) (v_{rs}-1)^{\widetilde{\gamma}_{rs}} \right) 
\left(\prod_{\substack{(rs) \in {\cal V}_{\rm OPE}\\s \neq n-1}} \int {d\gamma_{rs} \over 2\pi i} \,  \Gamma(\widetilde{\gamma}_{rs} + \gamma_{rs}) \right)
 \cr 
 &\times 
 \left[\left( \prod_{r=3}^{n-3} \int {d \gamma_{r(n-1)} \over 2\pi i} \right) \widetilde{L}_{n-1} \!\right] \!
\left(\!\prod_{j=1}^{{n\over 2}-2} \Gamma\!\left(\! \Delta_{\delta_{2j-1}\delta_{2j},\delta_{2j+1}} + k_{(2j-1)(2j),(2j+1)} - \!\!\!\!\! \sum_{\substack{(ab) \in {\cal V}_{\rm OPE} \\ a<b, b=2j+1 {\rm \ or\ }2j+2}} \!\!\!\!\!\!\!\! \gamma_{ab}\right) \right)
\cr 
&\times 
\Gamma\left(\widetilde{\gamma}_{1(n-1)} + \Delta_{1\delta_1,2} + k_1 -\sum_{j=3}^{n-2} \gamma_{1j} \right)
\Gamma\left( \Delta_{(n-2)\delta_{n-3}, (n-3)\delta_{n-5}} + k_{n-3,n-5}  + \sum_{j=1}^{n-4} \gamma_{j(n-3)}    \right)
\cr 
&\times 
\Gamma\left(\widetilde{\gamma}_{2(n-1)} +  \Delta_{2\delta_{n-3},1\delta_2\delta_4\ldots \delta_{n-4}} + k_{(n-3),24\ldots(n-4)} + \sum_{j=3}^{n-2} \gamma_{1j} + \sum_{\substack{3\leq i<j\leq n-2\\ (ij) \in {\cal V}_{\rm OPE}}} \gamma_{ij} \right) 
\cr 
&\times 
\Gamma\left( \sum_{i=1}^2 \widetilde{\gamma}_{i(n-1)}  +  \Delta_{(n-2)\delta_1, (n-3)\delta_{n-5}\delta_2\delta_4\ldots \delta_{n-4}} + k_{1,(n-5)24\ldots(n-4)} + \Delta_{\delta_{n-3}} + 2k_{n-3} \right. 
\cr 
&\qquad \left. + \sum_{j=1}^{n-4} \gamma_{j(n-3)}   + \! \sum_{\substack{3\leq i<j\leq n-2\\ (ij) \in {\cal V}_{\rm OPE}}} \gamma_{ij}  \right)^{-1} ,
}
\endgroup
where the integrand for the integrals over $\gamma_{3(n-1)},\ldots,\gamma_{(n-3)(n-1)}$ above has been collected into the following object, 
\begingroup\makeatletter\def\f@size{11}\check@mathfonts
\eqn{}{
& \widetilde{L}_{n-1} \cr 
&:= \left( \prod_{r=3}^{n-3}  \Gamma(\widetilde{\gamma}_{r(n-1)} + \gamma_{r(n-1)}) \right) 
\left(\prod_{j=1}^{{n\over 2}-2} \Gamma\left(\Delta_{(2j+1)\delta_{2j},(2j+2)} + k_{2j}   -\sum_{((2j+1)b) \in {\cal V}_{\rm OPE}} \gamma_{(2j+1)b} \right)  \right)
  \cr 
 & \times 
 \left(\prod_{j=1}^{{n\over 2}-3} \Gamma\left( \Delta_{(2j+2)\delta_{2j+1},(2j+1)\delta_{2j-1}} + k_{2j+1,2j-1}  + \sum_{\substack{(a(2j+1)) \in {\cal V}_{\rm OPE} \\ a<2j+1}} \gamma_{a(2j+1)}  -\sum_{\substack{((2j+2)b) \in {\cal V}_{\rm OPE}\\ b> 2j+2}}\!\!\! \gamma_{(2j+2)b} \!\right) \!\right)
\cr 
&\times 
\Gamma\left( \Delta_{(n-2)n\delta_1 \delta_{n-3}, (n-3)(n-1)\delta_{n-5}\delta_2\delta_4\ldots \delta_{n-4}} + k_{1(n-3),(n-5)24\ldots(n-4)} + \sum_{j=1}^{n-4} \gamma_{j(n-3)}   + \sum_{\substack{3\leq i<j\leq n-2\\ (ij) \in {\cal V}_{\rm OPE}}} \gamma_{ij} \right. \cr 
& \qquad \left. + \sum_{j=3}^{n-3} \gamma_{j(n-1)} \right)
\Gamma\left(\sum_{i=1}^2\widetilde{\gamma}_{i(n-1)}  + \Delta_{(n-1) \delta_{n-3},n} + k_{n-3} -\sum_{\substack{j=3}}^{n-3} \gamma_{j(n-1)}\right).
}
\endgroup
To evaluate the integrals over $\widetilde{L}_{n-1}$, we can use the inductive first Barnes lemma~\eno{MultiFirstBarnes} 
with the identifications
\eqn{}{
A_r &= \widetilde{\gamma}_{(r+2)(n-1)} \cr 
B_r &= \begin{cases}
       \Delta_{(r+2)\delta_{r+1},(r+3)} + k_{r+1}   -\displaystyle{\sum_{\substack{((r+2)b) \in {\cal V}_{\rm OPE}\\b\neq n-1}}} \gamma_{(r+2)b}  & (r {\rm \ odd}) \cr 
        \Delta_{(r+2)\delta_{r+1},(r+1)\delta_{r-1}} + k_{r+1,r-1}  + \displaystyle{\sum_{\substack{(a(r+1)) \in {\cal V}_{\rm OPE} \\ a<r+1}}} \gamma_{a(r+1)}  - \sum_{\substack{((r+2)b) \in {\cal V}_{\rm OPE}\\ b> r+2, b\neq n-1}} \gamma_{(r+2)b}  & (r {\rm \ even}) 
        \end{cases} \cr 
C &=  \Delta_{(n-2)n\delta_1 \delta_{n-3}, (n-3)(n-1)\delta_{n-5}\delta_2\delta_4\ldots \delta_{n-4}} + k_{1(n-3),(n-5)24\ldots(n-4)}  + \sum_{j=1}^{n-4} \gamma_{j(n-3)}   + \sum_{\substack{3\leq i<j\leq n-2\\ (ij) \in {\cal V}_{\rm OPE}}} \gamma_{ij} \cr 
D &= \sum_{i=1}^2\widetilde{\gamma}_{i(n-1)}  + \Delta_{(n-1) \delta_{n-3},n} + k_{n-3}
}
for $ 1 \leq r \leq n-5$.
Then  the contour integrals evaluate to
\begingroup\makeatletter\def\f@size{9}\check@mathfonts
\eqn{}{
 &\left( \prod_{r=3}^{n-3} \int {d \gamma_{r(n-1)} \over 2\pi i} \right)  \widetilde{L}_{n-1} \cr 
 &= { \Gamma\!\left(\!\Delta_{(n-2)\delta_1, (n-3)\delta_{n-5}\delta_2\delta_4\ldots \delta_{n-4}} + k_{1,(n-5)24\ldots(n-4)}  + \displaystyle{\sum_{j=1}^{n-4}} \gamma_{j(n-3)}   + \sum_{\substack{3\leq i<j\leq n-2\\ (ij) \in {\cal V}_{\rm OPE}}}\!\!\!\! \gamma_{ij}  + \sum_{i=1}^2\widetilde{\gamma}_{i(n-1)}  + \Delta_{ \delta_{n-3}} + 2 k_{n-3} \right)  
  \over
  \Gamma\left(\displaystyle{\sum_{r=1}^{n-5}} \widetilde{\gamma}_{(r+2)(n-1)}      + \sum_{i=1}^2\widetilde{\gamma}_{i(n-1)}  + \Delta_{\delta_{n-3}} + 2k_{n-3} \right) }  \cr 
 &\times  
 \left(\prod_{r=1,3,\ldots}^{n-5} \Gamma(\widetilde{\gamma}_{(r+2)(n-1)} + \Delta_{(r+2)\delta_{r+1},(r+3)} + k_{r+1}   -\displaystyle{\sum_{\substack{((r+2)b) \in {\cal V}_{\rm OPE}\\b\neq n-1}}} \gamma_{(r+2)b}) \right) \cr 
 &\times 
 \left(\prod_{r=2,4,\ldots}^{n-6} \Gamma(\widetilde{\gamma}_{(r+2)(n-1)} +  \Delta_{(r+2)\delta_{r+1},(r+1)\delta_{r-1}} + k_{r+1,r-1}  + \displaystyle{\sum_{\substack{(a(r+1)) \in {\cal V}_{\rm OPE} \\ a<r+1}}} \gamma_{a(r+1)}  - \sum_{\substack{((r+2)b) \in {\cal V}_{\rm OPE}\\ b> r+2, b\neq n-1}} \gamma_{(r+2)b} \right) \cr 
 &\times 
 \Gamma\left(  \Delta_{n \delta_{n-3}, (n-1)} + k_{(n-3)}  \right)
  \Gamma\left(\sum_{r=1}^{n-5} \widetilde{\gamma}_{(r+2)(n-1)} + \sum_{i=1}^2\widetilde{\gamma}_{i(n-1)}  + \Delta_{(n-1) \delta_{n-3},n} + k_{n-3}\right)
 }
\endgroup
where we used
\begingroup\makeatletter\def\f@size{11}\check@mathfonts
\eqn{}{
\sum_{r=1}^{n-5} B_r &= \Delta_{(n-3)\delta_2\delta_4\delta_6\ldots\delta_{n-6}\delta_{n-4}\delta_{n-5},(n-2)\delta_1} + k_{246\ldots(n-6)(n-4)(n-5),1}   - \sum_{\substack{3\leq i<j\leq n-2\\ (ij) \in {\cal V}_{\rm OPE}}}\!\!\!\! \gamma_{ij}  - \sum_{j=1}^{n-4} \gamma_{j(n-3)}\,.
 }
\endgroup
Putting this result back in $I$, we obtain
\begingroup\makeatletter\def\f@size{11}\check@mathfonts
\eqn{}{
I &=  \Gamma(\Delta_{12,\delta_1}-k_1) 
 \left(\prod_{j=1}^{{n\over 2}-2} \Gamma(\Delta_{(2j+1)(2j+2),\delta_{2j}}-k_{2j}) \right) 
\left(\prod_{(rs) \in {\cal V}_{\rm OPE}} \int  {d\widetilde{\gamma}_{rs} \over 2\pi i} \, \Gamma(-\widetilde{\gamma}_{rs}) (v_{rs}-1)^{\widetilde{\gamma}_{rs}} \right) 
\cr 
&\times
\left(\prod_{\substack{(rs) \in {\cal V}_{\rm OPE}\\s\neq n-1, n-2,n-3}} \int {d\gamma_{rs} \over 2\pi i} \,  \Gamma(\widetilde{\gamma}_{rs} + \gamma_{rs}) \right)
\left( \prod_{\substack{(rs) \in {\cal V}_{\rm OPE} \\ s=n-3, r<s, r \neq 2}} \int {d\gamma_{rs} \over 2\pi i} \right)
W_{n-3}
 \cr 
 &\times
\left(\prod_{j=1}^{{n\over 2}-3} \Gamma\left( \Delta_{\delta_{2j-1}\delta_{2j},\delta_{2j+1}} + k_{(2j-1)(2j),(2j+1)} - \sum_{\substack{(ab) \in {\cal V}_{\rm OPE} \\ a<b, b=2j+1 {\rm \ or\ }2j+2}} \gamma_{ab}\right) \right) \cr 
&\times 
\left( \prod_{\substack{(rs)\in {\cal V}_{\rm OPE} \\ s=n-2 {\rm \ or\ } (rs)=(2(n-3))}} \int {d\gamma_{rs} \over 2\pi i} \right) L_{n-2}\,,
}
\endgroup
where we have defined
 \begingroup\makeatletter\def\f@size{10}\check@mathfonts
\eqn{WnMinus3}{
& W_{n-3} \cr 
&:=   {\Gamma\left(\Delta_{(n-1)n,\delta_{n-3}}-k_{n-3}\right)  
   \Gamma\left(\Delta_{n  \delta_{n-3},  (n-1) } + k_{n-3}  
 \right)  
 \Gamma\left( \displaystyle{\sum_{(i(n-1)) \in {\cal V}_{\rm OPE}}} \widetilde{\gamma}_{i(n-1)}  + \Delta_{(n-1) \delta_{n-3},n} + k_{n-3}\right) 
\over 
 \Gamma\left( \displaystyle{\sum_{(i(n-1)) \in {\cal V}_{\rm OPE}}} \widetilde{\gamma}_{i(n-1)}    + \Delta_{ \delta_{n-3}} + 2k_{n-3}  
 \right) },
}
\endgroup
and
\begingroup\makeatletter\def\f@size{11}\check@mathfonts
\eqn{}{
& L_{n-2} \cr 
&:=   \left( \prod_{\substack{(rs) \in {\cal V}_{OPE}\\ r<s,s=n-2, n-3} } \Gamma(\widetilde{\gamma}_{rs} + \gamma_{rs}) \right) 
 \cr 
&\times 
 \Gamma\left(\widetilde{\gamma}_{1(n-1)} + \Delta_{1\delta_1,2} + k_1 -\sum_{j=3}^{n-2} \gamma_{1j}\right)
 \Gamma\left( \Delta_{\delta_{n-5}\delta_{n-4},\delta_{n-3}} + k_{(n-5)(n-4),(n-3)} - \!\!\!\!\! \sum_{\substack{(ab) \in {\cal V}_{\rm OPE} \\ a<b, b=n-3 {\rm \ or\ }n-2}}\!\!\!\! \gamma_{ab}\right) 
\cr 
&\times 
 \left( \prod_{j=1}^{{n\over 2}-2}  \Gamma(\widetilde{\gamma}_{(2j+1)(n-1)} + \Delta_{(2j+1)\delta_{2j},(2j+2)} + k_{2j}   - \sum_{\substack{((2j+1)b) \in {\cal V}_{\rm OPE} \\ b\neq n-1}} \gamma_{(2j+1)b}) \right)
 \cr 
 &\times 
 \left(\prod_{j=1}^{{n\over 2}-3} \Gamma\left( \widetilde{\gamma}_{(2j+2)(n-1)} + \Delta_{(2j+2)\delta_{2j+1},(2j+1)\delta_{2j-1}} + k_{2j+1,2j-1}  +\!\!\! \sum_{\substack{(a(2j+1)) \in {\cal V}_{\rm OPE} \\ a<2j+1}} \gamma_{a(2j+1)}   \right. \right. \cr 
 & \left. \left. - \!\!\! \sum_{\substack{((2j+2)b) \in {\cal V}_{\rm OPE}\\ b> 2j+2, b \neq n-1}} \!\!\!\!\gamma_{(2j+2)b} \right) \right)
%
\Gamma\!\left(\widetilde{\gamma}_{2(n-1)} +  \Delta_{2\delta_{n-3},1\delta_2\delta_4\ldots \delta_{n-4}} + k_{(n-3),24\ldots(n-4)} + \sum_{j=3}^{n-2} \gamma_{1j} + \sum_{\substack{3\leq i<j\leq n-2\\ (ij) \in {\cal V}_{\rm OPE}}} \!\!\! \gamma_{ij} \right)
\cr 
&\times 
\Gamma\left( \Delta_{(n-2)\delta_{n-3}, (n-3)\delta_{n-5}} + k_{n-3,n-5}  + \sum_{j=1}^{n-4} \gamma_{j(n-3)}    \right).
}
\endgroup
Now, we will integrate over $\gamma_{2(n-3)}, \gamma_{1(n-2)}, \gamma_{3(n-2)}, \gamma_{4(n-2)}, \ldots, \gamma_{(n-4)(n-2)}$. The dependence on all these variables has been packaged into $L_{n-2}$ above.
We first integrate over $\gamma_{2(n-3)}$ using the first Barnes lemma~\eno{FirstBarnes}, and then do a change of variables $\gamma_{i(n-3)} \to \gamma_{i(n-3)} - \gamma_{i(n-2)}$, for $i=1,3,4,\ldots,n-4$.
Doing this, we obtain,
\begingroup\makeatletter\def\f@size{11}\check@mathfonts
\eqn{}{
& \int {d\gamma_{2(n-3)} \over 2\pi i} \, L_{n-2} \cr 
& =
 \Gamma\left(\widetilde{\gamma}_{1(n-1)} + \Delta_{1\delta_1,2} + k_1 -\sum_{j=3}^{n-3} \gamma_{1j}\right)
 \left( \prod_{j=1}^{{n\over 2}-3}  \Gamma(\widetilde{\gamma}_{(2j+1)(n-1)} + \Delta_{(2j+1)\delta_{2j},(2j+2)} + k_{2j}  \right. \cr 
 &\quad \left. - \!\!\!\! \sum_{\substack{((2j+1)b) \in {\cal V}_{\rm OPE} \\ b\neq n-1,n-2}} \!\!\!\!\! \gamma_{(2j+1)b})   \right) 
 \!
 \Gamma\!\left(\!\widetilde{\gamma}_{2(n-1)} +  \Delta_{2\delta_{n-3},1\delta_2\delta_4\ldots \delta_{n-4}} + k_{(n-3),24\ldots(n-4)} + \sum_{j=3}^{n-3} \gamma_{1j} +\!\! \sum_{\substack{3\leq i<j\leq n-3\\ (ij) \in {\cal V}_{\rm OPE}}} \!\!\!\! \gamma_{ij} \right)
 \cr 
 &\times 
 \left(\prod_{j=1}^{{n\over 2}-3} \Gamma\left( \widetilde{\gamma}_{(2j+2)(n-1)} + \Delta_{(2j+2)\delta_{2j+1},(2j+1)\delta_{2j-1}} + k_{2j+1,2j-1}  +\!\!\! \sum_{\substack{(a(2j+1)) \in {\cal V}_{\rm OPE} \\ a<2j+1}} \gamma_{a(2j+1)}  \right. \right. \cr 
 &\quad \left. \left. -\!\!\!\!\sum_{\substack{((2j+2)b) \in {\cal V}_{\rm OPE}\\ b> 2j+2, b \neq n-1,n-2}} \gamma_{(2j+2)b} \right) \right)
 \Gamma\left( \widetilde{\gamma}_{(n-3)(n-1)} + \Delta_{\delta_{n-3}\delta_{n-4},\delta_{n-5}} + k_{(n-3)(n-4),(n-5)}   \right) 
\cr 
&\times 
{   \Gamma(\widetilde{\gamma}_{2(n-3)} + \Delta_{\delta_{n-5}\delta_{n-4},\delta_{n-3}} + k_{(n-5)(n-4),(n-3)} - \sum_{\substack{a=1,a\neq 2}}^{n-4} \gamma_{a(n-3)})
  \over 
  \Gamma( \widetilde{\gamma}_{2(n-3)} + \widetilde{\gamma}_{(n-3)(n-1)} + \Delta_{\delta_{n-4}} + 2k_{n-4} - \sum_{\substack{a=1,a\neq 2}}^{n-4} \gamma_{a(n-3)}  )
 }
  \; \widetilde{L}_{n-2}\,,
}
\endgroup
where $\widetilde{L}_{n-2}$  is defined to be
\eqn{MnMinus2}{
\widetilde{L}_{n-2} &:=  \left( \prod_{r=1, r \neq 2}^{n-4} \Gamma(\widetilde{\gamma}_{r(n-2)} + \gamma_{r(n-2)}) \right) 
\left( \prod_{r=1, r \neq 2 }^{n-4} \Gamma(\widetilde{\gamma}_{r(n-3)} + \gamma_{r(n-3)} - \gamma_{r(n-2)} ) \right) 
\cr 
&\times 
\Gamma\left(\widetilde{\gamma}_{2(n-3)} + \widetilde{\gamma}_{(n-3)(n-1)} + \Delta_{(n-3)\delta_{n-4},(n-2)} + k_{n-4}   - \sum_{b=1,b\neq 2}^{n-4} ( \gamma_{b(n-3)} - \gamma_{b(n-2)}) \right) 
\cr 
&\times 
\Gamma\left( \Delta_{(n-2)\delta_{n-4}, (n-3)} + k_{n-4}  - \sum_{j=1,j\neq 2}^{n-4} \gamma_{j(n-2)}  \right).
}
$\widetilde{L}_{n-2}$ collects the Mellin variable-dependent integrand for the integrals over $\gamma_{i(n-2)}$ for $i=1,3,4,\ldots,n-4$.
Using~\eno{MultiFirstBarnes} we can easily evaluate these integrals in one go to obtain
\begingroup\makeatletter\def\f@size{11}\check@mathfonts
\eqn{}{
& \left( \prod_{r=1,r \neq 2}^{n-4} \int {d\gamma_{r(n-2)} \over 2\pi i} \right)  \, M_{n-2} 
\cr 
&= 
 \left( \prod_{i=1, i\neq 2}^{n-4} \Gamma(\widetilde{\gamma}_{i(n-2)} + \widetilde{\gamma}_{i(n-3)} + \gamma_{i(n-3)} ) \right)
 \Gamma\left( \widetilde{\gamma}_{2(n-3)} + \widetilde{\gamma}_{(n-3)(n-1)} + \Delta_{\delta_{n-4}} + 2k_{n-4} \right. \cr 
 &\quad \left. - \sum_{\substack{a=1,a\neq 2}}^{n-4} \gamma_{a(n-3)}  \right) 
 { \Gamma\left(\sum_{i=1,i\neq 2}^{n-4} \widetilde{\gamma}_{i(n-2)}  + \Delta_{(n-2)\delta_{n-4}, (n-3)} + k_{n-4}  \right)
   \over 
   \Gamma\left( \sum_{i=1}^{n-4} \widetilde{\gamma}_{i(n-3)}   + \widetilde{\gamma}_{(n-3)(n-1)} + \sum_{i=1,i\neq 2}^{n-4} \widetilde{\gamma}_{i(n-2)} + \Delta_{\delta_{n-4}} + 2k_{n-4}   \right )
   }
   \cr 
   &\times 
    \Gamma\left( \sum_{i=1,i\neq 2}^{n-4} \widetilde{\gamma}_{i(n-3)} + \widetilde{\gamma}_{2(n-3)}   + \widetilde{\gamma}_{(n-3)(n-1)} + \Delta_{(n-3)\delta_{n-4},(n-2)} + k_{n-4}   \right) .
}
\endgroup
Substituting these results back in $I$, we obtain
\begingroup\makeatletter\def\f@size{11}\check@mathfonts
\eqn{IOPEAfterTwo}{
I &=  \Gamma(\Delta_{12,\delta_1}-k_1) 
 \left(\prod_{j=1}^{{n\over 2}-3} \Gamma(\Delta_{(2j+1)(2j+2),\delta_{2j}}-k_{2j}) \right) 
\left(\prod_{(rs) \in {\cal V}_{\rm OPE}} \int  {d\widetilde{\gamma}_{rs} \over 2\pi i} \, \Gamma(-\widetilde{\gamma}_{rs}) (v_{rs}-1)^{\widetilde{\gamma}_{rs}} \right) 
\cr 
&\times
\left(\prod_{\substack{(rs) \in {\cal V}_{\rm OPE}\\s\neq n-1, n-2,n-3}} \int {d\gamma_{rs} \over 2\pi i} \,  \Gamma(\widetilde{\gamma}_{rs} + \gamma_{rs}) \right)
\left( \prod_{r=1, r\neq 2}^{n-4} \int {d \gamma_{r(n-3)} \over 2\pi i}\, \Gamma(\sum_{b=n-3}^{n-2}\widetilde{\gamma}_{rb}  + \gamma_{r(n-3)} ) \right)
 \cr 
 &\times 
 W_{n-3} W_{n-4}
\left(\prod_{j=1}^{{n\over 2}-3} \Gamma\left( \Delta_{\delta_{2j-1}\delta_{2j},\delta_{2j+1}} + k_{(2j-1)(2j),(2j+1)} - \sum_{\substack{(ab) \in {\cal V}_{\rm OPE} \\ a<b, b=2j+1 {\rm \ or\ }2j+2}} \gamma_{ab}\right) \right)
\cr 
&\times 
  \Gamma\left(\widetilde{\gamma}_{2(n-3)} + \Delta_{\delta_{n-5}\delta_{n-4},\delta_{n-3}} + k_{(n-5)(n-4),(n-3)} - \sum_{\substack{a=1,a\neq 2}}^{n-4} \gamma_{a(n-3)}\right)
 \cr 
 &\times 
\Gamma\left( \widetilde{\gamma}_{(n-3)(n-1)} + \Delta_{\delta_{n-3}\delta_{n-4},\delta_{n-5}} + k_{(n-3)(n-4),(n-5)}   \right) 
 \Gamma\left(\widetilde{\gamma}_{1(n-1)} + \Delta_{1\delta_1,2} + k_1 -\sum_{j=3}^{n-3} \gamma_{1j}\right)
\cr 
&\times
\Gamma\left(\widetilde{\gamma}_{2(n-1)} +  \Delta_{2\delta_{n-3},1\delta_2\delta_4\ldots \delta_{n-4}} + k_{(n-3),24\ldots(n-4)} + \sum_{j=3}^{n-3} \gamma_{1j} + \sum_{\substack{3\leq i<j\leq n-3\\ (ij) \in {\cal V}_{\rm OPE}}} \gamma_{ij} \right)    
\cr 
&\times 
 \left( \prod_{j=1}^{{n\over 2}-3}  \Gamma(\widetilde{\gamma}_{(2j+1)(n-1)} + \Delta_{(2j+1)\delta_{2j},(2j+2)} + k_{2j}   - \sum_{\substack{((2j+1)b) \in {\cal V}_{\rm OPE} \\ b\neq n-1,n-2}} \gamma_{(2j+1)b})   \right)
 \cr 
 &\times 
 \left(\prod_{j=1}^{{n\over 2}-3} \Gamma\left( \widetilde{\gamma}_{(2j+2)(n-1)} + \Delta_{(2j+2)\delta_{2j+1},(2j+1)\delta_{2j-1}} + k_{2j+1,2j-1}  +\!\!\! \sum_{\substack{(a(2j+1)) \in {\cal V}_{\rm OPE} \\ a<2j+1}} \gamma_{a(2j+1)}   \right. \right. \cr 
 &\qquad \left. \left. -\!\sum_{\substack{((2j+2)b) \in {\cal V}_{\rm OPE}\\ b> 2j+2, b \neq n-1,n-2}} \gamma_{(2j+2)b} \right) \right),
}
\endgroup
where we have defined
 \begingroup\makeatletter\def\f@size{10}\check@mathfonts
\eqn{WnMinus4}{
W_{n-4} &:=
 { \Gamma(\Delta_{(n-3)(n-2),\delta_{n-4}}-k_{n-4}) 
  \Gamma\left( \displaystyle{\sum_{(i(n-2)) \in {\cal V}_{\rm OPE}}} \widetilde{\gamma}_{i(n-2)}  + \Delta_{(n-2)\delta_{n-4}, (n-3)} + k_{n-4}  \right) 
   \over 
   \Gamma\left(\displaystyle{\sum_{(i(n-3)) \in {\cal V}_{\rm OPE}}} \widetilde{\gamma}_{i(n-3)}   + \displaystyle{\sum_{(i(n-2)) \in {\cal V}_{\rm OPE}}} \widetilde{\gamma}_{i(n-2)} + \Delta_{\delta_{n-4}} + 2k_{n-4}    \right)
   }
   \cr 
   &\times 
    \Gamma\left(\displaystyle{\sum_{(i(n-3)) \in {\cal V}_{\rm OPE}}} \widetilde{\gamma}_{i(n-3)}   + \Delta_{(n-3)\delta_{n-4},(n-2)} + k_{n-4}  \right) .
 }
 \endgroup

\subsubsection{Integrals over a green-colored chain}
\label{APP:OPESECOND}

First, integrating $L_{n-2K-1}$, defined in~\eno{LnMinus2KMinus1}, over $\gamma_{1(n-2K-1)}$ we get
\begingroup\makeatletter\def\f@size{11}\check@mathfonts
\eqn{}{
&\int {d\gamma_{1(n-2K-1)} \over 2\pi i}\, L_{n-2K-1} \cr 
&= 
\Gamma\left(\widetilde{\gamma}_{1(n-2K)} + \widetilde{\gamma}_{1(n-2K-1)} + \sum_{j=n-2K+1}^{n-1} \widetilde{\gamma}_{1j}  +  \Delta_{1\delta_1,2} + k_1 -\sum_{j=3}^{n-2K-2} \gamma_{1j}\right) 
\cr
&\times 
\Gamma\left(\sum_{i=0}^{K-1} \widetilde{\gamma}_{2(n-2i-1)} + \widetilde{\gamma}_{2(n-2K-1)} +  \Delta_{2\delta_{n-2K-3},1\delta_2\delta_4\ldots \delta_{n-2K-4}} + k_{(n-2K-3),24\ldots(n-2K-4)}  \right. 
\cr 
&\quad \left. + \sum_{j=3}^{n-2K-2} \gamma_{1j} + \sum_{\substack{3\leq i<j\leq n-2K-2\\ (ij) \in {\cal V}_{\rm OPE} }} \gamma_{ij}     \right) 
\cr 
&\times 
\Gamma\left(\sum_{i=0}^{K} \widetilde{\gamma}_{2(n-2i-1)}  +   \sum_{j=n-2K-1}^{n-1} \widetilde{\gamma}_{1j}  +  \Delta_{\delta_{n-2K-3}\delta_1,\delta_2\delta_4\ldots \delta_{n-2K-4}}  + k_{(n-2K-3)1,24\ldots(n-2K-4)}   \right. 
\cr 
&\quad \left. + \sum_{\substack{3\leq i<j\leq n-2K-2\\ (ij) \in {\cal V}_{\rm OPE} }} \gamma_{ij}  \right)^{-1} 
\widetilde{L}_{n-2K-1} \,,
}
\endgroup
where 
\begingroup\makeatletter\def\f@size{11}\check@mathfonts
\eqn{}{
\widetilde{L}_{n-2K-1} &:= \left(\prod_{r=3}^{n-2K-2}  \Gamma(\widetilde{\gamma}_{r(n-2K)} + \widetilde{\gamma}_{r(n-2K-1)} + \gamma_{r(n-2K-1)}) \right) 
\cr 
&\times 
\left(\prod_{j=1}^{{n\over 2}-K-2} \Gamma\left( \sum_{a=n-2K+1}^{n-1} \widetilde{\gamma}_{(2j+1)a} + \Delta_{(2j+1)\delta_{2j},(2j+2)} + k_{2j} - \sum_{\substack{((2j+1)b) \in {\cal V}_{\rm OPE}\\ b\neq n-2K,\ldots, n-1}}  \gamma_{(2j+1)b} \right) \right)
\cr 
&\times 
\left(\prod_{j=1}^{{n\over 2}-K-2} \Gamma\left( \sum_{a=n-2K+1}^{n-1} \widetilde{\gamma}_{(2j+2)a} + \Delta_{(2j+2)\delta_{2j+1},(2j+1)\delta_{2j-1}} + k_{2j+1,2j-1} \right. \right. 
\cr 
&\quad 
 \left. \left.  + \sum_{\substack{(b(2j+1)) \in {\cal V}_{\rm OPE}\\ b<2j+1}} \gamma_{b(2j+1)} - \sum_{\substack{((2j+2)b) \in {\cal V}_{\rm OPE}\\ b>2j+2, b\neq n-2K,\ldots, n-1}}  \gamma_{(2j+2)b}  \right)  \right)
 \cr 
&\times 
\Gamma\left(\sum_{i=0}^{K-1} \widetilde{\gamma}_{2(n-2i-1)} + \sum_{j=n-2K+1}^{n-1} \widetilde{\gamma}_{1j}  +  \Delta_{\delta_{n-2K-1}\delta_1,\delta_2\delta_4\ldots \delta_{n-2K-2}} + k_{(n-2K-1)1,24\ldots(n-2K-2)} \right. 
\cr 
&\quad \left. + \sum_{\substack{3\leq i<j\leq n-2K-1\\ (ij) \in {\cal V}_{\rm OPE} }} \gamma_{ij}    \right) 
\cr
&\times
\Gamma\left(\widetilde{\gamma}_{1(n-2K)} + \widetilde{\gamma}_{1(n-2K-1)} + \widetilde{\gamma}_{2(n-2K-1)} +  \Delta_{\delta_{n-2K-3}\delta_{n-2K-2},\delta_{n-2K-1}} \right. \cr 
&\quad \left. + k_{(n-2K-3)(n-2K-2),(n-2K-1)}  - \sum_{a=3}^{n-2K-2} \gamma_{a(n-2K-1)}\right) .
 }
 \endgroup
Next, we will integrate $\widetilde{L}_{n-2K-1}$ over the Mellin variables $\gamma_{3(n-2K-1)}, \gamma_{4(n-2K-1)},\ldots, \gamma_{(n-2K-2)(n-2K-1)}$ using the inductive Barnes lemma~\eno{MultiFirstBarnes}.
The integrand $\widetilde{L}_{n-2K-1}$ matches the integrand of~\eno{MultiFirstBarnes} with the coefficient assignments:
 \eqn{}{
 A_r &=  \widetilde{\gamma}_{(r+2)(n-2K)} + \widetilde{\gamma}_{(r+2)(n-2K-1)} 
 \cr 
 B_r &= \begin{cases} 
         \displaystyle{\sum_{a=n-2K+1}^{n-1}} \widetilde{\gamma}_{(r+2)a} + \Delta_{(r+2)\delta_{r+1},(r+3)} + k_{r+1} - \sum_{\substack{((r+2)b) \in {\cal V}_{\rm OPE}\\ b\neq n-2K-1,\ldots, n-1}}  \gamma_{(r+2)b}  & r {\rm \ odd} 
         \cr 
         \displaystyle{\sum_{a=n-2K+1}^{n-1}} \widetilde{\gamma}_{(r+2)a} + \Delta_{(r+2)\delta_{r+1},(r+1)\delta_{r-1}} + k_{r+1,r-1} + \sum_{\substack{(b(r+1)) \in {\cal V}_{\rm OPE}\\ b<r+1}} \gamma_{b(r+1)} \cr 
         \quad - \displaystyle{\sum_{\substack{((r+2)b) \in {\cal V}_{\rm OPE}\\ b>r+2, b\neq n-2K-1,\ldots, n-1}}}  \gamma_{(r+2)b}  & r {\rm \ even}
         \end{cases} 
         \cr 
 C &=  \sum_{i=0}^{K-1} \widetilde{\gamma}_{2(n-2i-1)} + \sum_{j=n-2K+1}^{n-1} \widetilde{\gamma}_{1j}  +  \Delta_{\delta_{n-2K-1}\delta_1,\delta_2\delta_4\ldots \delta_{n-2K-2}} + k_{(n-2K-1)1,24\ldots(n-2K-2)} 
\cr 
&\quad  + \sum_{\substack{3\leq i<j\leq n-2K-2\\ (ij) \in {\cal V}_{\rm OPE} }} \gamma_{ij}  
\cr 
 D &= \widetilde{\gamma}_{1(n-2K)} + \widetilde{\gamma}_{1(n-2K-1)} + \widetilde{\gamma}_{2(n-2K-1)} +  \Delta_{\delta_{n-2K-3}\delta_{n-2K-2},\delta_{n-2K-1}} \cr 
 & \quad + k_{(n-2K-3)(n-2K-2),(n-2K-1)} \,,
 }
for $1 \leq r \leq n-2K-4$. Then applying~\eno{MultiFirstBarnes}, we obtain
\begingroup\makeatletter\def\f@size{10}\check@mathfonts
\eqn{}{
&\left(\prod_{r=1}^{n-2K-4} \int {d\gamma_{(r+2)(n-2K-1)} \over 2\pi i} \right)  \widetilde{L}_{n-2K-1} \cr 
&={
\Gamma\left( \displaystyle{\sum_{i=0}^{K}} \widetilde{\gamma}_{2(n-2i-1)} + \sum_{j=n-2K-1}^{n-1} \widetilde{\gamma}_{1j}  +  \Delta_{\delta_{n-2K-3}\delta_1,\delta_2\delta_4\ldots \delta_{n-2K-4}} + k_{(n-2K-3)1,24\ldots(n-2K-4)}  + \sum_{\substack{3\leq i<j\leq n-2K-2\\ (ij) \in {\cal V}_{\rm OPE} }} \gamma_{ij}   \right)
 \over 
 \Gamma\left( \sum_{i=1,i\neq 2}^{n-2K-2} \sum_{j=n-2K-1}^{n-1} \widetilde{\gamma}_{ij}  + \sum_{i=0}^{K} \widetilde{\gamma}_{2(n-2i-1)} + \Delta_{\delta_{n-2K-3}} + 2k_{n-2K-3} \right) }
\cr 
&\times 
 \left(\prod_{j=1}^{{n\over 2}-K-2} \Gamma\left(  \displaystyle{\sum_{a=n-2K-1}^{n-1}} \widetilde{\gamma}_{(2j+1)a} + \Delta_{(2j+1)\delta_{2j},(2j+2)} + k_{2j}  - \sum_{\substack{((2j+1)b) \in {\cal V}_{\rm OPE}\\ b\neq n-2K-1,\ldots, n-1}}  \gamma_{(2j+1)b} \right) \right) 
 \cr 
&\times 
 \left(\prod_{j=1}^{{n\over 2}-K-2} \Gamma( \displaystyle{\sum_{a=n-2K-1}^{n-1}} \widetilde{\gamma}_{(2j+2)a} + \Delta_{(2j+2)\delta_{2j+1},(2j+1)\delta_{2j-1}} + k_{2j+1,2j-1}  \right. \cr 
         &\quad \left.+ \sum_{\substack{(b(2j+1)) \in {\cal V}_{\rm OPE}\\ b<2j+1}} \gamma_{b(2j+1)}  - \displaystyle{\sum_{\substack{((2j+2)b) \in {\cal V}_{\rm OPE}\\ b>2j+2, b\neq n-2K-1,\ldots, n-1}}}  \gamma_{(2j+2)b}) \right) 
\cr 
&\times 
\Gamma\left(\sum_{i=1, i \neq 2}^{n-2K-2}  \widetilde{\gamma}_{i(n-2K)} + \sum_{i=1}^{n-2K-2}  \widetilde{\gamma}_{i(n-2K-1)}   +  \Delta_{\delta_{n-2K-3}\delta_{n-2K-2},\delta_{n-2K-1}} + k_{(n-2K-3)(n-2K-2),(n-2K-1)}  \right) 
\cr 
&\times 
\Gamma\left( \sum_{i=1,i\neq 2}^{n-2K-2} \sum_{j=n-2K+1}^{n-1} \widetilde{\gamma}_{ij} + \sum_{i=0}^{K-1} \widetilde{\gamma}_{2(n-2i-1)}  +  \Delta_{\delta_{n-2K-3}\delta_{n-2K-1},\delta_{n-2K-2}}  + k_{(n-2K-3)(n-2K-1),(n-2K-2)}  \right) 
 }
 \endgroup
 where we used
 \eqn{}{ 
 \sum_{r=1}^{n-2K-4} B_r &= \sum_{i=3}^{n-2K-2} \sum_{j=n-2K+1}^{n-1} \widetilde{\gamma}_{ij} + \Delta_{\delta_2\delta_4\delta_6\ldots\delta_{n-2K-4}\delta_{n-2K-3},\delta_1} + k_{246\ldots(n-2K-4)(n-2K-3),1}  \cr 
 &\quad - \sum_{\substack{3 \leq a < b \leq n-2K-2\\ (ab) \in {\cal V}_{\rm OPE}}} \gamma_{ab} \,.
 }
Substituting the results of these integrations back into $\widehat{I}_{n-2K-1}$ in~\eno{IhatDef}, we get
\begingroup\makeatletter\def\f@size{9.5}\check@mathfonts
\eqn{IhatAgainOne}{
\widehat{I}_{n-2K-1} &=  \Gamma(\Delta_{12,\delta_1}-k_1) 
 \left(\prod_{j=1}^{{n\over 2}-K-2} \Gamma(\Delta_{(2j+1)(2j+2),\delta_{2j}}-k_{2j}) \right) 
\left(\prod_{(rs) \in {\cal V}_{\rm OPE}} \int  {d\widetilde{\gamma}_{rs} \over 2\pi i} \, \Gamma(-\widetilde{\gamma}_{rs}) (v_{rs}-1)^{\widetilde{\gamma}_{rs}} \right) 
\cr 
&\times
\left( \prod_{j=n-2K-3}^{n-3} W_j  \right)
\left(\prod_{\substack{(rs) \in {\cal V}_{\rm OPE}\\s\neq n-2K-1, \ldots ,n-1}} \int {d\gamma_{rs} \over 2\pi i} \,  \Gamma(\widetilde{\gamma}_{rs} + \gamma_{rs}) \right)
 \cr 
 &\times 
\left(\prod_{j=1}^{{n\over 2}-K-2} \Gamma\left( \Delta_{\delta_{2j-1}\delta_{2j},\delta_{2j+1}} + k_{(2j-1)(2j),(2j+1)} - \sum_{\substack{(ab) \in {\cal V}_{\rm OPE} \\ a<b, b=2j+1 {\rm \ or\ }2j+2}} \gamma_{ab}\right) \right)
\cr 
&\times 
\Gamma\left(\widetilde{\gamma}_{1(n-2K)} + \widetilde{\gamma}_{1(n-2K-1)} + \sum_{j=n-2K+1}^{n-1} \widetilde{\gamma}_{1j}  +  \Delta_{1\delta_1,2} + k_1 -\sum_{j=3}^{n-2K-2} \gamma_{1j}\right) 
\cr
&\times 
\Gamma\left(\sum_{i=0}^{K-1} \widetilde{\gamma}_{2(n-2i-1)} + \widetilde{\gamma}_{2(n-2K-1)} +  \Delta_{2\delta_{n-2K-3},1\delta_2\delta_4\ldots \delta_{n-2K-4}} + k_{(n-2K-3),24\ldots(n-2K-4)} \right. 
\cr 
&\quad \left.  + \sum_{j=3}^{n-2K-2} \gamma_{1j} + \sum_{\substack{3\leq i<j\leq n-2K-2\\ (ij) \in {\cal V}_{\rm OPE} }} \gamma_{ij}     \right) 
\cr 
&\times 
 \left(\prod_{j=1}^{{n\over 2}-K-2} \Gamma\left(  \displaystyle{\sum_{a=n-2K-1}^{n-1}} \widetilde{\gamma}_{(2j+1)a} + \Delta_{(2j+1)\delta_{2j},(2j+2)} + k_{2j}  - \sum_{\substack{((2j+1)b) \in {\cal V}_{\rm OPE}\\ b\neq n-2K-1,\ldots, n-1}}  \gamma_{(2j+1)b} \right) \right) 
 \cr 
&\times 
 \left(\prod_{j=1}^{{n\over 2}-K-2} \Gamma( \displaystyle{\sum_{a=n-2K-1}^{n-1}} \widetilde{\gamma}_{(2j+2)a} + \Delta_{(2j+2)\delta_{2j+1},(2j+1)\delta_{2j-1}} + k_{2j+1,2j-1}  \right. \cr 
         &\quad \left.+ \sum_{\substack{(b(2j+1)) \in {\cal V}_{\rm OPE}\\ b<2j+1}} \gamma_{b(2j+1)}  - \displaystyle{\sum_{\substack{((2j+2)b) \in {\cal V}_{\rm OPE}\\ b>2j+2, b\neq n-2K-1,\ldots, n-1}}}  \gamma_{(2j+2)b}) \right),
}
\endgroup
where 
\begingroup\makeatletter\def\f@size{9}\check@mathfonts
\eqn{}{
& W_{n-2K-3} \cr 
&:= \Gamma\left(\sum_{i=1, i \neq 2}^{n-2K-2}  \widetilde{\gamma}_{i(n-2K)} + \sum_{i=1}^{n-2K-2}  \widetilde{\gamma}_{i(n-2K-1)}   +  \Delta_{\delta_{n-2K-3}\delta_{n-2K-2},\delta_{n-2K-1}} + k_{(n-2K-3)(n-2K-2),(n-2K-1)}  \right) 
\cr 
&\times 
\Gamma\left( \sum_{i=1,i\neq 2}^{n-2K-2} \sum_{j=n-2K+1}^{n-1} \widetilde{\gamma}_{ij} + \sum_{i=0}^{K-1} \widetilde{\gamma}_{2(n-2i-1)}  +  \Delta_{\delta_{n-2K-3}\delta_{n-2K-1},\delta_{n-2K-2}}  + k_{(n-2K-3)(n-2K-1),(n-2K-2)}  \right) 
\cr 
&\times
{
\Gamma\left(\displaystyle{\sum_{a=n-2K+1}^{n-1}} \widetilde{\gamma}_{(n-2K)a} +   \displaystyle{\sum_{a=n-2K+1}^{n-1}} \widetilde{\gamma}_{(n-2K-1
)a} + \Delta_{\delta_{n-2K-1}\delta_{n-2K-2},\delta_{n-2K-3}} + k_{(n-2K-1)(n-2K-2),(n-2K-3)}     \right) 
 \over 
 \Gamma\left( \sum_{i=1,i\neq 2}^{n-2K-2} \sum_{j=n-2K-1}^{n-1} \widetilde{\gamma}_{ij}  + \sum_{i=0}^{K} \widetilde{\gamma}_{2(n-2i-1)} + \Delta_{\delta_{n-2K-3}} + 2k_{n-2K-3} \right) },
}
\endgroup
which agrees with $W_j$ in~\eno{Wj} after setting $j=n-2K-3$ (i.e.\ for odd $j$).

\subsubsection{Integrals over a magenta-colored chain}
\label{APP:OPETHIRD}

Integrating~\eno{LnMinus2KMinus2} over $\gamma_{2(n-2K-3)}$ using the first Barnes lemma~\eno{FirstBarnes}, and shifting variables $\gamma_{i(n-2K-3)} \to \gamma_{i(n-2K-3)} - \gamma_{i(n-2K-2)}$ for $i=1,3,4,\ldots,n-2K-4$, we get
\begingroup\makeatletter\def\f@size{10}\check@mathfonts
\eqn{}{
& \int {d \gamma_{2(n-2K-3)} \over 2\pi i}\, L_{n-2K-2}  = 
\left(\prod_{j=1}^{{n\over 2}-K-3} \Gamma\left( \Delta_{\delta_{2j-1}\delta_{2j},\delta_{2j+1}} + k_{(2j-1)(2j),(2j+1)} - \sum_{\substack{(ab) \in {\cal V}_{\rm OPE} \\ a<b, b=2j+1 {\rm \ or\ }2j+2}} \gamma_{ab}\right) \right)
\cr 
&\times 
\Gamma\left(\widetilde{\gamma}_{1(n-2K)} + \widetilde{\gamma}_{1(n-2K-1)} + \sum_{j=n-2K+1}^{n-1} \widetilde{\gamma}_{1j}  +  \Delta_{1\delta_1,2} + k_1 -\sum_{j=3}^{n-2K-3} \gamma_{1j}\right) 
\cr
&\times 
\Gamma\left(\sum_{i=0}^{K-1} \widetilde{\gamma}_{2(n-2i-1)} + \widetilde{\gamma}_{2(n-2K-1)} +  \Delta_{2\delta_{n-2K-3},1\delta_2\delta_4\ldots \delta_{n-2K-4}} + k_{(n-2K-3),24\ldots(n-2K-4)} + \sum_{j=3}^{n-2K-3} \gamma_{1j} \right. 
\cr 
&\qquad \left. + \sum_{\substack{3\leq i<j\leq n-2K-3\\ (ij) \in {\cal V}_{\rm OPE} }} \gamma_{ij}     \right) 
\widetilde{L}_{n-2K-2}
\cr 
&\times 
 \left(\prod_{j=1}^{{n\over 2}-K-3} \Gamma\left(  \displaystyle{\sum_{a=n-2K-1}^{n-1}} \widetilde{\gamma}_{(2j+1)a} + \Delta_{(2j+1)\delta_{2j},(2j+2)} + k_{2j}  - \sum_{\substack{((2j+1)b) \in {\cal V}_{\rm OPE}\\ b\neq n-2K-2,\ldots, n-1}}  \gamma_{(2j+1)b} \right) \right) 
 \cr 
&\times 
 \left(\prod_{j=1}^{{n\over 2}-K-3} \Gamma\left( \displaystyle{\sum_{a=n-2K-1}^{n-1}} \widetilde{\gamma}_{(2j+2)a} + \Delta_{(2j+2)\delta_{2j+1},(2j+1)\delta_{2j-1}} + k_{2j+1,2j-1} + \sum_{\substack{(b(2j+1)) \in {\cal V}_{\rm OPE}\\ b<2j+1}} \gamma_{b(2j+1)}  \right. \right. \cr 
         &\qquad \left. \left.  - \displaystyle{\sum_{\substack{((2j+2)b) \in {\cal V}_{\rm OPE}\\ b>2j+2, b\neq n-2K-2,\ldots, n-1}}}  \gamma_{(2j+2)b}) \right)\right)
\cr 
&\times 
 \Gamma\left(\widetilde{\gamma}_{2(n-2K-3)} +  \Delta_{\delta_{n-2K-5}\delta_{n-2K-4},\delta_{n-2K-3}} + k_{(n-2K-5)(n-2K-4),(n-2K-3)} - \sum_{a=1,a\neq 2}^{n-2K-4} \gamma_{a(n-2K-3)}\right) 
\cr 
&\times 
 \Gamma\left(  \sum_{\substack{((n-2K-2)a) \in {\cal V}_{\rm OPE}\\ a>n-2K-2}} \!\!\!\!\!\! \widetilde{\gamma}_{(n-2K-2)a} +  \!\!\! \sum_{\substack{((n-2K-3)a) \in {\cal V}_{\rm OPE}\\ a>n-2K-2}} \!\!\!\!\!\!\! \widetilde{\gamma}_{(n-2K-3)a}  + \Delta_{\delta_{n-2K-3}\delta_{n-2K-4},\delta_{n-2K-5}} \right. \cr 
 &\qquad + k_{(n-2K-3)(n-2K-4),(n-2K-5)}     \Bigg)
 \cr 
 &\times
 \Gamma\left( \widetilde{\gamma}_{2(n-2K-3)} +   \sum_{\substack{((n-2K-2)a) \in {\cal V}_{\rm OPE}\\ a>n-2K-2}} \!\!\!\!\!\! \widetilde{\gamma}_{(n-2K-2)a} +  \!\!\! \sum_{\substack{((n-2K-3)a) \in {\cal V}_{\rm OPE}\\ a>n-2K-2}} \!\!\!\!\!\!\! \widetilde{\gamma}_{(n-2K-3)a}   + \Delta_{\delta_{n-2K-4}} + 2k_{n-2K-4}  \right. \cr 
 &\qquad \left. - \sum_{a=1,a\neq 2}^{n-2K-4} \gamma_{a(n-2K-3)} \right)^{-1},
 }
\endgroup
where  we have defined 
\begingroup\makeatletter\def\f@size{11}\check@mathfonts
\eqn{}{
& \widetilde{L}_{n-2K-2} \cr 
&:= \left(\prod_{r=1,r\neq 2}^{n-2K-4}   \Gamma(\widetilde{\gamma}_{r(n-2K-2)} + \gamma_{r(n-2K-2)}) \right)
\left(\prod_{r=1,r\neq 2}^{n-2K-4}   \Gamma(\widetilde{\gamma}_{r(n-2K-3)} + \gamma_{r(n-2K-3)} - \gamma_{r(n-2K-2)} ) \right)
\cr 
&  \times 
\Gamma\left(\widetilde{\gamma}_{2(n-2K-3)} +  \displaystyle{\sum_{a=n-2K-1}^{n-1}} \widetilde{\gamma}_{(n-2K-3)a} + \Delta_{(n-2K-3)\delta_{n-2K-4},(n-2K-2)} + k_{n-2K-4}   \right. \cr 
&\left. \qquad - \sum_{a=1,a\neq 2}^{n-2K-4}  \gamma_{a(n-2K-3)} + \sum_{a=1,a\neq 2}^{n-2K-4}  \gamma_{a(n-2K-2)} \right) 
 \cr 
 &\times 
 \Gamma\left(    \sum_{\substack{((n-2K-2)a) \in {\cal V}_{\rm OPE}\\ a>n-2K-2}} \!\!\!\!\!\!\!\!\! \widetilde{\gamma}_{(n-2K-2)a}  + \Delta_{(n-2K-2)\delta_{n-2K-4},(n-2K-3)} + k_{n-2K-4}  - \sum_{a=1,a\neq 2}^{n-2K-4} \gamma_{a(n-2K-2)}  \right).
}
\endgroup
Now using~\eno{MultiFirstBarnes}, we can evaluate the integral $\left( \prod_{r=1,r\neq 2}^{n-2K-4}\int {d \gamma_{r(n-2K-2)} \over 2\pi i} \right) \widetilde{L}_{n-2K-2}$, to get
\begingroup\makeatletter\def\f@size{10}\check@mathfonts
\eqn{}{
& \left( \prod_{r=1,r\neq 2}^{n-2K-4}\int {d \gamma_{r(n-2K-2)} \over 2\pi i} \right) \widetilde{L}_{n-2K-2} \cr  
&= {\left(\prod_{r=1,r\neq 2}^{n-2K-4} \Gamma(\widetilde{\gamma}_{r(n-2K-2)} + \widetilde{\gamma}_{r(n-2K-3)} + \gamma_{r(n-2K-3)}) \right) 
  \over 
 \Gamma\left( \displaystyle{\sum_{(r(n-2K-2)) \in {\cal V}_{\rm OPE} }} \widetilde{\gamma}_{r(n-2K-2)} + \sum_{(r(n-2K-3)) \in {\cal V}_{\rm OPE} } \widetilde{\gamma}_{r(n-2K-3)}  + \Delta_{\delta_{n-2K-4}} + 2k_{n-2K-4}    \right) } 
\cr 
&\times 
 \Gamma\left(\widetilde{\gamma}_{2(n-2K-3)} +   \sum_{\substack{((n-2K-2)a) \in {\cal V}_{\rm OPE}\\ a>n-2K-2}} \!\!\!\!\!\! \widetilde{\gamma}_{(n-2K-2)a} +  \!\!\! \sum_{\substack{((n-2K-3)a) \in {\cal V}_{\rm OPE}\\ a>n-2K-2}} \!\!\!\!\!\!\! \widetilde{\gamma}_{(n-2K-3)a} + \Delta_{\delta_{n-2K-4}} + 2k_{n-2K-4} \right. \cr 
&\qquad \left. - \sum_{a=1,a\neq 2}^{n-2K-4}  \gamma_{a(n-2K-3)} \right) 
\Gamma\left(\sum_{(r(n-2K-2)) \in {\cal V}_{\rm OPE} } \widetilde{\gamma}_{r(n-2K-2)}  + \Delta_{(n-2K-2)\delta_{n-2K-4},(n-2K-3)} + k_{n-2K-4}   \right)
\cr 
&\times 
 \Gamma\left(\sum_{(r(n-2K-3)) \in {\cal V}_{\rm OPE} } \widetilde{\gamma}_{r(n-2K-3)}  + \Delta_{(n-2K-3)\delta_{n-2K-4},(n-2K-2)}  + k_{n-2K-4} \right) 
\,.
}
\endgroup
Putting the results of these integrations back into $\widehat{I}_{n-2K-1}$ in~\eno{IhatTwo}, we get 
\begingroup\makeatletter\def\f@size{11}\check@mathfonts
\eqn{IhatThree}{
& \widehat{I}_{n-2K-1} \cr 
&=  \Gamma(\Delta_{12,\delta_1}-k_1) 
 \left(\prod_{j=1}^{{n\over 2}-K-3} \Gamma(\Delta_{(2j+1)(2j+2),\delta_{2j}}-k_{2j}) \right) 
\left(\prod_{(rs) \in {\cal V}_{\rm OPE}} \int  {d\widetilde{\gamma}_{rs} \over 2\pi i} \, \Gamma(-\widetilde{\gamma}_{rs}) (v_{rs}-1)^{\widetilde{\gamma}_{rs}} \right) 
\cr 
&\times
\left( \prod_{j=n-2K-4}^{n-3} W_j  \right)
\left(\prod_{\substack{(rs) \in {\cal V}_{\rm OPE}\\s\neq n-2K-3, \ldots ,n-1}} \int {d\gamma_{rs} \over 2\pi i} \,  \Gamma(\widetilde{\gamma}_{rs} + \gamma_{rs}) \right)
\left(\prod_{r=1, r\neq 2}^{n-2K-4} \int {d\gamma_{r(n-2K-3)} \over 2\pi i}  \right) 
\cr &\times 
\left(\prod_{j=1}^{{n\over 2}-K-3} \Gamma\left( \Delta_{\delta_{2j-1}\delta_{2j},\delta_{2j+1}} + k_{(2j-1)(2j),(2j+1)} - \sum_{\substack{(ab) \in {\cal V}_{\rm OPE} \\ a<b, b=2j+1 {\rm \ or\ }2j+2}} \gamma_{ab}\right) \right)
\cr 
&\times 
\Gamma\left(\widetilde{\gamma}_{1(n-2K)} + \widetilde{\gamma}_{1(n-2K-1)} + \sum_{j=n-2K+1}^{n-1} \widetilde{\gamma}_{1j}  +  \Delta_{1\delta_1,2} + k_1 -\sum_{j=3}^{n-2K-3} \gamma_{1j}\right) 
\cr
&\times 
\Gamma\left(\sum_{i=0}^{K-1} \widetilde{\gamma}_{2(n-2i-1)} + \widetilde{\gamma}_{2(n-2K-1)} +  \Delta_{2\delta_{n-2K-3},1\delta_2\delta_4\ldots \delta_{n-2K-4}} + k_{(n-2K-3),24\ldots(n-2K-4)} \right. 
\cr 
&\quad \left. + \sum_{j=3}^{n-2K-3} \gamma_{1j}  + \sum_{\substack{3\leq i<j\leq n-2K-3\\ (ij) \in {\cal V}_{\rm OPE} }} \gamma_{ij}     \right) 
\cr 
&\times 
 \left(\prod_{j=1}^{{n\over 2}-K-3} \Gamma\left(  \displaystyle{\sum_{a=n-2K-1}^{n-1}} \widetilde{\gamma}_{(2j+1)a} + \Delta_{(2j+1)\delta_{2j},(2j+2)} + k_{2j}  - \sum_{\substack{((2j+1)b) \in {\cal V}_{\rm OPE}\\ b\neq n-2K-2,\ldots, n-1}}  \gamma_{(2j+1)b} \right) \right) 
 \cr 
&\times 
 \left(\prod_{j=1}^{{n\over 2}-K-3} \Gamma\left( \displaystyle{\sum_{a=n-2K-1}^{n-1}} \widetilde{\gamma}_{(2j+2)a} + \Delta_{(2j+2)\delta_{2j+1},(2j+1)\delta_{2j-1}} + k_{2j+1,2j-1} + \sum_{\substack{(b(2j+1)) \in {\cal V}_{\rm OPE}\\ b<2j+1}} \gamma_{b(2j+1)}  \right. \right. \cr 
 &\quad \left. \left.  - \displaystyle{\sum_{\substack{((2j+2)b) \in {\cal V}_{\rm OPE}\\ b>2j+2, b\neq n-2K-2,\ldots, n-1}}}  \gamma_{(2j+2)b} \right)\right)
\cr 
&\times 
 \Gamma\left(\widetilde{\gamma}_{2(n-2K-3)} +  \Delta_{\delta_{n-2K-5}\delta_{n-2K-4},\delta_{n-2K-3}} + k_{(n-2K-5)(n-2K-4),(n-2K-3)} - \sum_{a=1,a\neq 2}^{n-2K-4} \gamma_{a(n-2K-3)}\right) 
\cr 
&\times 
 \Gamma\left(  \sum_{\substack{((n-2K-2)a) \in {\cal V}_{\rm OPE}\\ a>n-2K-2}} \!\!\!\!\!\! \widetilde{\gamma}_{(n-2K-2)a} +  \!\!\! \sum_{\substack{((n-2K-3)a) \in {\cal V}_{\rm OPE}\\ a>n-2K-2}} \!\!\!\!\!\!\! \widetilde{\gamma}_{(n-2K-3)a} + \Delta_{\delta_{n-2K-3}\delta_{n-2K-4},\delta_{n-2K-5}} \right. \cr  
 &\quad  + k_{(n-2K-3)(n-2K-4),(n-2K-5)}     \Bigg)
\left(\prod_{r=1,r\neq 2}^{n-2K-4} \Gamma(\widetilde{\gamma}_{r(n-2K-2)} + \widetilde{\gamma}_{r(n-2K-3)} + \gamma_{r(n-2K-3)}) \right) ,
}
\endgroup
where
\eqn{WnMinus2KMinus4}{
& W_{n-2K-4} \cr 
&:= { \Gamma(\Delta_{(n-2K-3)(n-2K-2),\delta_{n-2K-4}}-k_{n-2K-4})
  \over 
 \Gamma\left( \displaystyle{\sum_{(r(n-2K-2)) \in {\cal V}_{\rm OPE} }} \widetilde{\gamma}_{r(n-2K-2)} + \sum_{(r(n-2K-3)) \in {\cal V}_{\rm OPE} } \widetilde{\gamma}_{r(n-2K-3)}  + \Delta_{\delta_{n-2K-4}} + 2k_{n-2K-4}    \right) } 
\cr 
&\times 
\Gamma\left(\sum_{(r(n-2K-2)) \in {\cal V}_{\rm OPE} } \widetilde{\gamma}_{r(n-2K-2)}  + \Delta_{(n-2K-2)\delta_{n-2K-4},(n-2K-3)} + k_{n-2K-4}   \right)
\cr 
&\times 
 \Gamma\left(\sum_{(r(n-2K-3)) \in {\cal V}_{\rm OPE} } \widetilde{\gamma}_{r(n-2K-3)}  + \Delta_{(n-2K-3)\delta_{n-2K-4},(n-2K-2)}  + k_{n-2K-4} \right) ,
}
which agrees with $W_j$ in~\eno{Wj} if we set $j=n-2K-4$ (i.e.\ for even $j$).

\bibliographystyle{ssg}
\bibliography{main}

\end{document}